\documentclass[paperpaper,superscriptaddress,english,floatfix, showpacs, amsfonts, amssymb]{revtex4}
\usepackage{epsfig,graphicx,psfrag,amsmath,amssymb,float}
\usepackage[T1]{fontenc}
\usepackage[latin9]{inputenc}
\usepackage{amsmath}
\usepackage{amssymb}
\usepackage{bbold}
\usepackage{amscd}
\usepackage{bm}
\usepackage{psfrag}
\usepackage{epsfig}
\usepackage{graphicx}
\usepackage{dcolumn}
\usepackage{bm}
\usepackage{subfigure}

\usepackage{babel}

\begin{document}
     
\title{The role of $q$-spin singlet pairs of physical spins in the dynamical properties of the spin-$1/2$ Heisenberg-Ising $XXZ$ chain}
\author{Jos\'e M. P. Carmelo}
\affiliation{Center of Physics of University of Minho and University of Porto, P-4169-007 Oporto, Portugal}
\affiliation{Department of Physics, University of Minho, Campus Gualtar, P-4710-057 Braga, Portugal}
\affiliation{Boston University, Department of Physics, 590 Commonwealth Ave, Boston, MA 02215, USA}
\author{Pedro D. Sacramento}
\affiliation{CeFEMA, Instituto Superior T\'ecnico, Universidade de Lisboa, Av. Rovisco Pais, P-1049-001 Lisboa, Portugal}

\date{31 July 2021}

\begin{abstract}
Dynamical correlation functions contain important physical information on correlated spin models.
Here a dynamical theory suitable suitable to the $\Delta =1$ isotropic spin-$1/2$ Heisenberg chain 
in a longitudinal magnetic field is extended to anisotropy $\Delta > 1$. The aim of this paper is the study 
of the $(k,\omega)$-plane line shape of the spin dynamical structure factor components $S^{+-} (k,\omega)$, $S^{-+} (k,\omega)$, 
and $S^{zz} (k,\omega)$ of the $\Delta >1$ spin-$1/2$ Heisenberg-Ising chain in a longitudinal magnetic field
near their $(k,\omega)$-plane sharp peaks. However, the extension
of the theory to anisotropy $\Delta > 1$ requires as a first step the clarification of the nature of a specific type of 
elementary magnetic configurations in terms of physical spins $1/2$. 
To reach that goal, that the spin $SU(2)$ symmetry of the isotropic $\Delta =1$ point
is in the case of anisotropy $\Delta > 1$ replaced by a continuous quantum group deformed $q$-spin 
$SU_q(2)$ symmetry plays a key role. For $\Delta >1$, spin projection $S^z$ remains a good 
quantum number whereas spin $S$ is not, being replaced by the $q$-spin $S_q$ in the eigenvalue 
of the Casimir generator of the continuous $SU_q(2)$ symmetry. Based on the isomorphism between
the irreducible representations of the $\Delta =1$ spin $SU(2)$ symmetry and
$\Delta >1$ continuous $SU_q(2)$ symmetry and on their relation to the occupancy configurations
of the Bethe-ansatz quantum numbers one finds that the elementary magnetic configurations under 
study are $q$-spin neutral. This determines the form of $S$ matrices on which the
extended dynamical theory relies. They are found in this paper to describe the scattering of 
{\it $n$-particles} whose relation to configurations of physical spins $1/2$ is established. 
Specifically, their internal degrees of freedom refer to unbound $q$-spin singlet pairs of physical spins $1/2$
described by $n=1$ real single Bethe rapidities and to $n=2,3,...$ bound such $q$-spin singlet pairs described by Bethe $n$-strings for
$n>1$. There is a relationship between the negativity and the length of the momentum $k$ interval
of the $k$ dependent exponents that control the power-law line shape of the spin dynamical structure factor components
near the lower threshold of a given $(k,\omega)$-plane continuum
and the amount of spectral weight over the latter. Using such a relationship, one finds that in the thermodynamic 
limit the significant spectral weight contributions from Bethe $n$-strings at a finite longitudinal magnetic field
refer to specific two-parametric $(k,\omega)$-plane gapped continua in the spectra of the spin dynamical structure 
factor components $S^{+-} (k,\omega)$ and $S^{zz} (k,\omega)$. 
In contrast to the $\Delta =1$ isotropic chain, excited energy eigenstates including up to $n=3$ Bethe $n$-strings
lead to finite spectral-weight contributions to $S^{+-} (k,\omega)$.
Most spectral weight stems though from excited energy eigenstates whose $q$-spin singlet 
$S^z = S_q = 0$ pairs of physical spins $1/2$ are all unbound. It is associated with $(k,\omega)$-plane continua in the spectra of the 
spin dynamical structure factor components that are gapless at some specific momentum values.
We derive analytical expressions for the line shapes of $S^{+-} (k,\omega)$, $S^{-+} (k,\omega)$, and $S^{zz} (k,\omega)$ 
valid in the vicinity of $(k,\omega)$-plane lines of sharp peaks. Those are mostly located at and just above lower thresholds 
of $(k,\omega)$-plane continua associated with both states 
with only unbound $q$-spin singlet pairs and states populated by such pairs and a single $n=2$ or $n=3$
$n$-particle. Our results provide physically interesting and important information on the microscopic processes that determine 
the dynamical properties of the non-perturbative spin-$1/2$ Heisenberg-Ising chain in a longitudinal magnetic field. 
\end{abstract}


\maketitle

\section{Introduction}
\label{SECI}

The spin-$1/2$ Heisenberg chain $XXZ$ is a physically interesting quantum problem whose
integrability was shown by R. Orbach \cite{Orbac_58}. The ground-state and the simplest excited states were studied by 
J. des Cloizeaux and M. Gaudin \cite{Cloizeaux_66}. In a series of three papers, C.N. Yang and C.P. Yang exhaustively 
discussed the ground-state properties of the model \cite{Yang_66}.

Dynamical correlation functions contain important physical information on correlated spin models.
In the case of the spin-$1/2$ Heisenberg chain $XXZ$, most previous studies on dynamical properties 
referred to anisotropy $\Delta <1$ \cite{Caux_05,Pereira_06,Pereira_07,Pereira_08,Caux_11,Caux_12}. 
Therefore, here we consider the spin-$1/2$ Heisenberg chain 
\cite{Takahashi_71,Gaudin_71,Takahashi_72,Takahashi_99,Gaudin_14} 
with anisotropy $\Delta >1$ \cite{Gaudin_71,Gaudin_14}, the so called spin-$1/2$ Heisenberg-Ising chain.
That model has a gapped spin-insulating quantum phase at magnetic field $h=0$ and
spin density $m=0$. Previous studies on its dynamical correlation functions refer mostly to that 
quantum phase \cite{Caux_08}. 

More recently, there has been a study on the dynamical properties
of the model for $\Delta >1$ at finite magnetic fields by a finite-size method that relied on the
algebraic Bethe ansatz formalism \cite{Yang_19}. The dynamical theory used in the
studies of this paper rather refers to the thermodynamic limit and provides analytical
expressions for the dynamical correlation functions valid in that limit. In spite of some 
technical similarities of the $\Delta<1$ and $\Delta>1$ Bethe-ansatz solutions, that 
dynamical theory has a different specific form for each of them.

The dynamical structure factor components $S^{xx} (k,\omega)$, $S^{yy} (k,\omega)$ e $S^{zz} (k,\omega)$
are given by,
\begin{eqnarray}
S^{aa} (k,\omega) & = & \sum_{j=1}^N e^{-ik j}\int_{-\infty}^{\infty}dt\,e^{-i\omega t}\langle GS\vert\hat{S}^{a}_j (t)\hat{S}^{a}_j (0)\vert GS\rangle 
\nonumber \\
& = & \sum_{\nu}\vert\langle \nu\vert\hat{S}^{a}_k\vert GS\rangle\vert^2
\delta (\omega - \omega^{aa}_{\nu} (k)) \hspace{0.40cm}{\rm for}\hspace{0.40cm}a =x,y,z \,  .
\label{SDSF}
\end{eqnarray}
Here and in this paper we use the notation $k$ for the excitation momentum of energy eigenstates relative
to a reference ground state. In Eq. (\ref{SDSF}), the spectra read $\omega^{aa}_{\nu} (k) = (E_{\nu}^{aa} - E_{GS})$, $E_{\nu}^{aa}$ refers to
the energies of the excited energy eigenstates that contribute to the 
dynamical structure factors components $aa= xx,yy,zz$, $\sum_{\nu}$ is the sum over such states,
$E_{GS}$ is the initial ground state energy, and $\hat{S}^{a}_k$ are for $a =x,y,z$ the 
Fourier transforms of the usual local $a =x,y,z$ spin operators $\hat{S}^{a}_j$, respectively. 

The components $S^{+-} (k,\omega)$ and $S^{-+} (k,\omega)$ of the spin dynamical structure factor are directly related to 
the two transverse components $S^{xx} (k,\omega)$ and $S^{yy} (k,\omega)$, Eq. (\ref{SDSF})
for $aa=xx,yy$, which are identical,  $S^{xx} (k,\omega)=S^{yy} (k,\omega)$, as follows,
\begin{equation}
S^{xx} (k,\omega) = S^{yy} (k,\omega) = {1\over 4}\left(S^{+-} (k,\omega)+S^{-+} (k,\omega)\right) \, .
\label{xxPMMP}
\end{equation}

One can then address the properties of $S^{xx} (k,\omega)=S^{yy} (k,\omega)$
in terms of those of $S^{+-} (k,\omega)$ and $S^{-+} (k,\omega)$. 
The main goal of this paper is the study of the power-law line shape of the spin dynamical structure factor 
components $S^{+-} (k,\omega)$, $S^{-+} (k,\omega)$, and $S^{zz} (k,\omega)$ of the $\Delta >1$ spin-$1/2$ 
Heisenberg-Ising chain in a longitudinal magnetic field near their $(k,\omega)$-plane lines of sharp peaks.

The dynamical theory used in Refs. \onlinecite{Carmelo_20,Carmelo_15A} for the isotropic point, $\Delta =1$,
refers to the thermodynamic limit. It provides analytical expressions 
for the momentum dependent exponents that control the power-law line shape of such spin dynamical structure factor 
components near their sharp peaks. 
In the isotropic case, a necessary condition for the validity of that theory is that the $S$ matrices associated with the 
scattering of {\it $n$-particles} are dimension-one scalars of a particular form given below in Sec. \ref{SECV}. 
At $\Delta =1$ a $n$-particle internal degrees of freedom refer to a single unbound singlet pair of physical 
spins $1/2$ for $n=1$ and to a number $n$ of bound such pairs for $n>1$ \cite{Carmelo_15,Carmelo_17}.
The phase shifts in the the $S$ matrices expressions can be extracted from the exact Bethe-ansatz solution. 
The momentum dependent exponents that control the power-law line shape of the spin dynamical structure factor components 
under consideration near their sharp peaks are simple combinations of such phase shifts \cite{Carmelo_20,Carmelo_15A}.

The form of the dimension-one scalar $S$ matrices under consideration follows in the isotropic case from the 
{\it spin-neutral} nature of both the unbound singlet pairs of physical spins $1/2$ described by $n=1$ real single Bethe 
rapidities and $n>1$ bound such pairs described by $n$-stings associated with complex Bethe-ansatz rapidities
\cite{Carmelo_15,Carmelo_17}. The extension of the dynamical theory of  Refs. \onlinecite{Carmelo_20,Carmelo_15A} to $\Delta >1$ requires that
the corresponding $S$ matrices have a similar form, which for $\Delta >1$ remains an unsolved problem.

The Hamiltonian of the spin-$1/2$ Heisenberg-Ising chain describes $N=\sum_{\sigma =\uparrow,\downarrow}N_{\sigma}$ 
physical spins $1/2$ of projection $\sigma =\uparrow,\downarrow$. For anisotropy parameter range $\Delta=\cosh\eta\geq 1$ and thus $\eta\geq 0$,
spin densities $m = (N_{\uparrow}-N_{\downarrow})/N \in [0,1]$, exchange integral $J$, and length $L\rightarrow\infty$
for finite $N/L$, that Hamiltonian in a longitudinal magnetic field $h$ reads,
\begin{equation}
\hat{H} = \hat{H}_{\Delta} + g\mu_B\,h\sum_{j=1}^{N} \hat{S}_j^z
\hspace{0.40cm}{\rm where}\hspace{0.40cm}
\hat{H}_{\Delta} = J\sum_{j=1}^{N}\left({\hat{S}}_j^x{\hat{S}}_{j+1}^x + {\hat{S}}_j^y{\hat{S}}_{j+1}^y + 
\Delta\,{\hat{S}}_j^z{\hat{S}}_{j+1}^z\right) \, .
\label{Hphi}
\end{equation}
In such expressions, $\hat{\vec{S}}_{j}$ is the spin-$1/2$ operator at site $j=1,...,N$ with components $\hat{S}_j^{x,y,z}$,
$g$ is the Land\'e factor, and $\mu_B$ is the Bohr magneton. 
The present study uses natural units of lattice spacing and Planck constant one.

It is well known that the Hamiltonian $\hat{H}_{\Delta}$ in Eq. (\ref{Hphi})
has spin $SU(2)$ symmetry at $\Delta = \cosh\eta=1$ and thus $\eta =0$. For $\Delta >1$ and $\eta >0$ it has 
a related continuous quantum group deformed symmetry 
$SU_q(2)$ associated with the $q$-spin $S_q$ in the eigenvalue of the Casimir generator
\cite{Pasquier_90,Prosen_13}. In either case, the $2^{N}$ irreducible representations of the corresponding
symmetry refer to the $2^{N}$ energy eigenstates of the Hamiltonian
$\hat{H} = \hat{H}_{\Delta} + g\mu_B\,h\sum_{j=1}^{N} \hat{S}_j^z$, Eq. (\ref{Hphi}).

For $\eta >0$, spin projection $S^z$ remains a good quantum number whereas spin is not, spin $S$ being replaced by 
the $q$-spin $S_q$. Inside the $\eta>0$ many-particle system, each of the $N$ physical spins $1/2$ have
$q$-spin $S_q=1/2$ and spin projection $S^z =\pm 1/2$. Hence the present
designation {\it physical spin $1/2$} refers to $S=1/2$ and $S_q=1/2$ for $\eta=0$
and $\eta>0$, respectively. 

A first step of the extension of the dynamical theory used in Refs. \onlinecite{Carmelo_20,Carmelo_15A} for the isotropic 
case to anisotropy $\Delta > 1$ refers to showing that for the corresponding spin-$1/2$ $XXZ$ chain,
the $n=1$ real single Bethe rapidities and the $n>1$ Bethe $n$-stings describe unbound
$q$-spin singlet $S^z =S_q = 0$ pairs of physical spins $1/2$ and bound states of a number $n>1$ 
of such $q$-spin singlet pairs, respectively. 
This ensures that the $S$ matrices associated with the scattering of the corresponding $n$-particles introduced 
in this paper are for $\eta >0$ dimension-one scalars whose form is similar to that for the $\eta =0$ isotropic case. 
The phase shifts associated with the $S$ matrices are indeed found to differ from those of the isotropic case only in their 
dependence on $\eta$. 

A key step of our study to reach the results for $\eta>0$ is made possible by explicitly accounting for
the isomorphism of the $\eta=0$ spin $SU(2)$ and $\eta>0$ $q$-spin $SU_q(2)$ symmetries's irreducible representations.
Due to such an isomorphism, $q$-spin $S_q$ has exactly the same values as spin $S$. 
Specifically, accounting for the one-to-one correspondence between the configurations 
of the $N$ physical spins $1/2$ described by the Hamiltonian, Eq. (\ref{Hphi}), that generate such representations 
and the occupancy configurations of the quantum numbers of the exact Bethe-ansatz solution that generate
the corresponding energy eigenstates. 

The confirmation that the dimension-one scalar $S$ matrices under consideration have for $\Delta >1$ 
the same form as for the isotropic $\Delta =1$ case, renders the extension
to anisotropy $\Delta >1$ of the dynamical theory suitable to $\Delta =1$
a straightforward procedure. The use of that dynamical theory provides
useful information on the $(k,\omega)$-plane spectral-weight distributions 
of $S^{+-} (k,\omega)$, $S^{xx} (k,\omega)$, and $S^{zz} (k,\omega)$.
Most of such spectral weight stems from specific classes of excited energy eigenstates
whose spectra are associated with that distribution.

Analytical expressions for these dynamical correlation function components valid at and just above their 
$(k,\omega)$-plane lines of sharp peaks are derived. Our results provide interesting and important information on 
the non-perturbative microscopic processes that control the physics of the spin-$1/2$ Heisenberg-Ising chain 
in a longitudinal magnetic field.

\section{A suitable representation for the spin-$1/2$ Heisenberg-Ising chain in terms of physical spins $1/2$ configurations}
\label{SECII}

\subsection{General representation for the whole Hilbert space}
\label{SECIIA}

The structure of the Bethe $n$-strings depends on the system size.
The thermodynamic limit in which that structure simplifies
\cite{Takahashi_71,Gaudin_71,Gaudin_14} is that more relevant for the description of physical spin systems.
The $n=1$ real single Bethe rapidities and the Bethe $n$-strings 
are defined for $\eta\geq 0$ in terms of general Bethe-ansatz rapidities whose form in that
limit is given by,
\begin{equation}
\Lambda_j^{n,l} = \Lambda_j^n + i(n + 1 -2l)\hspace{0.40cm}{\rm where}\hspace{0.40cm}l=1,...,n \, .
\label{LambdaIm}
\end{equation}
Here $j = 1,...,L_n$, $n=1,...,\infty$, and we have used the notation of Ref. \onlinecite{Takahashi_71},
whose relation to the $\eta >0$ rapidities $\varphi_{n,j}$ of Ref. \onlinecite{Gaudin_71} is 
$\Lambda_j^n = - \varphi_{n,j}/\eta$. The real numbers $\Lambda_j^n$ on the right-hand 
side of Eq. (\ref{LambdaIm}) are defined by a number $n=1,...,\infty$ of coupled Bethe-ansatz 
equations \cite{Takahashi_71,Gaudin_71,Gaudin_14}. Those are given in Eq. (\ref{BAEqDeltal1}) of Appendix \ref{A}
in a useful functional form.

For $n=1$, the rapidity $\Lambda_j^{n,l}$ is real and otherwise 
its imaginary part is finite, except at $l= (n+1)/2$ when $l$ is odd.
A $n$-string is a group of $n>1$ rapidities with the same real part $\Lambda_j^n$.
Below in Sec. \ref{SECIII} it is confirmed that a $n=1$ real single Bethe rapidity and a $n>1$ Bethe $n$-string
describe one unbound $q$-spin singlet $S^z=S_q=0$ pair of physical spins $1/2$
and $n>1$ bound such pairs, respectively.

One associates a {\it Bethe-ansatz $n$-band} with each $n=1,...,\infty$ value. It
contains $L_n = N_n + N^h_{n}$ discrete momentum values $q_j \in [q_n^-,q_n^+]$ 
with spacing $q_{j+1}-q_j = 2\pi/L$ where $j=1,...,L_n$. In this paper we use
the notation $q$ for the momentum values of such $n$-bands whereas, as reported 
above in Sec. \ref{SECI}, $k$ refers to the excitation momentum of an energy eigenstate 
relative to a reference ground state. 

In Appendix \ref{A} it is found that the $n$-band limiting momentum values read
$q_n^{\pm} = \pm {\pi\over L}(L_n - 1 \mp \delta L_n^{\eta})$.
Here $L_n$ and $\delta L_n^{\eta}$ are independent and dependent on $\eta$, respectively. 
In that Appendix the former numbers are found to read,
\begin{equation}
L_n = N_n + N^h_{n}\hspace{0.40cm}{\rm where}\hspace{0.40cm}
N^h_{n} = 2S_q +\sum_{n'=n+1}^{\infty}2(n'-n)N_{n'}
\hspace{0.40cm}{\rm for}\hspace{0.40cm}n=1,...,\infty \, ,
\label{Ln}
\end{equation}
whereas $\delta L_n^{\eta}$ is given below.

The Bethe-ansatz quantum numbers $I_j^n$ are the $n$-band momentum values $q_j = {2\pi\over L}I_j^n$
in units of ${2\pi\over L}$. They are given by $I_j^n = 0,\pm 1,...,\pm {L_n -1\over 2}$ for $L_n$ odd
and $I_j^n =\pm 1/2,\pm 3/2,...,\pm {L_n -1\over 2}$ for $L_n$ even. 
Such numbers and thus the set of $n$-band discrete momentum values $\{q_j\}$ have a Pauli-like 
occupancy of zero and one, respectively.

The corresponding numbers $N_n$ and $N^h_{n}$ are those of occupied and unoccupied $n$-band discrete 
momentum values $q_j$, respectively, of an energy eigenstate. Such a state $n$-band momentum distribution 
thus reads $N_{n} (q_{j}) =1$ and $N_{n} (q_{j}) =0$ for the $N_n$ and $N^h_{n}$ momentum values $q_{j}$ 
that are occupied and unoccupied, respectively.
Hence $\sum_{j=1}^{L_n}N_{n} (q_{j})= N_n$ and  $\sum_{j=1}^{L_n}N_{n}^h (q_{j})= N^h_n$
where $N_{n}^h (q_{j}) = 1 - N_{n} (q_{j})$. The solution of Bethe-ansatz equations for the Hamiltonian, Eq. (\ref{Hphi}), given in
functional form in Eq. (\ref{BAEqDeltal1}) of Appendix \ref{A}, provides the
real part, $\Lambda_j^n = \Lambda^n (q_j)$, of the $n=1,2,...$ rapidities, Eq. (\ref{LambdaIm}). 
We call them $n=1,2,...$ rapidity functions $\Lambda^n (q_j)$ of an energy eigenstate.

Each such a state is uniquely defined by the set of $n$-band momentum distributions
$\{N_{n} (q_{j})\}$ for $n=1,2,...$ and $j = 1,...,L_n$. 
The Bethe-ansatz solution refers to subspaces spanned by energy eigenstates for which $S^z = -S_q$.
As confirmed below in Sec. \ref{SECIIIB}, the remaining $S_q>0$ energy eigenstates for which $S^z = -S_q + n_z$
where $n_z = 1,...,2S_q$ are also accounted for by the present general representation.

Consistently with the $q_j$'s being momentum values,
the momentum eigenvalues are in functional form given by,
\begin{equation}
P = \pi \sum_{n=1}^{\infty}N_n + \sum_{n=1}^{\infty}\sum_{j=1}^{L_n}N_{n} (q_{j})\,q_j \, .
\label{P}
\end{equation}

Importantly, both the numbers $L_n$, Eq. (\ref{Ln}), and the momentum eigenvalues are
independent of the anisotropy parameter $\Delta = \cosh\eta$. In contrast, the energy eigenvalues and the numbers
$\delta L_n^{\eta}$ in $q_n^{\pm} = \pm {\pi\over L}(L_n - 1 \mp \delta L_n^{\eta})$
depend on $\eta$ and on the $n$-band momentum values $q_j$ through the rapidity functions $\Lambda^n (q_j)$. 
Within the present functional representation the former read,
\begin{equation}
E = - \sum_{n=1}^{\infty}\sum_{j=1}^{L_n}N_{n} (q_{j})\,{J\over n}
\left({\sinh^2 (n\,\eta)\over \cosh (n\,\eta) - \cos (\Lambda^{n} (q_{j})\,\eta)} - C_n (\eta)\right)
+ g\mu_B\,h\,S^z \, .
\label{Energy}
\end{equation}
Here the constant $C_n (\eta)$ is given by,
\begin{equation}
C_n (\eta) = \left(1 - {n\,\sinh (\eta)\over\sinh (n\,\eta)}\right)\left(1 + \cosh (n\,\eta)\right)  \, .
\label{Cneta}
\end{equation}
It suitably defines the zero-energy level of the related $n$-band energy dispersions given in Appendix \ref{B}. 
The magnetic field $h=h (m)$ in Eq. (\ref{Energy}) is at $m=0$ and for $m\in [0,1]$ given by \cite{Takahashi_99},
\begin{eqnarray}
h (0) & \in & [0,h_{c1}]\hspace{0.40cm}{\rm and}
\nonumber \\
h (m) & = &{1\over g\mu_B}{J\sinh(\eta)\,2\pi\sigma_1 (B)\over \xi_{1\,1}}\in [h_{c1},h_{c2}] \, ,
\label{magcurve}
\end{eqnarray}
respectively. Here $2\pi\sigma_1 (B)$ is the distribution $2\pi\sigma_1 (\varphi)$,
Eq. (\ref{equA10}) of Appendix \ref{B}, the parameter $B$ is given in Eq. (\ref{QB-r0rs}) of
that Appendix, and the parameter $\xi_{1\,1}$ is defined in Eq. (\ref{xi1all}) of Appendix \ref{C}.
The critical magnetic fields read \cite{Takahashi_99},
\begin{equation}
h_{c1} = {2J\over\pi\,g\mu_B}\sinh (\eta) K (u_{\eta})\,\sqrt{1 - u_{\eta}^2}
\hspace{0.40cm}{\rm and}\hspace{0.40cm}
h_{c2} = {J (\Delta +1 )\over g\mu_B} = {J (\cosh \eta +1 )\over g\mu_B} \, .
\label{hc1hc2}
\end{equation}
The critical field $h_{c1}$ refers to the transition from the $m=0$ spin-insulating quantum phase
to the $0<m<1$ spin-conducting quantum phase and $h_{c2}$ refers to that from the latter phase to the 
fully-polarized ferromagnetic quantum phase. 
The complete elliptic integral $K (u_{\eta})$ appearing in the $h_{c1}$ expression is defined as,
\begin{equation}
K (u_{\eta}) = \int_0^{\pi\over 2}d\theta {1\over \sqrt{1 - u_{\eta}^2\sin^2\theta}} 
\hspace{0.40cm}{\rm and}\hspace{0.40cm}
\eta = \pi\,{K (u_{\eta}')\over K (u_{\eta})} =
\pi\left({\int_0^{\pi\over 2}d\theta {1\over \sqrt{1 - (1-u_{\eta}^2)\sin^2\theta}}\over 
\int_0^{\pi\over 2}d\theta {1\over \sqrt{1 - u_{\eta}^2\sin^2\theta}}}\right) \, .
\label{uphi}
\end{equation}  
The $\eta$-dependent function $u_{\eta} = u_{\eta} (\eta)$ has been defined 
in this equation by the corresponding inverse function.

The critical field $h_{c1}$ has limiting behaviors $h_{c1} \rightarrow 0$ for $\eta\rightarrow 0$
and $h_{c1} \approx J (\Delta - 2)/(g\mu_B)$ for $\eta\gg 1$. That $h_{c1}\rightarrow 0$
as $\eta\rightarrow 0$ is imposed by the emergence of the isotropic point's spin $SU (2)$ symmetry.
It follows that the field interval $h\in [-h_{c1},h_{c1}]$ of the $\Delta >1$ spin-insulating quantum phase shrinks to $h=0$ at $\Delta =1$.
In this paper we consider magnetic fields $h\geq 0$ for which $N_{\uparrow}\geq N_{\downarrow}$,
so that such a $\eta>0$ quantum phase occurs for $h\in [0,h_{c1}]$

In Appendix \ref{A} the functional representation of the numbers $\delta L_n^{\eta}$ in $q_n^{\pm} = \pm {\pi\over L}(L_n - 1 \mp \delta L_n^{\eta})$
is found to read,
\begin{equation}
\delta L_n^{\eta} = \sum_{n'=1}^{\infty}F_{n,n'} (\eta) 
\sum_{j=1}^{L_{n'}}N_{n'} (q_j)\left({\eta\,\Lambda^{n'} (q_{j})\over\pi}\right) 
= - \sum_{n'=1}^{\infty}F_{n,n'} (\eta) 
\sum_{j=1}^{L_{n'}}N_{n'} (q_j)\left({\varphi_{n'} (q_{j})\over\pi}\right) \, .
\label{deltaLn}
\end{equation}
Here $F_{n,n'} (\eta)$ has a long expression not needed for our studies 
whose limiting behaviors are,
\begin{eqnarray}
F_{n,n'} (\eta)  & = & n\,n'\,\eta\hspace{0.40cm}{\rm for}\hspace{0.40cm}\eta\ll 1
\nonumber \\
& = & n + n' - \vert n-n'\vert - \delta_{n',n}
\hspace{0.40cm}{\rm for}\hspace{0.40cm}\eta\gg 1 \, .
\label{Fnn}
\end{eqnarray}
Since $\Lambda^{n'} (q_{j}) = \Lambda_j^{n'}\in [-\pi/\eta,\pi/\eta]$
and $\varphi_{n'} (q_{j}) = \varphi_{n',j} \in [-\pi,\pi]$, one has in Eq. (\ref{deltaLn}) that
$\eta\,\Lambda^{n'} (q_{j})/\pi = \varphi_{n'} (q_{j})/\pi\in [-1,1]$.
For $\eta >0$ ground states and excited energy eigenstates 
contributing to the dynamical properties studied in this paper, the numbers $\delta L_n^{\eta}$ either vanish
or are of order $1/L$ and also vanish in the thermodynamic limit.

Within our representation in terms of the $N$ physical spins $1/2$ described by the
Hamiltonian, Eq. (\ref{Hphi}), the designation {\it $n$-particles} refers both to
{\it $1$-particles} and {\it $n$-string-particles} for $n>1$: 

- The internal degrees of freedom of a $1$-particle correspond to a single $n=1$
unbound pair of physical spins $1/2$ whose neutral $q$-spin nature is confirmed in Sec. \ref{SECIII}. Its translational degrees of freedom refer to
the $1$-band momentum $q_j \in [q_1^-,q_1^+]$ carried by each of the $N_1$ $1$-particles
of an energy eigenstate;

- The internal degrees of freedom of a 
$n$-string-particle refer to the $n>1$ physical spins $1/2$ pairs bound within
a configuration described by the corresponding $n$-string whose neutral $q$-spin nature is 
confirmed in Sec. \ref{SECIII}
to be the same as that of unbound pairs. Its translational degrees of freedom correspond to
the $n>1$ $n$-band momentum $q_j \in [q_n^-,q_n^+]$ carried by each of the $N_n$ 
$n$-string-particles of an energy eigenstate for which $N_n>0$.

\subsection{Representation for the subspaces of interest for the dynamical properties}
\label{SECIIB}

For ground states and their excited energy eigenstates associated with a significant amount of spin 
dynamical structure factor spectral weight, one has in the thermodynamic
limit and except for corrections of $1/L$ order that the limiting $n$-band
numbers $q_n^{\pm} = \pm {\pi\over L}(L_n - 1 \mp \delta L_n^{\eta})$ simplify to
$q_1^{\pm} = \pm k_{F\uparrow}$ for $n=1$
and to $q_n^{\pm} = \pm (k_{F\uparrow}-k_{F\downarrow})$ for $n>1$.
Here $k_{F\uparrow} = {\pi\over 2}(1+m)$ and $k_{F\downarrow} = {\pi\over 2}(1-m)$.
Ground states refer in that limit and again neglecting 
corrections of $1/L$ order to a $1$-band Fermi sea $q_j \in [-k_{F\downarrow},k_{F\downarrow}]$
and empty $n$-bands for $n>1$. In the thermodynamic limit, the $n$-band discrete momentum values
$q_j$ such that $q_{j+1}-q_j = 2\pi/L$ are often replaced by a continuous
variable $q$. 

The following ground-state analysis is consistent with both the unbound pair 
in each $1$-particle corresponding to a $q$-spin singlet $S^z =S_q = 0$ 
configuration and the occurrence of {\it unpaired physical spins $1/2$} in some
energy eigenstates, as confirmed below in Sec. \ref{SECIII}.
(Ground states are not populated by $n>1$ $n$-string-particles.)

In the spin-insulating quantum phase that for $\eta >0$ occurs for
spin density $m=0$ and magnetic fields $h\in [0,h_{c1}]$, the $1$-band is full in the 
ground state, $q \in [-\pi/2,\pi/2]$. For $N$ even all $N$ physical spins $1/2$
are paired within a number $N/2$ of $1$-particles each containing a
single pair. This is a gapped quantum phase for $0\leq h<h_{c1}$.
\begin{figure}
\begin{center}
\subfigure{\includegraphics[width=5.65cm]{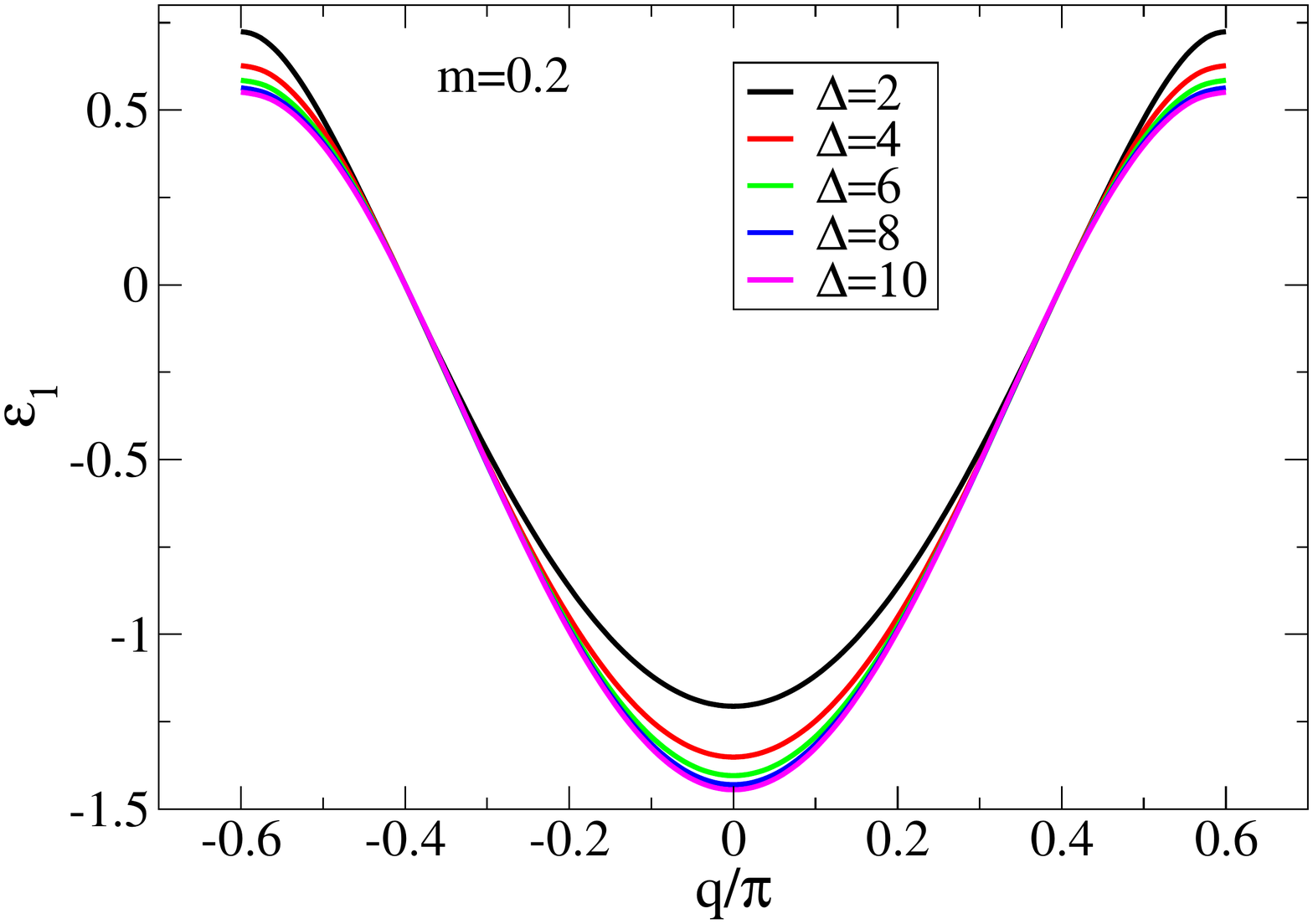}}
\hspace{0.25cm}
\subfigure{\includegraphics[width=5.65cm]{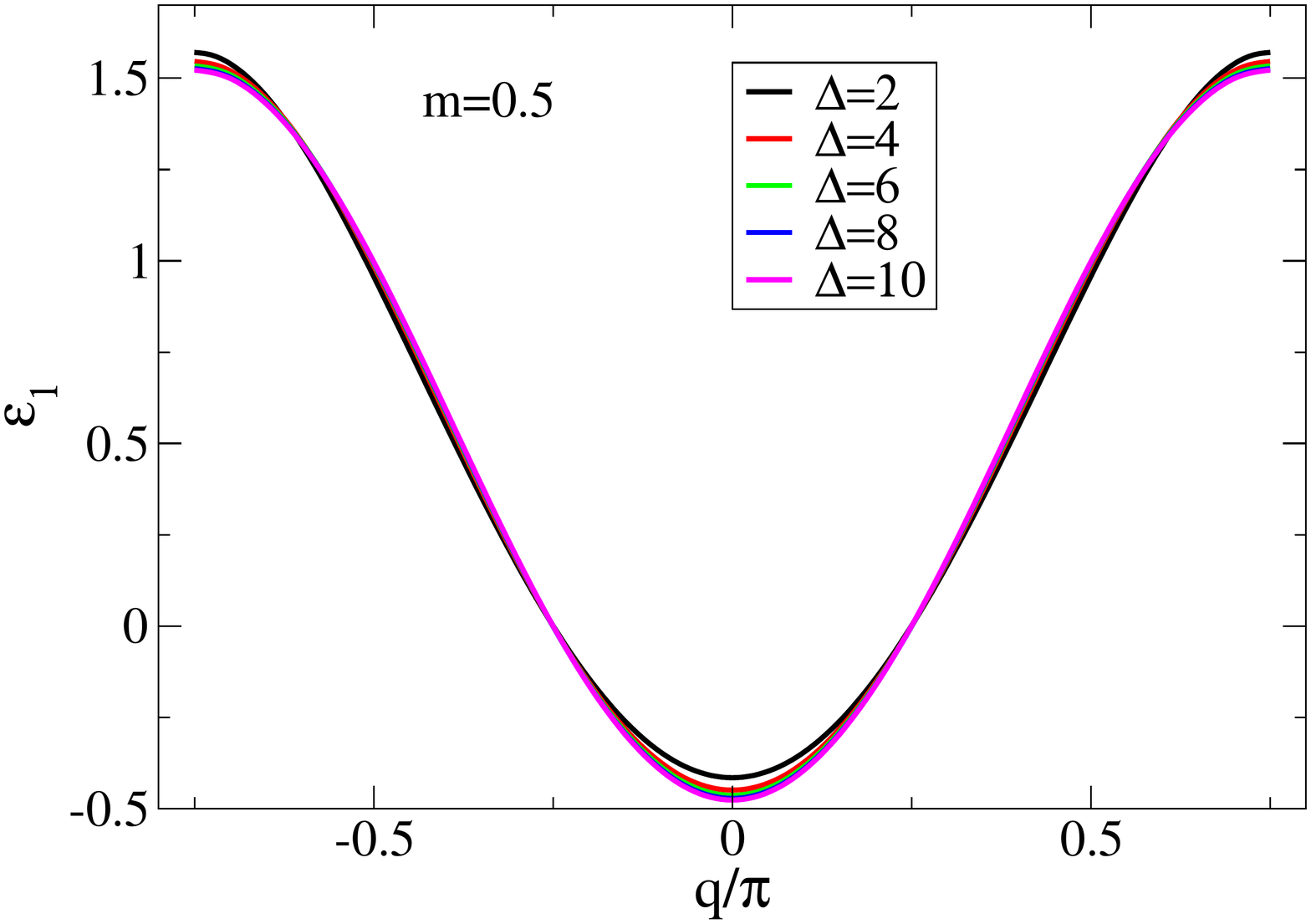}}
\hspace{0.25cm}
\subfigure{\includegraphics[width=5.65cm]{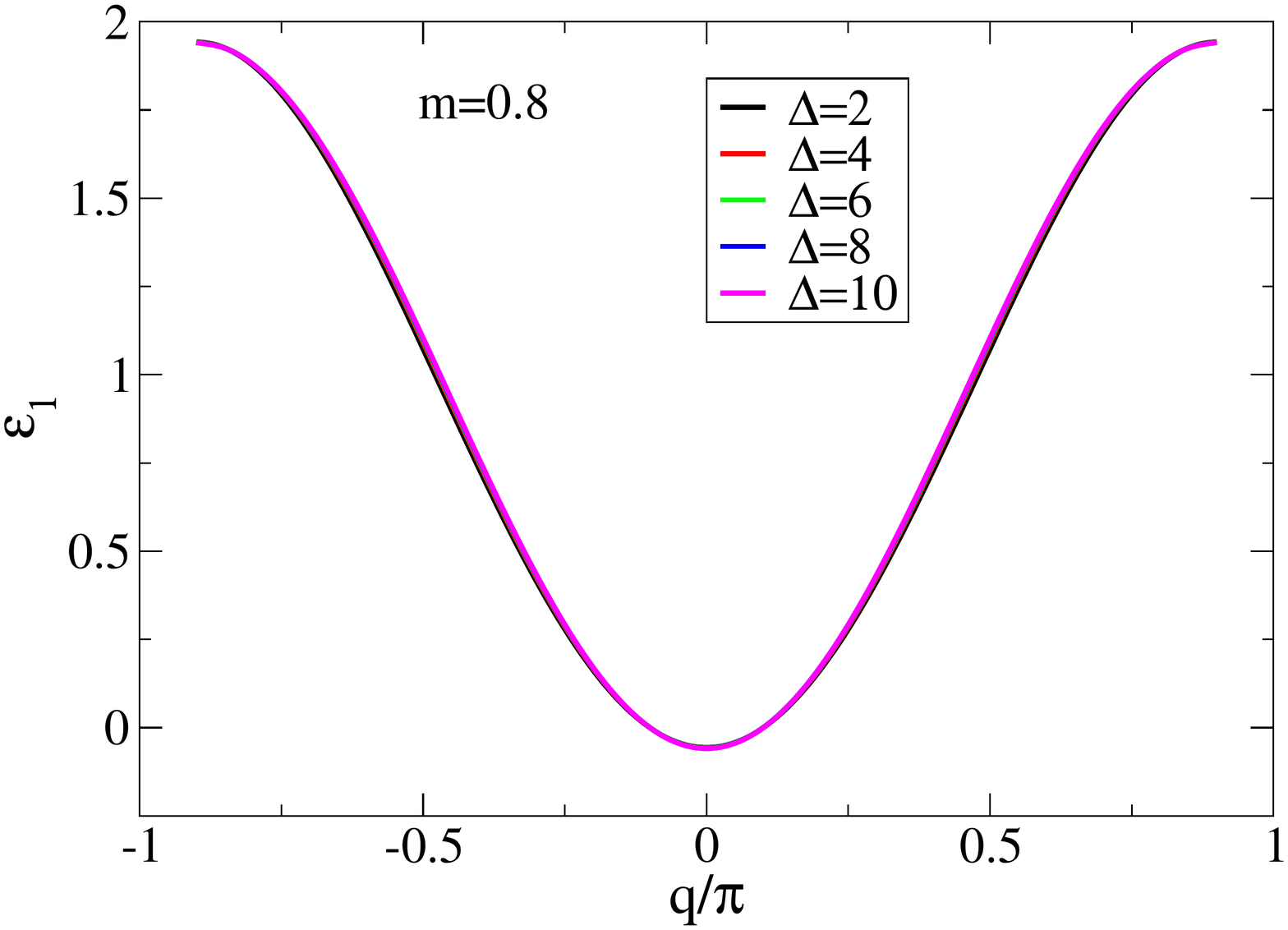}}
\caption{The $1$-particle energy dispersion $\varepsilon_{1} (q)$ in units of $J$
defined by Eqs. (\ref{equA4}) and (\ref{equA4B}) of Appendix \ref{B} plotted as
a function of the $1$-band momentum $q \in [-k_{F\uparrow},k_{F\uparrow}]$ for spin densities
$m=0.2$, $m=0.5$, and $m=0.8$ and several anisotropy
$\Delta$ values.}
\label{figure1NPB}
\end{center}
\end{figure}

The magnetic field range $h\in [h_{c1},h_{c2}]$ refers to a 
spin-conducting quantum phase in which the $1$-band is occupied in the ground state
for $q \in [-k_{F\downarrow},k_{F\downarrow}]$ and unoccupied
for $\vert q_j\vert \in [k_{F\downarrow},k_{F\uparrow}]$.
The ground state contains a number $N_1 = N_{\downarrow}$ of $1$-particles
each referring to a pair of physical spins $1/2$ of opposite projection.
The remaining $N-2N_{\downarrow}=N_{\uparrow}-N_{\downarrow}$ physical
spins $1/2$ are unpaired. 

Finally, in the fully-polarized ferromagnetic quantum phase occurring for
$h>h_{c2}$, there are no $n$-particles in the ground state both for $n=1$
and $n>1$. Indeed, in that state all $N$ physical spins $1/2$
are unpaired and have the same spin projection.

The $n$-particle energy dispersions $\varepsilon_{n} (q)$ defined by Eqs. (\ref{equA4})-(\ref{equA4n10}) of Appendix \ref{B}
play an important role in our study. The $1$-particle energy dispersion refers to the energy 
$\varepsilon_{1} (q)$ associated with creation onto a ground state of one $1$-particle 
at a $1$-band momentum in the interval $\vert q\vert\in [k_{F\downarrow},k_{F\uparrow}]$.
That dispersion also corresponds to the energy $-\varepsilon_{1} (q)$ associated with creation onto such a state of 
one $1$-hole at a $1$-band momentum in the interval $q\in [-k_{F\downarrow},k_{F\downarrow}]$.
The $n>1$ $n$-string-particle energy dispersion refers to the energy $\varepsilon_{n} (q)$
associated with creation onto a ground state of one $n$-particle at a $n$-band momentum in the interval 
$q \in [- (k_{F\uparrow}-k_{F\downarrow}), (k_{F\uparrow}-k_{F\downarrow})]$.

The $1$-particle, $2$-string-particle, and $3$-string-particle energy dispersions 
are plotted in units of $J$ in Figs. \ref{figure1NPB}, \ref{figure2NPB}, and \ref{figure3NPB}, respectively,
as a function of the corresponding band momentum $q$ for spin densities
$m=0.2$, $m=0.5$, and $m=0.8$ and several anisotropy $\Delta$ values.
Their simple limiting analytical expressions for $h \in [0,h_{c1}]$ and $m=0$ and 
for $h=h_{c2}$ and $m=1$ are in Appendix \ref{B} found to read,
\begin{eqnarray}
\varepsilon_{1} (q) & = & - {J\over\pi}\sinh (\eta) K (u_{\eta}) \sqrt{1 - u_{\eta}^2\sin^2 q} 
+ {1\over 2}\,g\mu_B\,h\hspace{0.40cm}
{\rm for}\hspace{0.40cm}q \in [-\pi/2,\pi/2]\hspace{0.40cm}{\rm and}\hspace{0.40cm}h \in [0,h_{c1}] 
\nonumber \\
& = & - {J\over\pi}\sinh (\eta) K (u_{\eta})\left(\sqrt{1 - u_{\eta}^2\sin^2 q} - \sqrt{1 - u_{\eta}^2}\right)
\hspace{0.40cm}{\rm for}\hspace{0.40cm}q \in [-\pi/2,\pi/2]\hspace{0.40cm}{\rm and}\hspace{0.40cm}h = h_{c1}
\nonumber  \\
\varepsilon_{1} (q) & = & J (1 - \cos q)
\hspace{0.40cm}{\rm for}\hspace{0.40cm}q \in [-\pi,\pi]\hspace{0.40cm}{\rm and}\hspace{0.40cm}h = h_{c2} \, ,
\label{vareband1m0m1}
\end{eqnarray}
and
\begin{eqnarray}
\varepsilon_{n} (0) & = & (n-1)\,g\mu_B\,h \hspace{0.40cm}
{\rm for}\hspace{0.40cm}q = 0\hspace{0.40cm}{\rm and}\hspace{0.40cm}h \in [0,h_{c1}] 
\nonumber \\
& = & {2(n-1)\,J\over\pi}\sinh (\eta) K (u_{\eta})\sqrt{1 - u_{\eta}^2}
\hspace{0.40cm}{\rm for}\hspace{0.40cm}q=0\hspace{0.40cm}{\rm and}\hspace{0.40cm}h = h_{c1}
\nonumber \\
\varepsilon_{n} (q) & = & {J\over n}(1 - \cos q) + n\,J\,(1+ \cosh (\eta)) 
\nonumber  \\
& - & J {\sinh (\eta)\over\sinh (n\,\eta)}\left(1 + \cosh (n\,\eta)\right) 
\hspace{0.40cm}{\rm for}\hspace{0.40cm} q \in [-\pi,\pi] \hspace{0.40cm}{\rm and}\hspace{0.40cm}h = h_{c2} \, .
\label{qdepvarepsilonm0m1}
\end{eqnarray}

The excitation energies $\delta E = E_{\nu} - E_{GS}$ where $E_{\nu}$ and $E_{GS}$ are the energy
eigenvalues of the excited state and ground state, respectively, considered below in Sec. \ref{SECIV}
are expressed in terms of simple combinations of such $1$-particle and $n$-string-particle energy dispersions. 

In the literature, fractionalized particles such as one-dimensional spin waves \cite{Faddeev_81}
usually called spinons \cite{Caux_06} at $m=0$ and psinons and antipsinons for $m>0$ \cite{Karbach_02,Karbach_00}
are often used. However, their transformation laws under both the spin operators in the Hamiltonian expression,
Eq. (\ref{Hphi}), and the generators of the $\eta =0$ $SU (2)$ symmetry and 
$\eta >0$ continuous $SU_q(2)$ symmetry given below in Sec. \ref{SECIIIA}
remain unknown. In contrast, the present $n$-particle representation refers to $1$-particles and
$n$-string-particles whose relation to configurations of physical spins $1/2$ 
is confirmed below in Sec. \ref{SECIII} to be more direct. Importantly, the dynamical properties
of the spin-$1/2$ Heisenberg-Ising chain with anisotropy $\Delta >1$ in a longitudinal
magnetic field studied below in Sec. \ref{SECV} are naturally and directly described by the scattering of 
such $n$-particles.

Spinons, psinons, and antipsinons are well defined in subspaces with no $n$-strings or with a vanishing
density of $n$-strings in the thermodynamic limit. Spinons are $1$-band holes within excited 
energy eigenstates of the $m=0$ ground state. Psinons and antipsinons 
are $1$-holes that emerge or are moved to inside the $1$-band Fermi sea and 
$1$-particles that emerge or are moved to outside that sea, respectively, in excited energy eigenstates of ground states 
corresponding to the $\eta>0$ spin-conducting quantum phase that occurs for $h_{c1}<h<h_{c2}$. 

\section{The nature of the bound and unbound elementary magnetic configurations in terms of physical spins $1/2$}
\label{SECIII}

The results of this section refer to the quantum problem defined by the Hamiltonian $\hat{H}$, Eq. (\ref{Hphi}), 
acting on its whole Hilbert space.

\subsection{The isomorphism of the $SU(2)$ and $SU_q (2)$ symmetries irreducible representations}
\label{SECIIIA}

Since the Hilbert space of the present quantum problem is the same for $\eta =0$ and $\eta >0$, 
there is a uniquely defined unitary transformation that relates the two sets of $2^N$
$\eta =0$ and $\eta >0$ energy eigenstates, $\{\left\vert l_{\rm r},S,S^z,0\right\rangle\}$ and
$\{\left\vert l_{\rm r},S_q,S^z,\eta\right\rangle\}$, respectively. Indeed, both of them
refer to sets of complete and orthonormal energy eigenstates for the same Hilbert space
of dimension $2^N$. In such a state notation, $l_{\rm r}$ stands for all quantum numbers other than 
spin $S$ for $\eta =0$, $q$-spin $S_q$ for $\eta >0$, $S^z$ for $\eta \geq 0$, and $\eta$ itself needed to specify 
$\vert l_{\rm r},S_q,S^z,\eta\rangle$ for {\it any} fixed $\eta\geq 0$ value. 
In Appendix \ref{A} it is shown that the quantum numbers described by $l_{\rm r}$
are independent of $\eta$.
\begin{figure}
\begin{center}
\subfigure{\includegraphics[width=5.65cm]{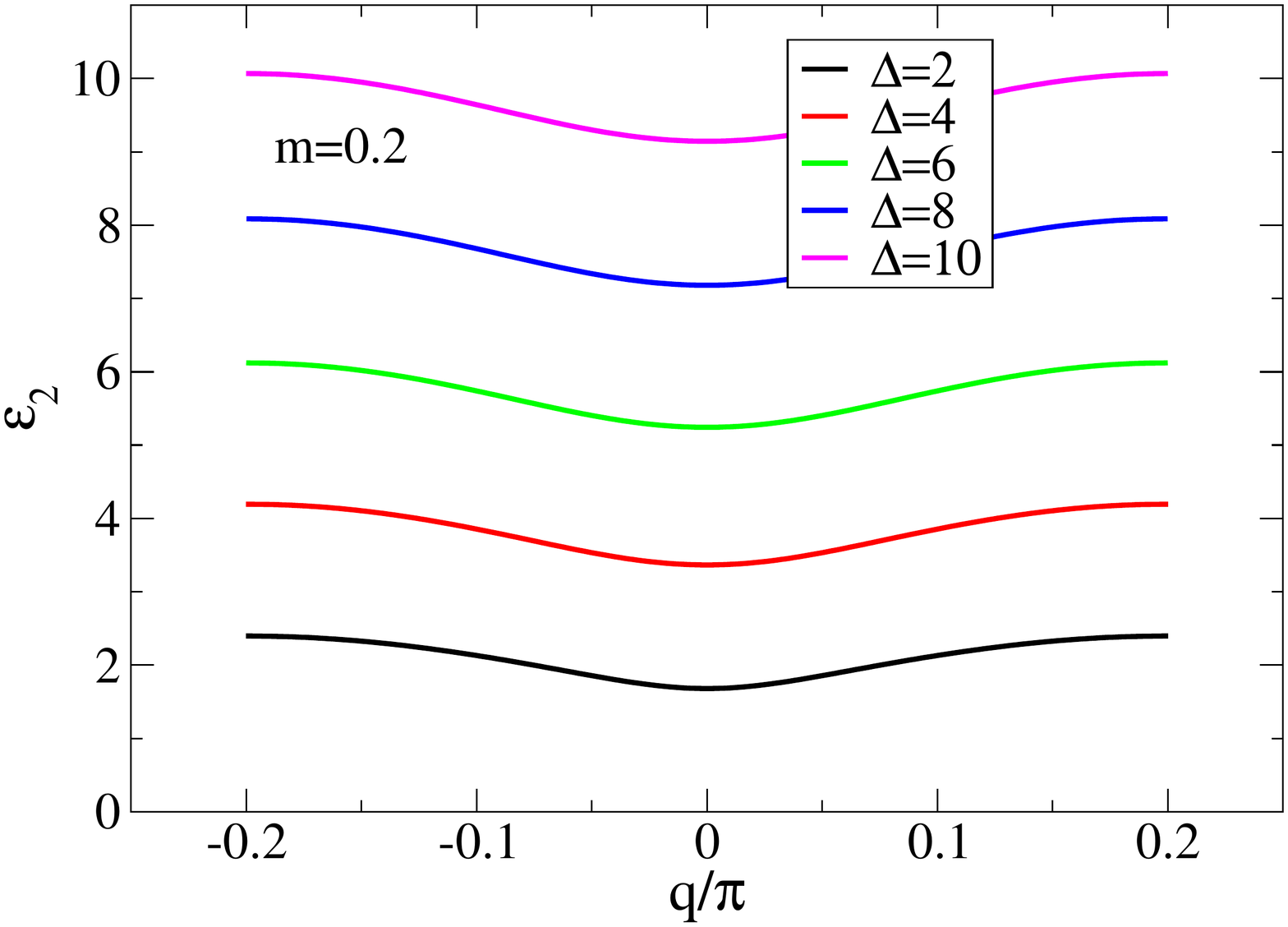}}
\hspace{0.25cm}
\subfigure{\includegraphics[width=5.65cm]{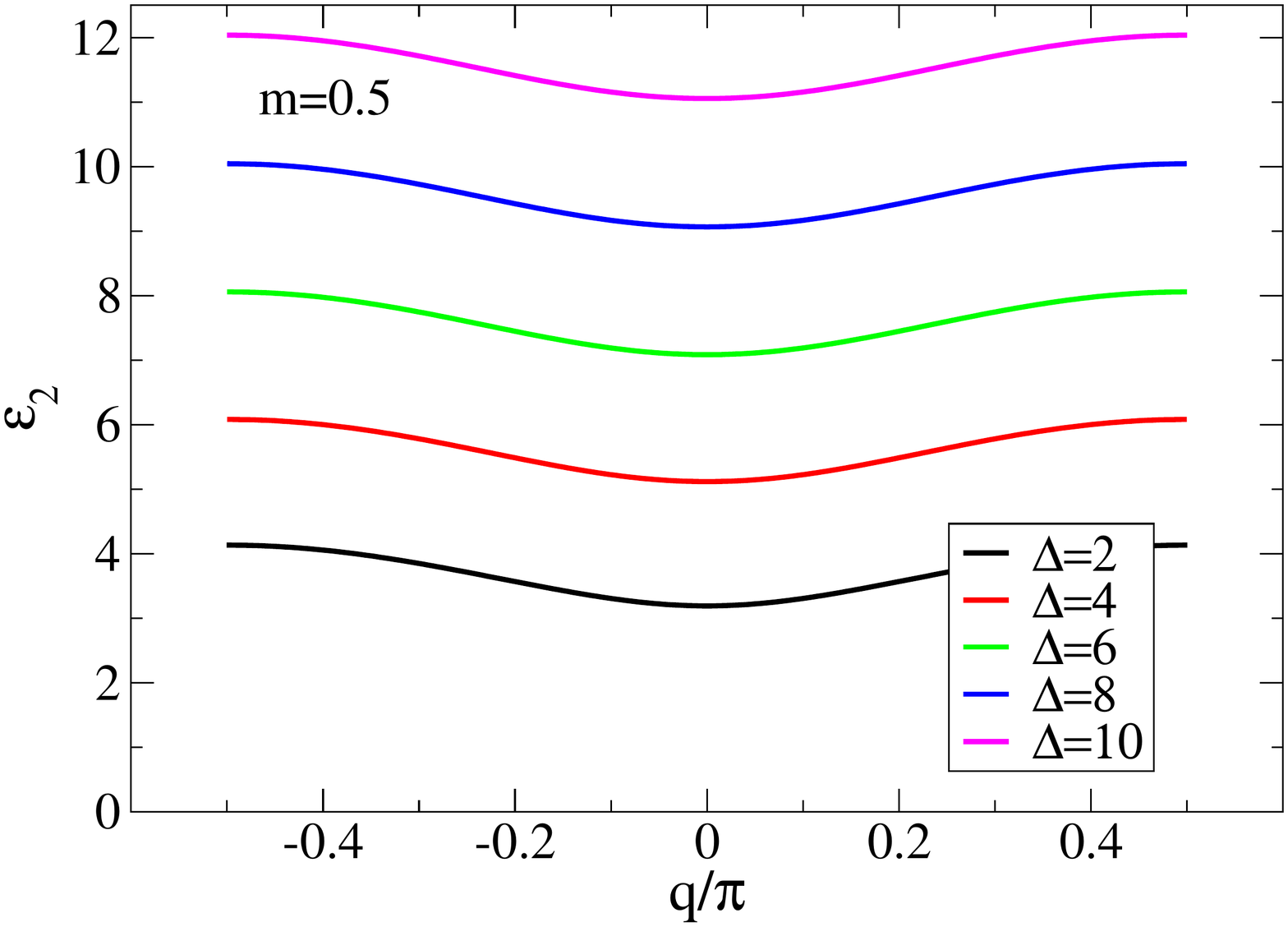}}
\hspace{0.25cm}
\subfigure{\includegraphics[width=5.65cm]{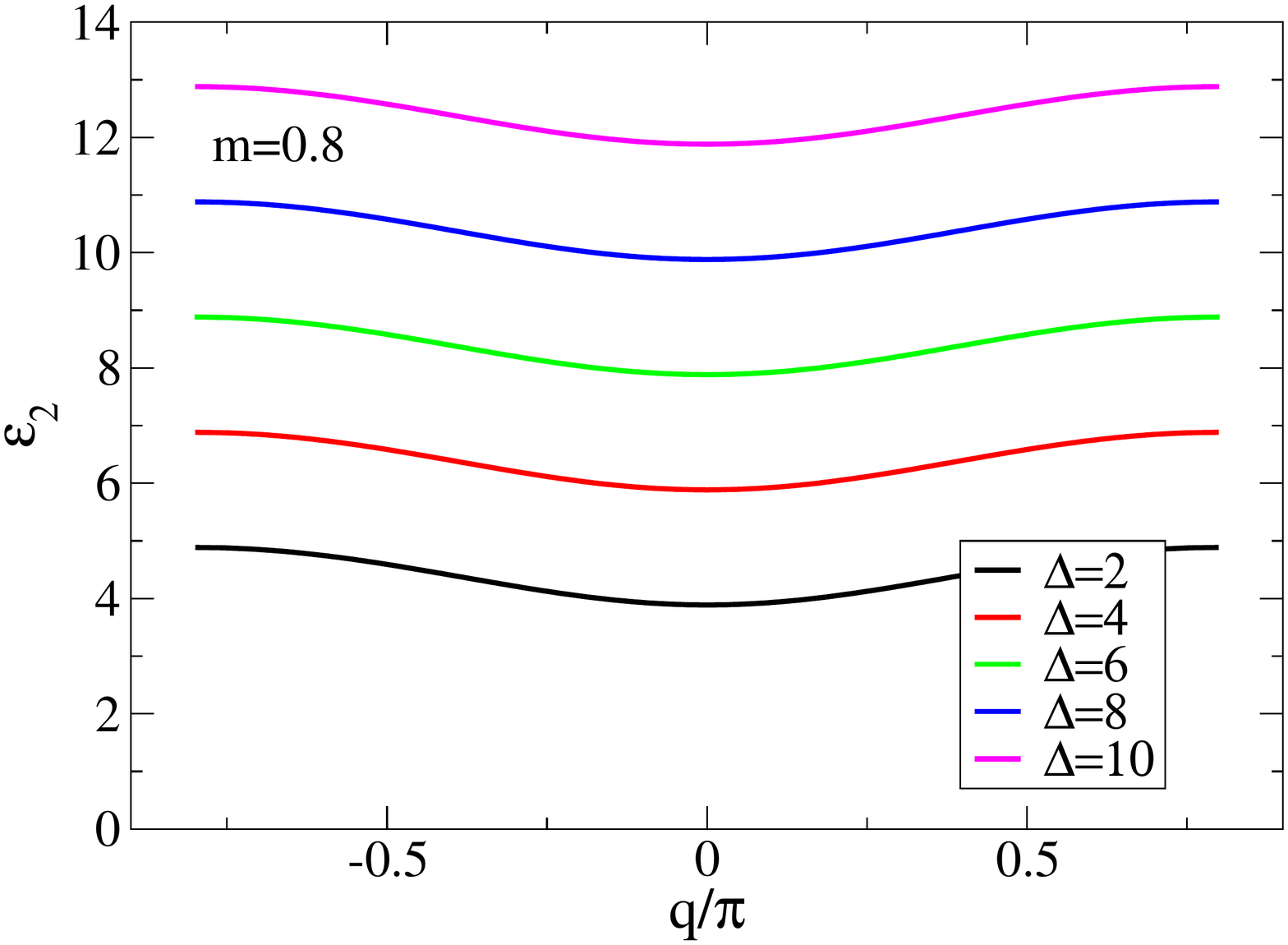}}
\caption{The $2$-string-particle energy dispersion $\varepsilon_{2} (q)$ in units of $J$
defined by Eqs. (\ref{equA4n}) and (\ref{equA4nb}) of Appendix \ref{B} for $n=2$ plotted as
a function of the $2$-band momentum $q \in [-(k_{F\uparrow}-k_{F\downarrow}),(k_{F\uparrow}-k_{F\downarrow})]$
for spin densities $m=0.2$, $m=0.5$, and $m=0.8$ and several anisotropy
$\Delta$ values.}
\label{figure2NPB}
\end{center}
\end{figure}

That unitary transformation defines a one-to-one relation between the $2^N$ $\eta=0$ and $2^N$ $\eta >0$ energy eigenstates,
respectively, of the Hamiltonian, Eq. (\ref{Hphi}). It is associated with unitary operators 
$\hat{U}_{\eta}^{\pm}$ such that,
\begin{equation}
\left\vert l_{\rm r},S_q,S^z,\eta\right\rangle = \hat{U}_{\eta}^+\left\vert l_{\rm r},S,S^z,0\right\rangle 
\hspace{0.40cm}{\rm and}\hspace{0.40cm}\left\vert l_{\rm r},S,S^z,0\right\rangle = \hat{U}_{\eta}^-\left\vert l_{\rm r},S_q,S^z,\eta\right\rangle \, .
\label{Ophi}
\end{equation}
Fortunately, the specific involved form of the unitary operators $\hat{U}_{\eta}^{\pm}$ 
is not needed for our studies. Only that they are uniquely defined and
the transformations they generate, Eq. (\ref{Ophi}), are needed.

As discussed below in Sec. \ref{SECIIIB}, in Appendix \ref{A} it is confirmed that the values of $l_{\rm r}$, $S_q$, and $S^z$ in 
$\left\vert l_{\rm r},S_q,S^z,\eta\right\rangle$ actually remain invariant 
under $\hat{U}_{\eta}^{\pm}$, yet only at $\eta =0$ the number $S_q$ is the spin $S$.
That invariance implies that for any two $\eta > 0$ and $\eta = 0$ energy eigenstates,
respectively, related as  in Eq. (\ref{Ophi}),
the $q$-spin $S_q$ has for $\eta >0$ exactly the same values as the spin $S$ at $\eta =0$.
($N$ is even and odd when the states $S_q$ is an integer and half-odd integer number, respectively.) 
 
That the irreducible representations of the $\eta >0$ continuous $SU_q(2)$ symmetry \cite{Pasquier_90,Prosen_13}
are isomorphic to those of the $\eta=0$ $SU(2)$ symmetry, requires that,
\begin{eqnarray}
\hat{S}^{\pm}_{\eta}\left\vert l_{\rm r},S_q,S^z,\eta\right\rangle & \propto &
\hat{U}_{\eta}^+\hat{S}^{\pm}\hat{U}_{\eta}^-\left\vert l_{\rm r},S_q,S^z,\eta\right\rangle
\nonumber \\
(\hat{\vec{S}}_{\eta})^2\left\vert l_{\rm r},S_q,S^z,\eta\right\rangle & \propto &
\hat{U}_{\eta}^+(\hat{\vec{S}})^2\hat{U}_{\eta}^-\left\vert l_{\rm r},S_q,S^z,\eta\right\rangle \, .
\label{SSSS}
\end{eqnarray}

To confirm that this requirement is fulfilled, the generators of the continuous $SU_q(2)$ symmetry 
are in the following expressed in terms of those of the spin symmetry $SU(2)$ symmetry. Besides transformations 
generated by the unitary operators $\hat{U}_{\eta}^{\pm}$ in Eqs. (\ref{Ophi}) and (\ref{SSSS}), this includes state summations,
\begin{equation}
\sum_{l_{\rm r},S_q,S^z} =  \sum_{l_{\rm r}}\sum_{S_q=0\hspace{0.1cm}{\rm or}\hspace{0.1cm}1/2\hspace{0.1cm}}^{N/2}
\sum_{S^z=-S_q}^{S_q} \, .
\label{sums}
\end{equation}
Here (i) $\sum_{S_q=0}^{N/2}$ and (ii) $\sum_{S_q=1/2}^{N/2}$ refers to
(i) $N$ even and $S_q$ integer and (ii) $N$ odd and $S_q$ half integer, respectively.
The summation $\sum_{l_{\rm r}}$ gives the number 
of fixed-$S_q$ energy eigenstates that is found below to equal 
the number ${\cal{N}}_{\rm singlet} (S_q)$ of corresponding $q$-spin 
singlet configurations at fixed $S_q$.

The following expressions refer to the required exact relations,
\begin{eqnarray}
\hat{S}^{\pm}_{\eta} & = & \sum_{l_{\rm r},S_q}
\sum_{S^z=-S_q+{(1\mp 1)\over 2}}^{S_q-{(1\pm 1)\over 2}}
\sqrt{{\sinh^2 (\eta\,(S_q+1/2)) - \sinh^2 (\eta\,(S^z \pm 1/2))\over
((S_q+1/2)^2 - (S^z \pm 1/2)^2)\sinh^2 \eta}}
\nonumber \\
& \times & \left\langle l_{\rm r},S_q,S^z\pm 1,\eta\right\vert
\hat{U}_{\eta}^+\hat{S}^{\pm}\hat{U}_{\eta}^-
\left\vert l_{\rm r},S_q,S^z,\eta\right\rangle
\left\vert l_{\rm r},S_q,S^z\pm 1,\eta\right\rangle\left\langle l_{\rm r},S_q,S^z,\eta\right\vert \, ,
\label{OneSOphipm}
\end{eqnarray}
and
\begin{eqnarray}
&& (\hat{\vec{S}}_{\eta})^2 = \sum_{l_{\rm r},S_q,S^z}
{\sinh^2 (\eta\,(S_q+1/2)) - \sinh^2 (\eta/2)\over
((S_q+1/2)^2 - 1/4)\sinh^2 \eta}
\nonumber \\
& \times & \left\langle l_{\rm r},S_q,S^z,\eta\right\vert
\hat{U}_{\eta}^+(\hat{\vec{S}})^2\hat{U}_{\eta}^-
\left\vert l_{\rm r},S_q,S^z,\eta\right\rangle
\left\vert l_{\rm r},S_q,S^z,\eta\right\rangle\left\langle l_{\rm r},S_q,S^z,\eta\right\vert \, ,
\label{CasimirO}
\end{eqnarray}
respectively. Combining this latter summation over the operators $\left\vert l_{\rm r},S_q,S^z\right\rangle\left\langle l_{\rm r},S_q,S^z\right\vert$ 
with the expression of $\hat{S}^+_{\eta}\hat{S}^-_{\eta}$ obtained from the use of Eq. (\ref{OneSOphipm}),
straightforwardly leads to the known $(\hat{\vec{S}}_{\eta})^2$'s expression
\cite{Pasquier_90,Prosen_13},
\begin{equation}
(\hat{\vec{S}}_{\eta})^2 = \hat{S}^+_{\eta}\hat{S}^-_{\eta} - {\sinh^2 (\eta/2)\over\sinh^2 \eta} 
+ {\sinh^2 (\eta\,(\hat{S}^z+1/2))\over\sinh^2 \eta}  \, .
\label{CasimirOknown}
\end{equation}
\begin{figure}
\begin{center}
\subfigure{\includegraphics[width=5.65cm]{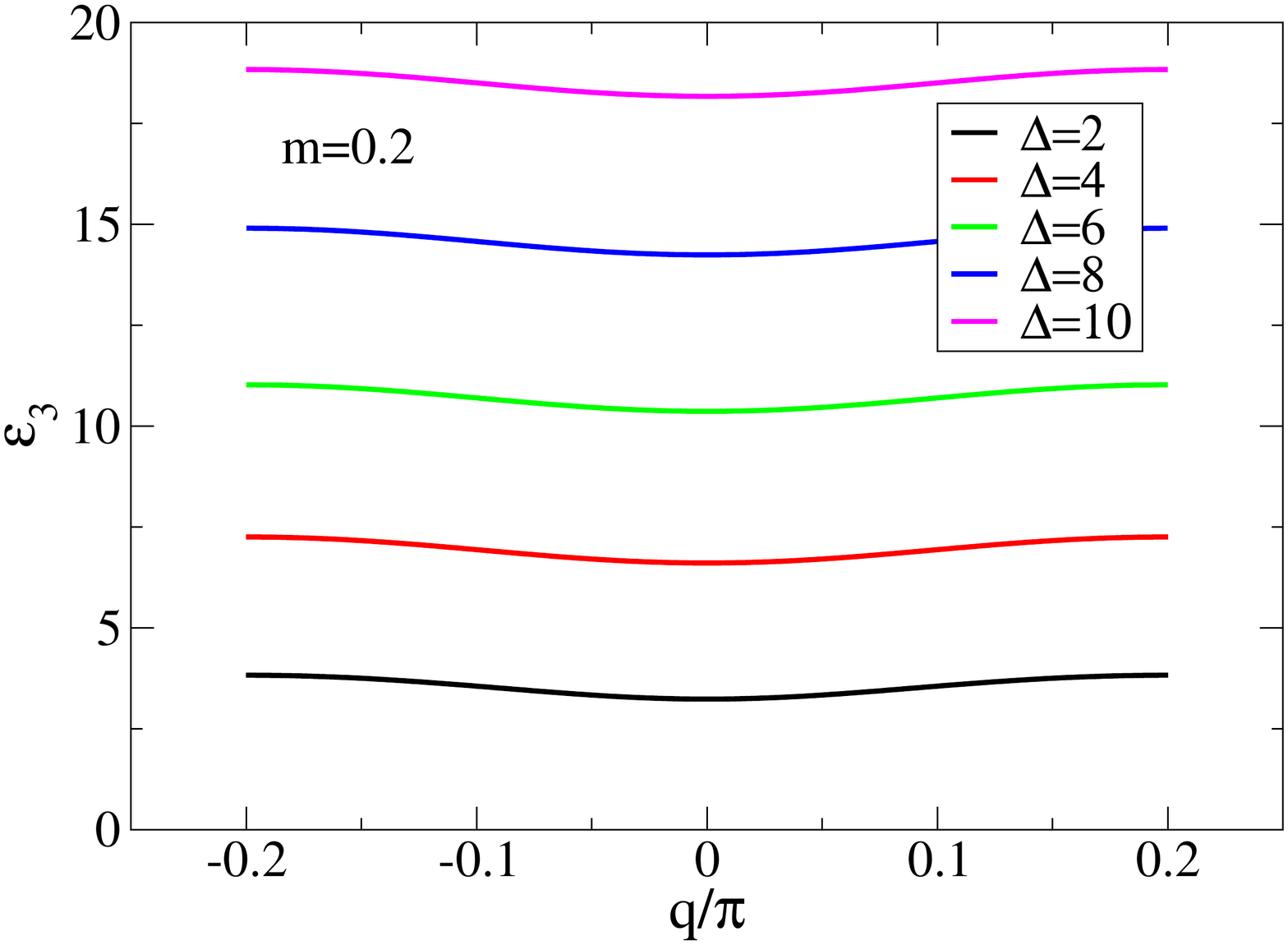}}
\hspace{0.25cm}
\subfigure{\includegraphics[width=5.65cm]{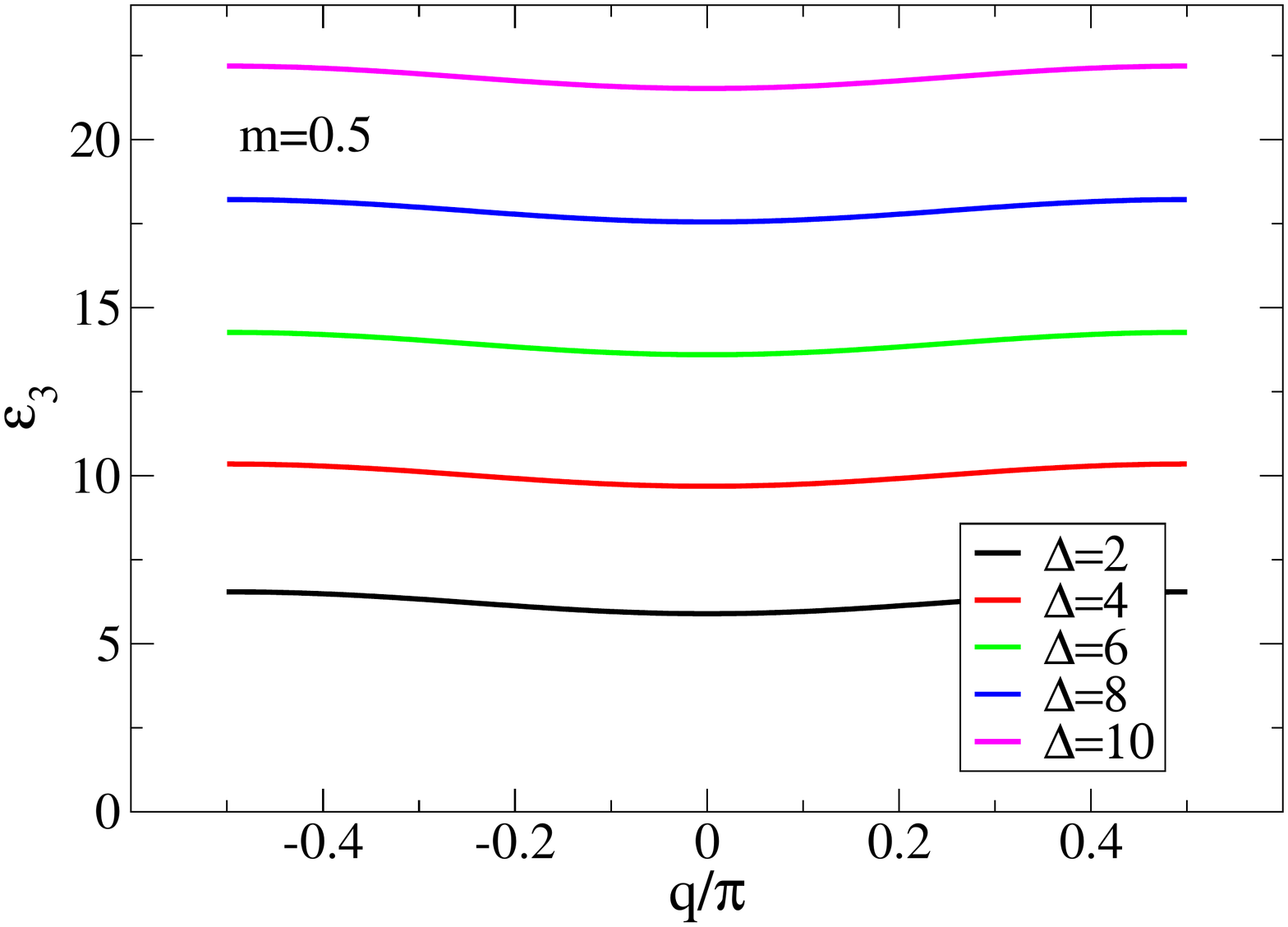}}
\hspace{0.25cm}
\subfigure{\includegraphics[width=5.65cm]{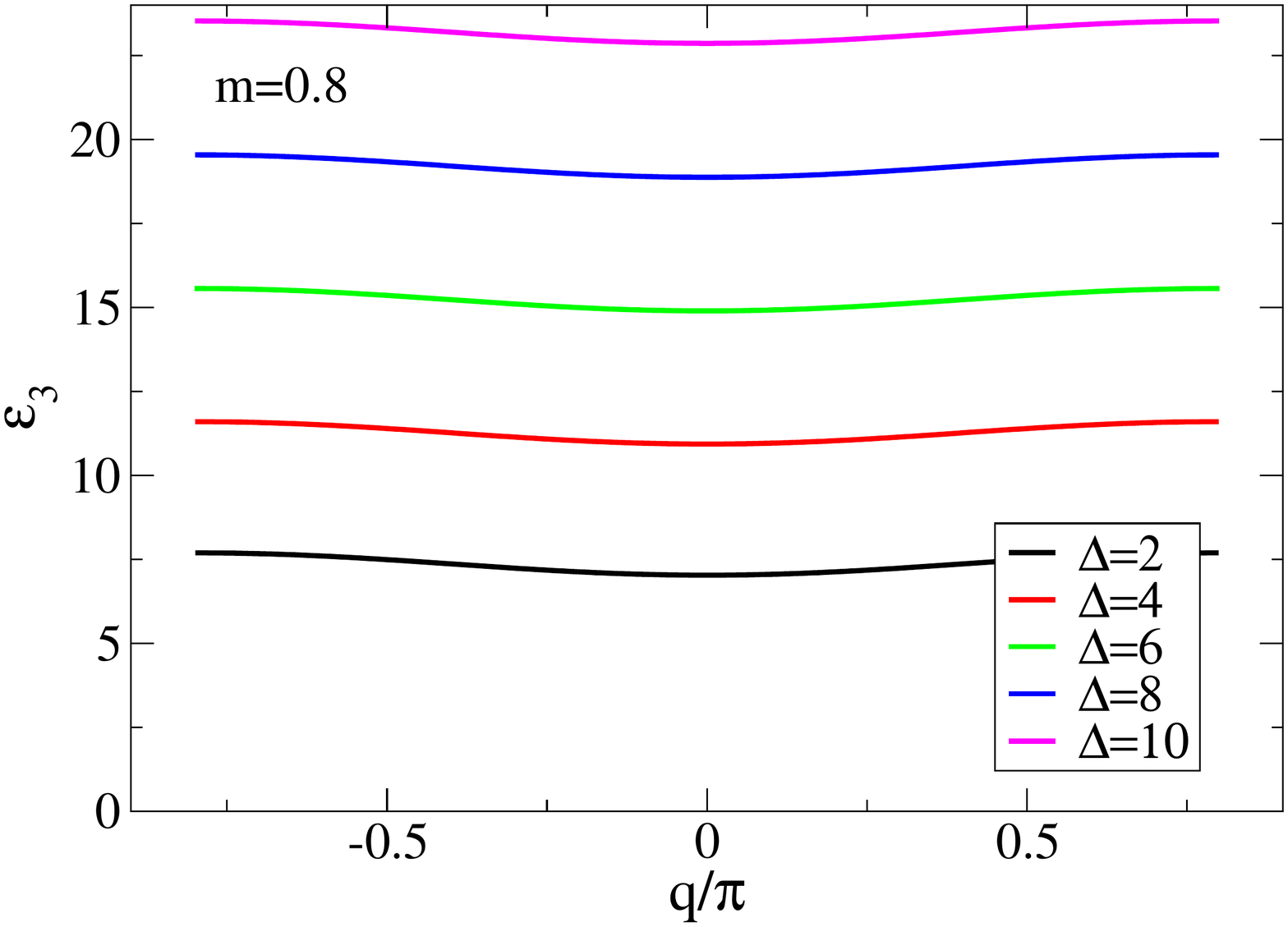}}
\caption{The same as in Fig. \ref{figure2NPB} for the $3$-string-particle energy dispersion $\varepsilon_{3} (q)$.}
\label{figure3NPB}
\end{center}
\end{figure}

That of the commutator $[\hat{S}^{+}_{\eta},\hat{S}^{-}_{\eta}]$
in terms of the operator $\hat{U}_{\eta}^+[\hat{S}^{+},\hat{S}^{-}]\hat{U}_{\eta}^-$ is thus given by,
\begin{eqnarray}
[\hat{S}^{+}_{\eta},\hat{S}^{-}_{\eta}] & = & \sum_{l_{\rm r},S_q,S^z}
{\sinh (\eta\, 2S^z)\over 2S^z \sinh \eta}
\left\langle l_{\rm r},S_q,S^z,\eta\right\vert
\hat{U}_{\eta}^+[\hat{S}^{+},\hat{S}^{-}]\hat{U}_{\eta}^-
\left\vert l_{\rm r},S_q,S^z,\eta\right\rangle
\left\vert l_{\rm r},S_q,S^z,\eta\right\rangle\left\langle l_{\rm r},S_q,S^z,\eta\right\vert 
\nonumber \\
& = & {\sinh (\eta\,2\hat{S}^z)\over\sinh \eta} \, .
\label{commPPhiDelta1}
\end{eqnarray}

Important symmetries are associated with the following commutations involving
the momentum operator and $\hat{S}^z$,
\begin{equation}
[\hat{P},\hat{U}_{\eta}^{\pm}]=0 \, ; \hspace{0.40cm}
[\hat{S}^z,\hat{U}_{\eta}^{\pm}]=0 \, .
\label{commPSzU}
\end{equation}
It follows that both the momentum eigenvalues $P$, Eq. (\ref{P}), and spin projection 
$S^z$ are good quantum numbers independent of $\eta$ for the whole $\eta \geq 0$ range.
Consistently, the commutator $[\hat{S}^z,\hat{S}^{\pm}_{\eta}]$ has the known simple form,
\begin{equation}
[\hat{S}^z,\hat{S}^{\pm}_{\eta}] = \pm \hat{S}^{\pm}_{\eta} \, .
\label{commSzPhiDelta1}
\end{equation}

The expressions of the $q$-spin $SU_q (2)$ symmetry
generators, Eqs. (\ref{OneSOphipm}) and (\ref{CasimirO}), continuously evolve into
those of the spin $SU(2)$ symmetry as $\eta$ is continuously decreased to zero. 
And the $SU_q(2)$ symmetry itself also continuously evolves into the $SU(2)$ symmetry 
as $\eta$ is adiabatically turned off to zero.

That the irreducible representations of the $\eta >0$ continuous $SU_q(2)$ symmetry 
are isomorphic to those of the $\eta=0$ $SU(2)$ symmetry, with the spin
$S$ being replaced by the $q$-spin $S_q$, refers to an important symmetry.
It is behind the use of the unified notation $\left\vert l_{\rm r},S_q,S^z,\eta\right\rangle$
of the energy eigenstates for $\eta\geq 0$. Within it, one has that $S_q=S$ at $\eta =0$. 
The whole analysis reported in the following also applies to the isotropic point under the
replacement of $q$-spin $S_q$ by spin $S$.

\subsection{Confirmation of the $n=1$ unbound pairs and $n>1$ bound pairs $q$-spin neutral nature}
\label{SECIIIB}

For $\eta >0$, a singlet means a $S_q=0$ configuration of physical spins $1/2$.
Standard counting of spin $SU(2)$ symmetry irreducible representations also applies to
those of the continuous $q$-spin $SU_q (2)$ symmetry. One then finds from the use of the two isomorphic algebras that 
$\sum_{l_{\rm r}}={\cal{N}}_{\rm singlet} (S_q)$ for the model in each fixed-$S_q$ subspace. Here,
\begin{eqnarray}
{\cal{N}}_{\rm singlet}  (S_q) = {N\choose N/2-S_q}-{N\choose N/2-S_q-1} \, ,
\label{Nsinglet}
\end{eqnarray}
is that subspace number of independent singlet configurations of physical spins $1/2$.
Including multiplet configurations when $S_q>0$, the dimension is larger and given by 
${\cal{N}}(S_q) = (2S_q+1)\,{\cal{N}}_{\rm singlet}  (S_q)$. Consistently, 
\begin{equation}
\sum_{S_q}\,{\cal{N}}(S_q) = 2^{N} \, ,
\label{SumRuleSphi}
\end{equation}
gives the total number of both irreducible representations and energy and momentum eigenstates. 

For $S_q>0$, one finds from the use of the $SU(2)$ and $SU_q(2)$ symmetries algebras
that each energy eigenstate is populated by physical spins $1/2$ in two types of configurations.
A set of $M=2S_q$ physical spins $1/2$ 
participate in a multiplet configuration. A complementary set of even number $2\Pi=N-2S_q$ physical
spins $1/2$ participate in singlet configurations. All the states with the same $S_q$ value
whose number is ${\cal{N}}(S_q) = (2S_q+1)\,{\cal{N}}_{\rm singlet}  (S_q)$ 
have the same Casimir operator's eigenvalue, $[\sinh^2 (\eta (S_q +1/2)) - \sinh^2 (\eta/2)]/\sinh^2 \eta$.

It follows that the energy eigenstates are superpositions of such configuration terms.
In them, each term is characterized by a different partition of $N$ physical spins $1/2$. 
$M=2S_q$ such spins participate in a $2S_q+1$ multiplet. A product of singlets involves the remaining
even number $2\Pi=N-2S_q$ of physical spins $1/2$. Those form a tensor product of singlet states. 

The {\it unpaired spins $1/2$} and {\it paired spins $1/2$} are the members of such two sets of $M = 2S_q$ and 
$2\Pi = N-2S_q$ physical spins $1/2$, respectively. 

The $SU_q(2)$ symmetry quantum number $S_q$ naturally emerges within
the Bethe-ansatz solution through the $n=1,2,...$ numbers $N^h_{n} = 2S_q +\sum_{n'=n+1}^{\infty}2(n'-n)N_{n'}$, Eq. (\ref{Ln}).
In the following it is confirmed that the dimension of each subspace spanned
by $q$-spin lowest-weight-states (LWSs) with fixed values of $S_q$ is in terms of Bethe-ansatz quantum numbers
occupancy configurations also given by ${\cal{N}}_{\rm singlet} (S_q)$, Eq. (\ref{Nsinglet}).
This is a needed major step towards showing that $n$-strings and $n=1$ real single Bethe rapidities
describe bound states of a number $n=2,...,\infty$ of $S^z=S_q=0$ physical spins $1/2$ pairs 
and a single unbound such a pair, respectively.

The Bethe-ansatz solution refers to subspaces spanned by $SU_q(2)$ symmetry LWSs  
$\left\vert l_{\rm r},S_q,-S_q,\eta\right\rangle$.
For such states, all the $M=2S_q$ unpaired physical spins $1/2$ 
have the same spin projection, $S^z = -S_q$.
(For the highest-weight-states (HWSs) 
of the $SU_q(2)$ symmetry algebra, $S^z = S_q$.)
Similarly to the bare ladder spin operators $\hat{S}^{\pm}$, the 
action of the ladder operators $\hat{S}^{\pm}_{\eta}$, Eq. (\ref{OneSOphipm}), 
onto $S_q > 0$ energy eigenstates flips a physical spin $1/2$ projection. 
A number $2S_q$ of $SU_q(2)$ symmetry non-LWSs outside the Bethe-ansatz 
solution are generated from the corresponding
$n_z = S_q + S^z=0$ LWS $\left\vert l_{\rm r},S_q,-S_q,\eta\right\rangle$ as,
\begin{equation} 
\left\vert l_{\rm r},S_q,S^z,\eta\right\rangle = 
{1\over \sqrt{{\cal{C}}_{\eta}}}({\hat{S}}^{+}_{\eta})^{n_z}\left\vert l_{\rm r},S_q,-S_q,\eta\right\rangle \, .
\label{state}
\end{equation} 
Here $n_z\equiv S_q + S^z = 1,...,2S_q$ so that $S^z = -S_q + n_z$ and,
\begin{equation}
{\cal{C}}_{\eta} = 
\prod_{l=1}^{n_z}{\sinh^2 (\eta\,(S_q+1/2)) - \sinh^2 (\eta\,(l - S_q - 1/2))\over\sinh^2 \eta} 
\hspace{0.40cm}{\rm for}\hspace{0.40cm}n_z= 1,...,2S_q \, .
\label{nonBAstatesDelta1}
\end{equation}

One has for LWSs of the Bethe-ansatz solution
that $-2S^z = 2S_q = N - \sum_{n'=1}^{\infty}2n'\,N_{n'}$. 
For the set of non-LWSs of the same tower, the value of $S^z = -S_q + n_z$ changes upon varying that of the number $n_z= 1,...,2S_q$. 
However, the values of both $N - \sum_{n'=1}^{\infty}2n'\,N_{n'}$ 
and $2S_q$ remain unchanged for all such non-LWSs. It then follows from the boundary-condition relation,
$2S_q = N - \sum_{n'=1}^{\infty}2n'\,N_{n'}$ at $S^z = -S_q$, that it
remains valid for the whole tower of $2S_q+1$ states. 

According to the $SU(2)$ and $SU_q(2)$ symmetries algebras, 
$M = 2S_q$ and $2\Pi = N-2S_q$ are the
numbers of unpaired physical spins $1/2$ and paired physical spins $1/2$, respectively. 
The exact relation $2S_q = N - \sum_{n=1}^{\infty}2n\,N_{n}$ then implies that
the number of paired physical spins $1/2$
within an energy eigenstate with $\Pi= \sum_{n=1}^{\infty}n\,N_{n}$ singlet pairs 
is given by $2\Pi = \sum_{n=1}^{\infty}2n\,N_{n}$. 

It is due to the isomorphism between the irreducible representations of the 
$SU(2)$ and $SU_q (2)$ symmetries, that the numbers $L_n$, Eq. (\ref{Ln}), are independent of $\eta$.
As justified in Appendix \ref{A}, that independence ensures that the set of quantum numbers described 
by $l_{\rm r}$ that label the energy eigenstates, Eq. (\ref{Ophi}),
are independent of that anisotropy parameter.
This is a needed confirmation for the set of quantum numbers described by $l_{\rm r}$ 
that label such states being exactly the same for the two
$\eta >0$ and $\eta=0$ energy eigenstates related as
$\vert l_{\rm r},S_q,S^z,\eta\rangle = \hat{U}_{\eta}^+\vert l_{\rm r},S,S^z,0\rangle$, Eq. (\ref{Ophi}). 

The exact relation, $2S_q = N - \sum_{n=1}^{\infty}2n\,N_{n}$, reveals
that for each subspace with fixed $S_q$ value, the allowed numbers $N_n$ 
can be given by zero and {\it all} positive integers that obey the exact sum rule, 
$\sum_{n=1}^{\infty}2n\,N_{n} = N - 2S_q = 2\Pi$. 
Such a sum rule is a necessary condition for the validity of statement (i), that $n$-strings describe bound states of $n>1$ out of the
$\Pi = N/2-S_q$ physical spins $1/2$ singlet pairs of an energy eigenstate;
and for the validity of statement (ii), that $1$-particles refer to one unbound pair out of such 
$\Pi = N/2-S_q$ pairs.

Importantly, the independence of $\eta$ of the numbers $L_n$, Eq. (\ref{Ln}), implies
that an exact relation proved in Appendix A of Ref. \onlinecite{Takahashi_71} 
for $S = -S^z$ LWSs of the $\eta=0$ spin chain holds for all $2^N$ energy eigenstates of
the $\eta\geq 0$ chain provided that $N_{\downarrow}$ (denoted by $M$ in that
reference, which here rather denotes the number of unpaired physical spins $1/2$
of an energy eigenstate, $M=2S_q$) is replaced in it by $N/2 - S$ for $\eta =0$
and by $N/2 - S_q$ for $\eta >0$. In the following we introduce the corresponding general 
{\it exact relation} that confirms the validity of statements (i) and (ii). 

The occupancy configurations of $1$-particles and $n>1$ $n$-string-particles
of a LWS and of the non-LWSs of the same tower are {\it exactly the same}.
Indeed, such $2S_q + 1$ states only differ in the projections of the $M=2S_q$ unpaired physical spins $1/2$. 
Due to the Pauli-like occupancy of the $n$-bands, for a
subspace with fixed $S_q$ and $\{N_n\}$ values, there is in each $n$-band with finite 
occupancy $N_n>0$ a number ${L_n\choose N_n}$ of such occupancy configurations. Each of
them is associated with a different LWS and corresponding tower of non-LWSs.

For a larger fixed-$S_q$ subspace containing several subspaces with a different set
of fixed $\{N_n\}$ values, this holds for each such a set 
that obeys the exact sum rule, $\sum_{n=1}^{\infty}2n\,N_{n} = N - 2S_q = 2\Pi$. 
We recall that the dimension ${\cal{N}}(S_q) = (2S_q+1)\,{\cal{N}}_{\rm singlet} (S_q)$ of such
a larger fixed-$S_q$ subspace corresponds to $2S_q+1$ multiplet configurations and a number 
${\cal{N}}_{\rm singlet} (S_q)$ of singlet configurations given in Eq. (\ref{Nsinglet}). 
Indeed, for the non-LWSs generated from the LWSs as given in Eq.
(\ref{state}), the value of $2S_q = N - \sum_{n=1}^{\infty}2n\,N_{n}$ 
remains unchanged, which is equivalent to the sum rule $\sum_{n=1}^{\infty}2n\,N_{n} = N - 2S_q = 2\Pi$.

From the relation under consideration 
given in Appendix A of Ref. \onlinecite{Takahashi_71}, one then straightforwardly finds the following 
more general {\it exact relation},
\begin{equation}
{N\choose N/2-S_q}-{N\choose N/2-S_q-1} 
= \sum_{\{N_{n}\}}\,\prod_{n =1}^{\infty} {L_n\choose N_n} \, .
\label{Nsinglet-MM}
\end{equation}
The left-hand side of this equation is the subspace singlet dimension ${\cal{N}}_{\rm singlet} (S_q)$
in ${\cal{N}}(S_q) = (2S_q+1)\,{\cal{N}}_{\rm singlet} (S_q)$, Eq. (\ref{Nsinglet}),
in terms of physical spins $1/2$ independent configurations.
On its right-hand side, $\sum_{\{N_{n}\}}$ is a sum over all sets $\{N_{n}\}$ obeying
the two equivalent exact Bethe-ansatz sum rules $2S_q = N - \sum_{n=1}^{\infty}2n\,N_{n}$ 
and $2\Pi = \sum_{n=1}^{\infty}2n\,N_{n} = N - 2S_q$.
It follows that in each fixed-$S_q$ subspace, the number of independent configurations of the 
$1$-particles and $n$-string-particles for $n>1$
associated with the set of $\Pi = \sum_{n=1}^{\infty}n\,N_n = N/2-S_q$ singlet pairs of physical spins
$1/2$ exactly equals that of physical spins $1/2$ singlet configurations, {\it i. e.}
$\sum_{\{N_{n}\}}\,\prod_{n =1}^{\infty} {L_n\choose N_n}={\cal{N}}_{\rm singlet} (S_q)$.

This holds for the whole Hilbert space. It then confirms that each $n$-string-particle contains 
$n>1$ pairs out of a number $\Pi$ of singlet $S^z=S_q=0$ pairs associated with the $2\Pi = N-2S_q$ 
paired physical spins $1/2$ of an energy eigenstate and each $1$-particle contains one of such $\Pi$ pairs. 

This is consistent with the transformation, Eq. (\ref{state}), of 
the $S_q>0$ energy eigenstates $\vert l_{\rm r},S_q,S^z,\eta\rangle$ under
the ladder operators $\hat{S}^{\pm}_{\eta}$. Indeed,
the operator ${\hat{S}}^{+}_{\eta}$ in Eq. (\ref{state})
only flips the $n_z=1,...,2S_q=M$ unpaired physical spins $1/2$.
It leaves invariant the singlet configuration of the $2\Pi = N-2S_q$ paired physical spins $1/2$. 
That all $2\Pi = N-2S_q$ paired physical spins $1/2$ of an energy eigenstate
participate only in singlet configurations is also confirmed by the transformation laws of the $S_q=0$ energy 
eigenstates $\vert l_{\rm r},0,0,\eta\rangle$ with no unpaired physical spins $1/2$ that
are populated by $2\Pi = N$ paired physical spins $1/2$. Those transform as
$\hat{S}^{\pm}_{\eta}\vert l_{\rm r},0,0,\eta\rangle =0$.
\begin{figure}
\begin{center}
\subfigure{\includegraphics[width=8.5cm]{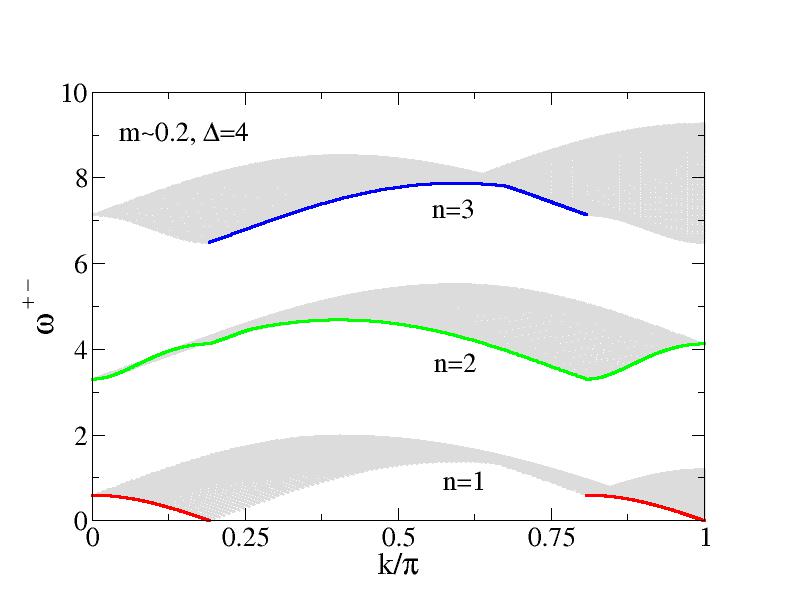}}
\hspace{0.50cm}
\subfigure{\includegraphics[width=8.5cm]{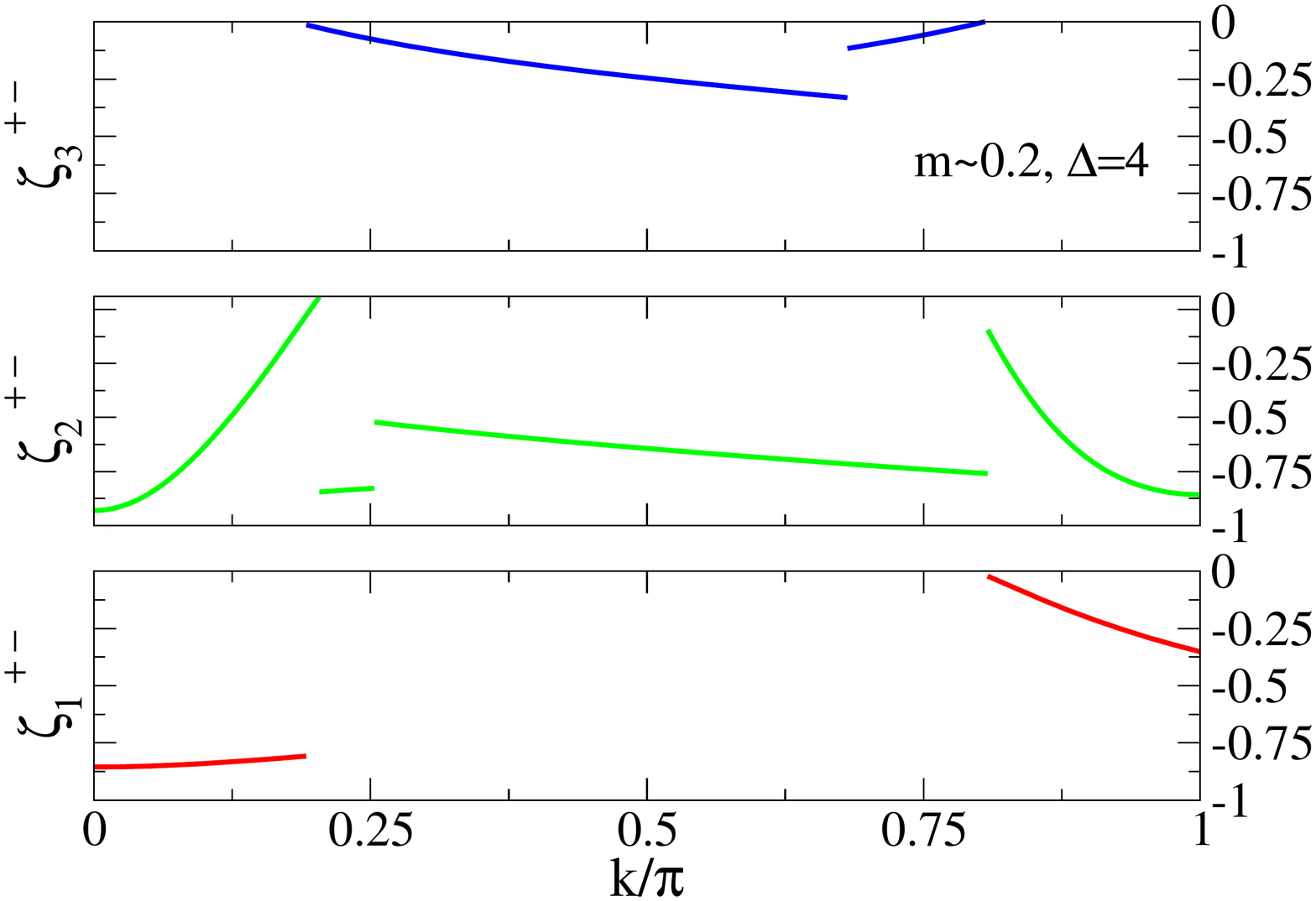}}
\subfigure{\includegraphics[width=8.5cm]{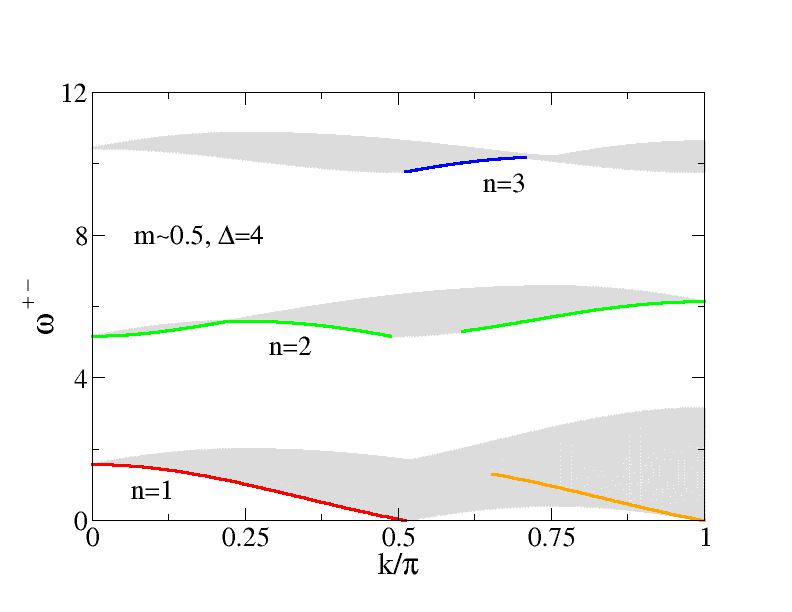}}
\hspace{0.50cm}
\subfigure{\includegraphics[width=8.5cm]{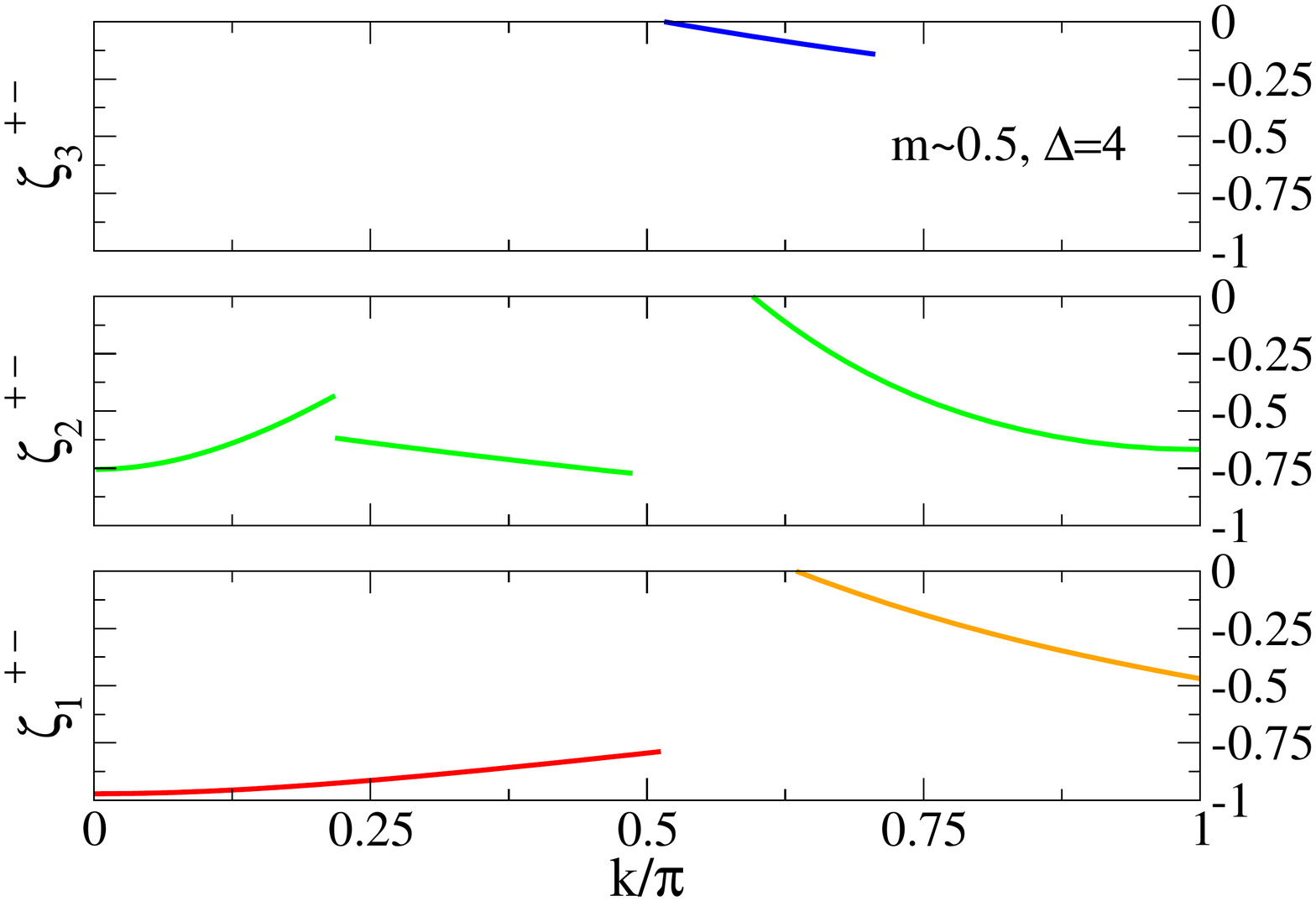}}
\subfigure{\includegraphics[width=8.5cm]{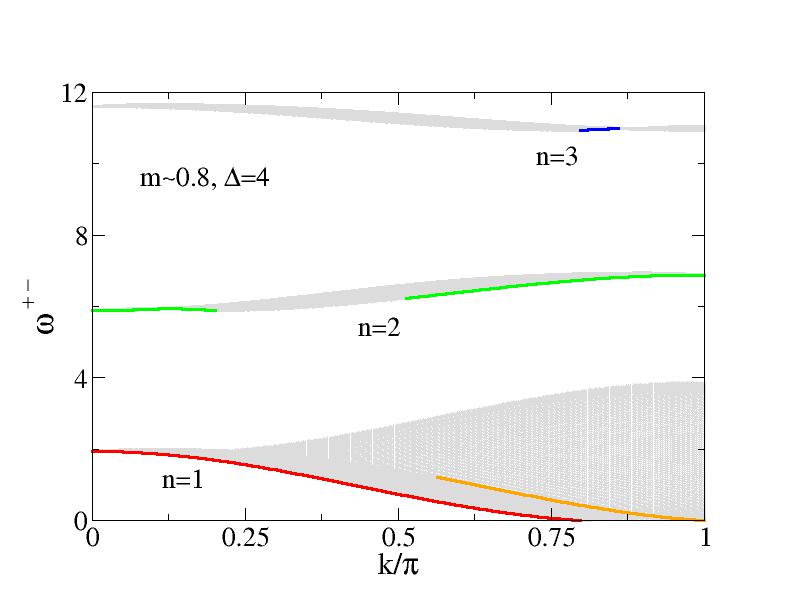}}
\hspace{0.50cm}
\subfigure{\includegraphics[width=8.5cm]{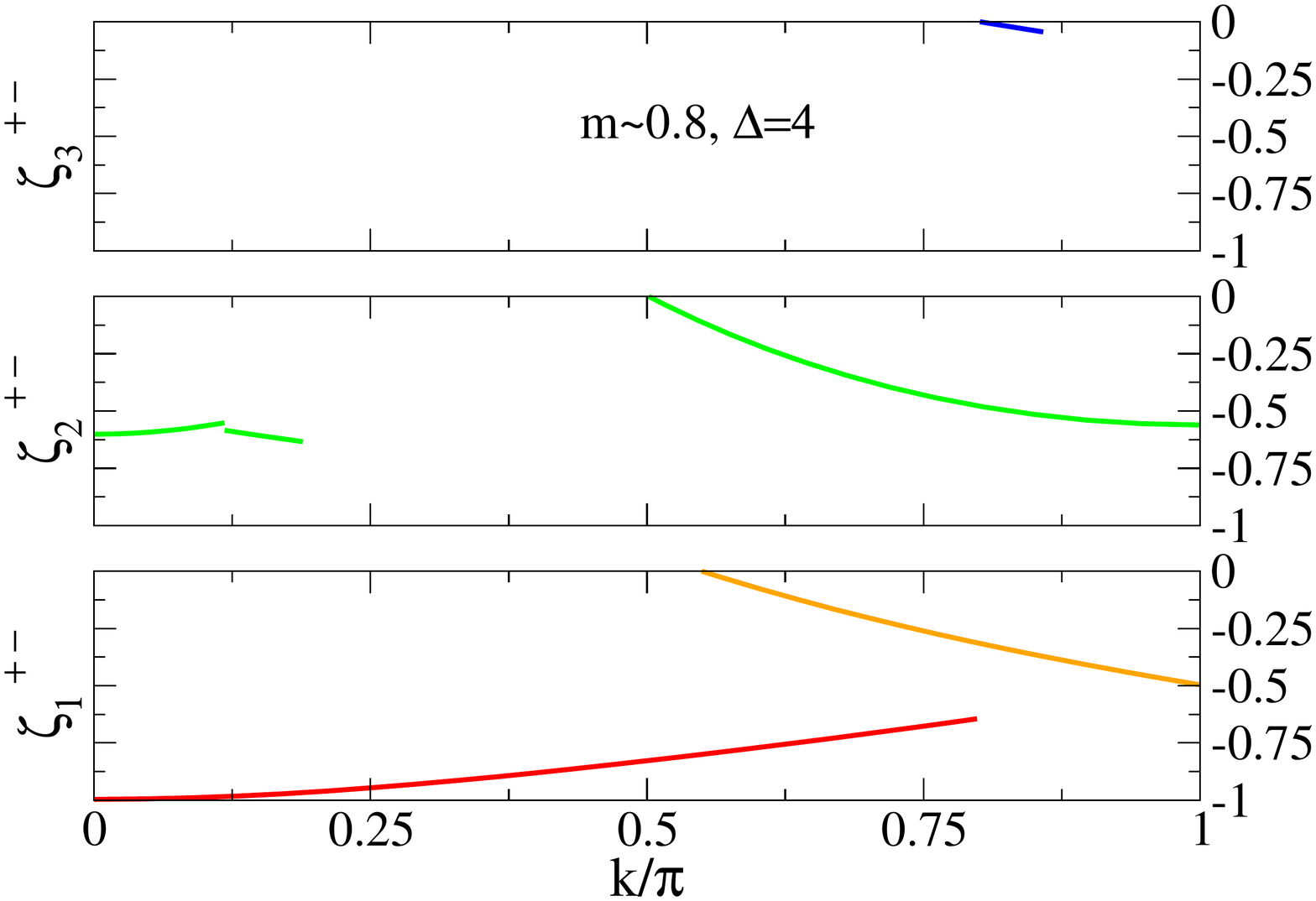}}
\caption{The $(k,\omega)$-plane continua where there is more
spectral weight in $S^{+-} (k,\omega)$ at $\Delta = 4$ for 
$m=0.1920\approx 0.2$, $m=0.5125\approx 0.5$, and $m=0.7985\approx 0.8$ (left)
and the corresponding negative $k$ dependent exponents that
control the line shape $S^{+-} (k,\omega)\propto (\omega - \omega^{+-}_n (k))^{\zeta^{+-}_n (k)}$
in the $k$ intervals near the lower thresholds of the $n=1$, $n=2$, and $n=3$ 
continua marked in the spectra (right.) In them $S^{+-} (k,\omega)$ displays sharp peaks.}
\label{figure4NPB}
\end{center}
\end{figure}

The $n>1$ pairs of physical spins $1/2$ bound within a configuration described by a $n$-string 
that corresponds to the internal degrees of freedom of one $n$-string-particle
and the unbound pair of physical spins $1/2$ that refers to the internal degrees of freedom 
of one $1$-particle only exist within the many-body system. 
Our results show that for $\eta = 0$ and $\eta >0$ such pairs have $S^z = S = 0$
and $S^z = S_q = 0$, respectively.  Hence as was known for the isotropic case \cite{Carmelo_15,Carmelo_17},
at $\eta =0$ they have zero spin. On the other hand, for $\eta >0$ spin {\it is not} a good quantum number 
and thus cannot be used to label such pairs. We have found they have zero $q$-spin $S_q$ under the 
generators of the continuous $SU_q(2)$ symmetry algebra. 

The result of this section that $n=1$ real single Bethe rapidities describe 
unbound singlet $S^z=S_q=0$ pairs of physical spins $1/2$ and $n>1$ Bethe $n$-strings
describe a number $n=2,3,...$ of bound such pairs renders the extension of the
$\Delta =1$ dynamical theory used in the studies of Refs. \onlinecite{Carmelo_20,Carmelo_15A}
to $\Delta >1$ a simple and direct procedure.

Finally, that the results reported in this section refer to the thermodynamic limit is consistent with the use of the 
ideal $n$-strings of form, Eq. (\ref{LambdaIm}), considered in Refs. \onlinecite{Takahashi_71} and 
\onlinecite{Gaudin_71}. Deformations from the ideal $n$-strings 
that occur in large finite-size systems \cite{Takahashi_03}, such as the collapse of narrow pairs,  
preserve the total number $\Pi = N/2-S_q$ of singlet $S^z=S_q=0$ pairs of physical spins $1/2$. 
They only lead to a different distribution of such pairs within the exact sum rule, 
$\sum_{n=1}^{\infty}n\,N_n = \Pi$. Such finite-size deformations do not affect our results as
the equalities in Eq. (\ref{Nsinglet-MM}) remain unchanged.

Before addressing the above mentioned dynamical theory extension and the use of the corresponding 
extended dynamical theory for $\Delta >1$, in the following the issue
where in the Bethe-ansatz solution is stored the information on the unpaired physical spins $1/2$
is shortly discussed.

\subsection{Where in the Bethe-ansatz solution are the unpaired physical spins $1/2$?}
\label{SECIIIC}

Out of the $2\Pi = \sum_{n=1}^{\infty}2n\,N_n = N-2S_q$ paired physical spins $1/2$ of an energy eigenstate,
two are contained in each of its $N_1$ $1$-particles described by the $n=1$ real single Bethe rapidities
and $2n$ in each of its $N_n$ $n$-string-particles are for $n>1$ described by the Bethe $n$-strings.
In the case of $S_q>0$ energy eigenstates, the question is then where in the Bethe-ansatz 
solution's quantities is stored the information on the $M=2S_q$ unpaired physical spins $1/2$ leftover? 

The clarification of this issue involves a squeezed space 
construction \cite{Kruis_04}. 
Each Bethe-ansatz $n$-band is found to correspond in the thermodynamic limit
to a $n$-squeezed lattice with $L_n = N_n + N^h_{n}$
sites. A number $N_n^h$ of such sites are not occupied by $1$-particles for $n=1$
and by $n$-string-particles for $n>1$. Hence they are the $n$-squeezed lattice unoccupied sites.

Out of the $N$ sites of the original lattice, the occupancies of 
$2S_q$ such sites by the $M=2S_q$ unpaired physical spins $1/2$ remain invariant under
the squeezed space construction. This important symmetry refers to each of the local configurations 
whose superposition generates  
an energy eigenstate.

Information on the $M = 2S_q$ unpaired physical spins $1/2$ is then found to be stored by the Bethe-ansatz solution
in the $n$-squeezed lattice's $N^h_{n}$ unoccupied sites. Those are directly related 
to the $N^h_{n}$ $n$-holes. Out of the $N^h_{n} = 2S_q +\sum_{n'=n+1}^{\infty}2(n'-n)N_{n'}$, Eq. (\ref{Ln}),
unoccupied sites of each $n$-squeezed lattice, a number $2S_q$ of such sites is in the original lattice
singly occupied by unpaired physical spins $1/2$. 

For energy eigenstates without $n$-strings, one has that $N^h_{1}=2S_q$. In that case all 
$N^h_{1}=2S_q$ $1$-holes describe the translational degrees of freedom of the $M=2S_q$ 
unpaired physical spins $1/2$. Such $1$-holes are often in the literature identified with spinons, 
our results clarifying their relation to the unpaired physical spins $1/2$. 

\section{Excitation spectra of the dynamical structure factor components}
\label{SECIV}

The results of this paper reported in this section and in Sec. \ref{SECV} refer to
the spin-conducting quantum phase for magnetic fields $h_{c1}<h<h_{c2}$.
The spectra and and the dynamical properties are qualitatively different in the spin-insulating 
quantum phase associated with magnetic fields $0\leq h<h_{c1}$. The latter problem was 
studied previously in the literature \cite{Caux_08} and is not addressed in this paper.
However, the limiting behaviors for $0\leq h<h_{c1}$ and $h=h_{c2}$ of several important
quantities are given in Appendix \ref{B}.
\begin{figure}
\begin{center}
\subfigure{\includegraphics[width=8.5cm]{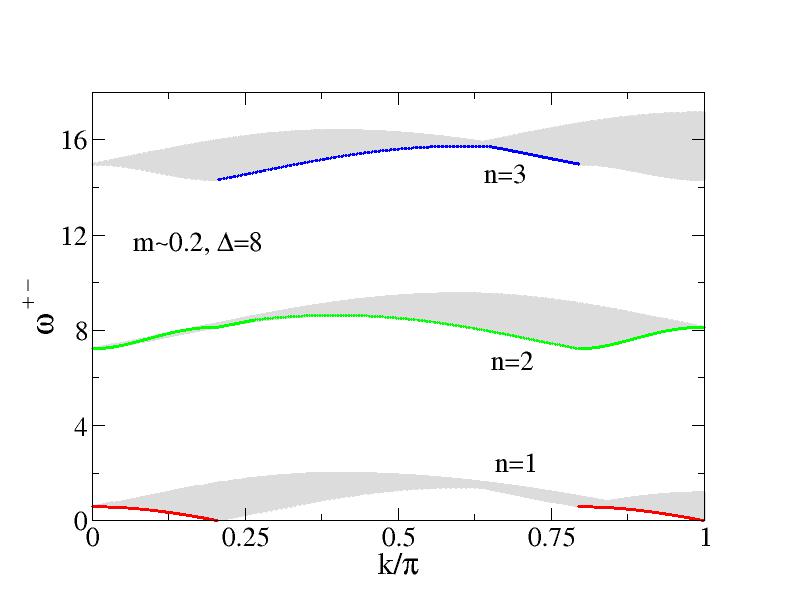}}
\hspace{0.50cm}
\subfigure{\includegraphics[width=8.5cm]{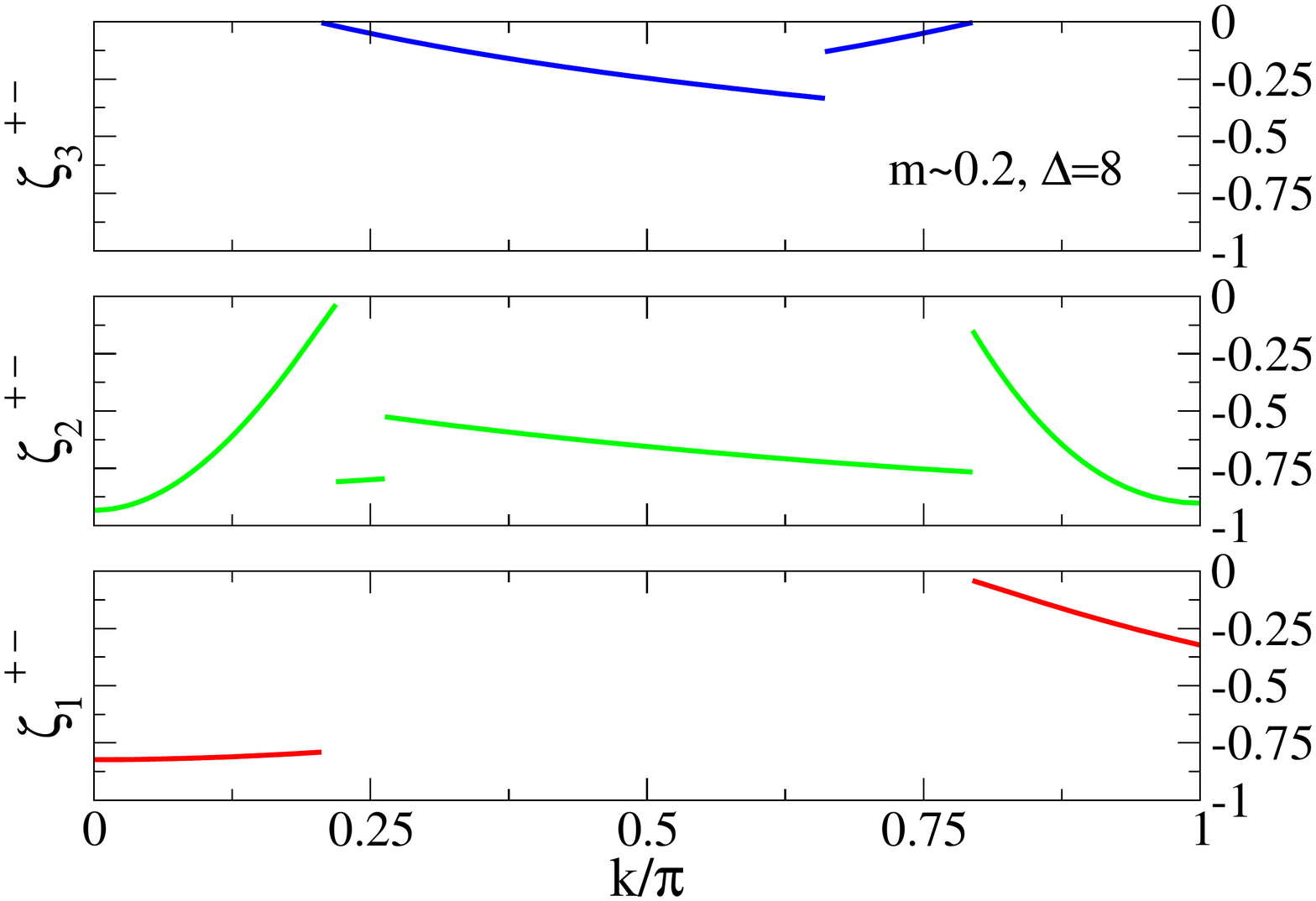}}
\subfigure{\includegraphics[width=8.5cm]{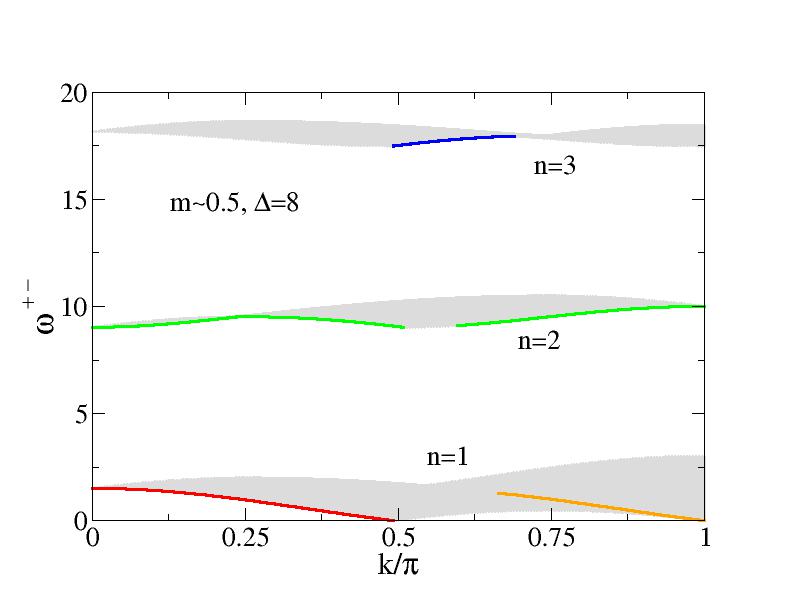}}
\hspace{0.50cm}
\subfigure{\includegraphics[width=8.5cm]{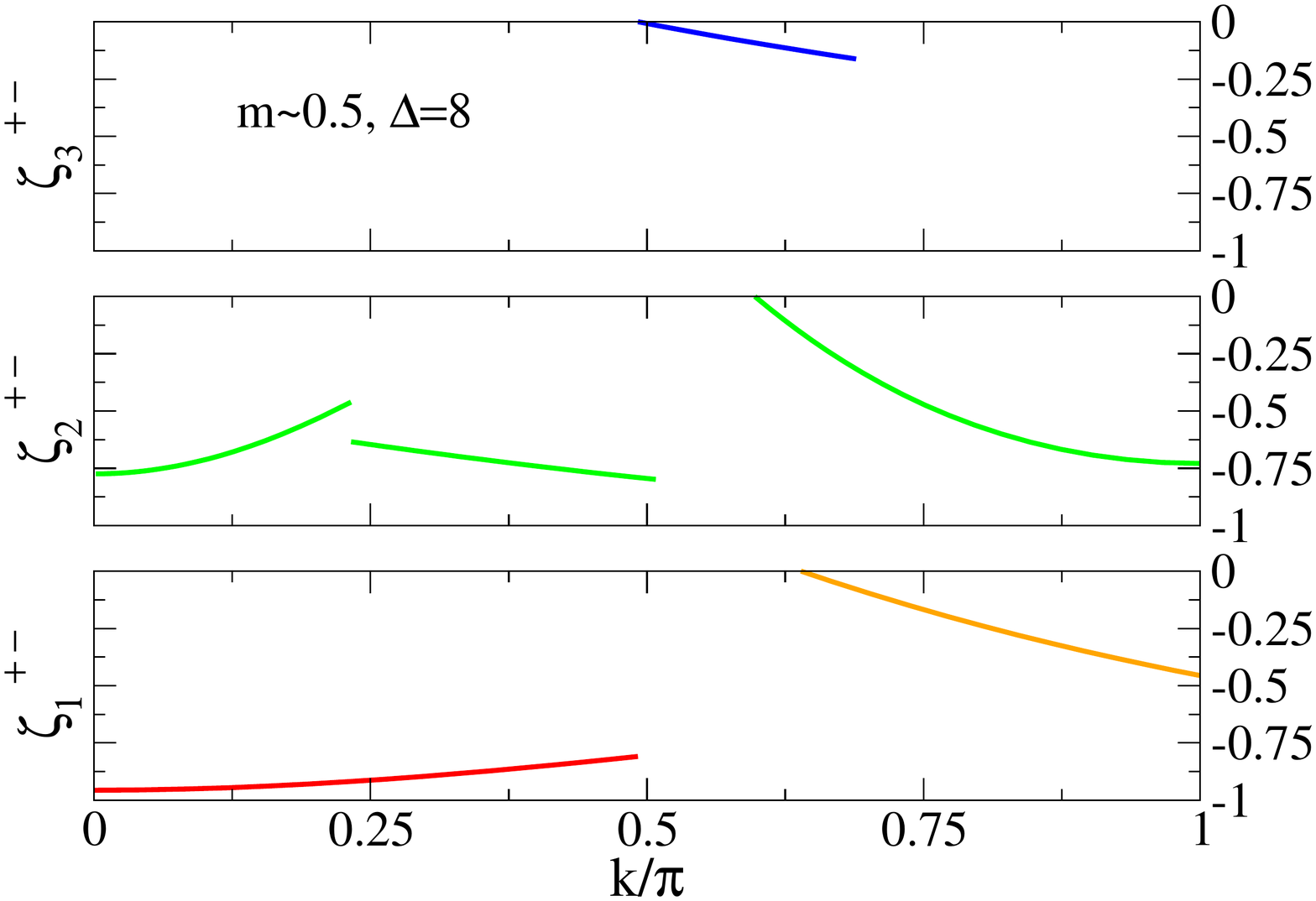}}
\subfigure{\includegraphics[width=8.5cm]{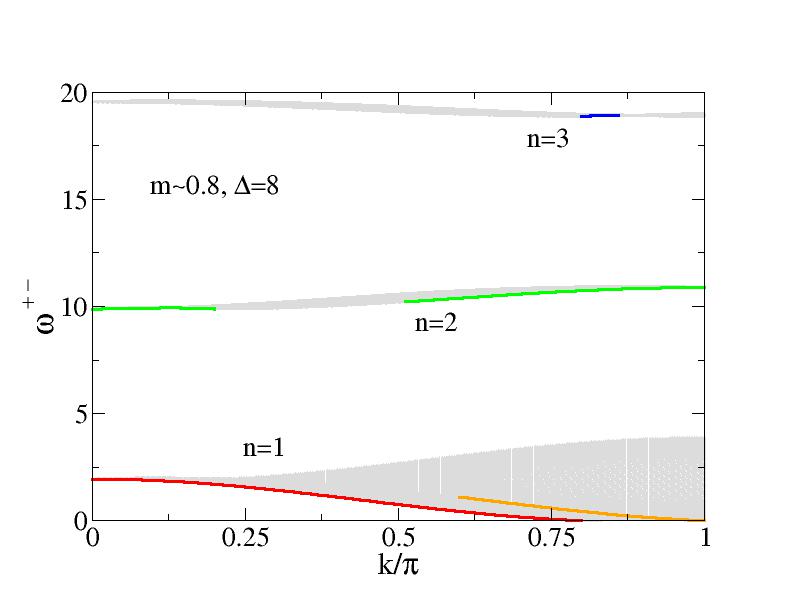}}
\hspace{0.50cm}
\subfigure{\includegraphics[width=8.5cm]{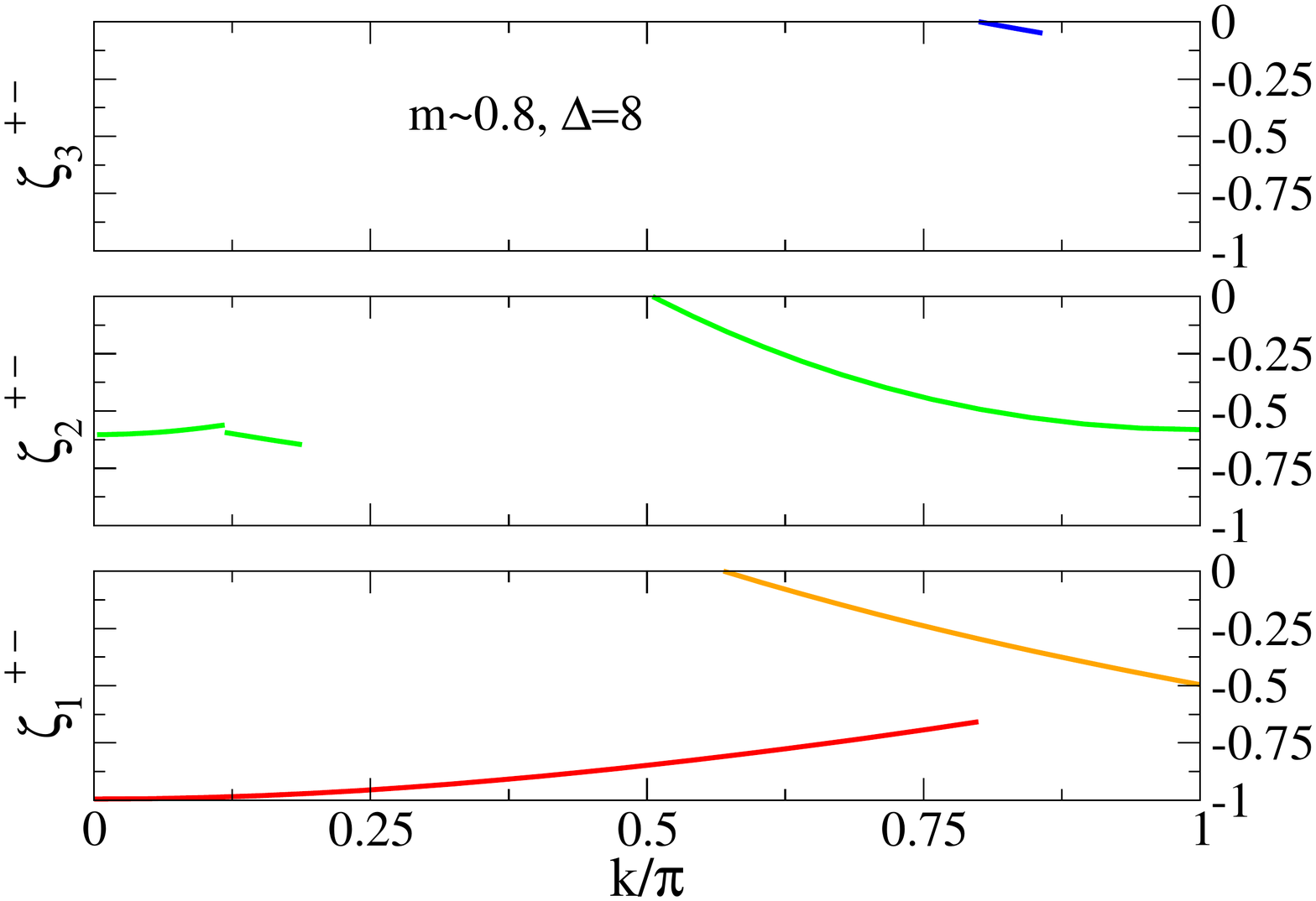}}
\caption{The same as in Fig. \ref{figure4NPB} for spin densities
$m=0.2056\approx 0.2$, $m=0.4918\approx 0.5$, and $m=0.7997\approx 0.8$ 
and anisotropy $\Delta = 8$.}
\label{figure5NPB}
\end{center}
\end{figure}

\subsection{Classes of excited states that lead to most $(k,\omega)$-plane spectral weight for $m>0$}
\label{SECIVA}

A consequence of the singlet $S^z=S_q=0$ nature of the elementary magnetic configurations
under study in Sec. \ref{SECIII}, is that the general dimension-one scalar $S$ matrices whose expression 
is given below in Sec. \ref{SECV} have for the spin-conducting phase at fields $h_{c1}<h<h_{c2}$
the same form as for the isotropic case for $0<h<h_{c2}$. Such $S$ matrices refer to
the scattering of the $q$-spin neutral $1$-particles and $q$-spin neutral $n$-string-particles 
that controls the line shape near the sharp peaks 
in $S^{+-} (k,\omega)$, $S^{-+} (k,\omega)$, and $S^{zz} (k,\omega)$.

To properly identify the $(k,\omega)$-plane lines of sharp peaks under study in Sec. \ref{SECV}, in this section we provide the expressions 
in the thermodynamic limit of the spectra of the classes of excited states that contribute to a significant amount
of spin dynamical structure factor's spectral weight. As justified in that section,
the spectra of the corresponding selected classes of excited states lead to the $(k,\omega)$-plane continua
shown in Figs. \ref{figure4NPB},\ref{figure5NPB},\ref{figure6NPB},\ref{figure7NPB},\ref{figure8NPB},\ref{figure9NPB}.
For technical reasons, such figures refer to spin densities $m=0.1920\approx 0.2$, $m=0.5125\approx 0.5$, and $m=0.7985\approx 0.8$ for
anisotropy $\Delta =4$ and spin densities $m=0.2056\approx 0.2$, $m=0.4918\approx 0.5$, and $m=0.7997\approx 0.8$ for
anisotropy $\Delta =8$. 

All $n=1$ continua in these figures
refer to spectra of excited energy eigenstates for which $N_1 = N_{\downarrow}$ and $N_n = 0$ for $n>1$ where
the number $N_{\downarrow}$ of down physical spins refers here and for all other continua to the excited states under consideration.
The $n=2$ continua in Figs. \ref{figure4NPB},\ref{figure5NPB},\ref{figure8NPB},\ref{figure9NPB} correspond to 
spectra of excited energy eigenstates for which $N_1 = N_{\downarrow}-2$, $N_2 = 1$, and $N_n = 0$ for $n>2$.
Finally, the $n=3$ continua in Figs. \ref{figure4NPB},\ref{figure5NPB} refer to spectra of excited energy eigenstates for which 
$N_1 = N_{\downarrow}-3$, $N_2 = 0$, $N_3 = 1$, and $N_n = 0$ for $n>3$. Hence, as justified in Sec. \ref{SECV}, the $n$-string states
that contribute to a significant amount of spectral weight are populated only by either a single $2$-string-particle
or a single $3$-string-particle.

In the present thermodynamic limit, the spectra of the classes of selected excited states 
have simple expressions in terms of the $1$-particle energy dispersion $\varepsilon_{1} (q)$ 
plotted in Fig. \ref{figure1NPB} and $n=2$ and $n=3$ $n$-string-particle energy dispersions $\varepsilon_{n} (q)$ 
plotted in Figs. \ref{figure2NPB} and \ref{figure3NPB}, respectively. 

\subsection{The excitation spectra with most spectral weight for $m>0$}
\label{SECIVB}

\begin{figure}
\begin{center}
\subfigure{\includegraphics[width=8.5cm]{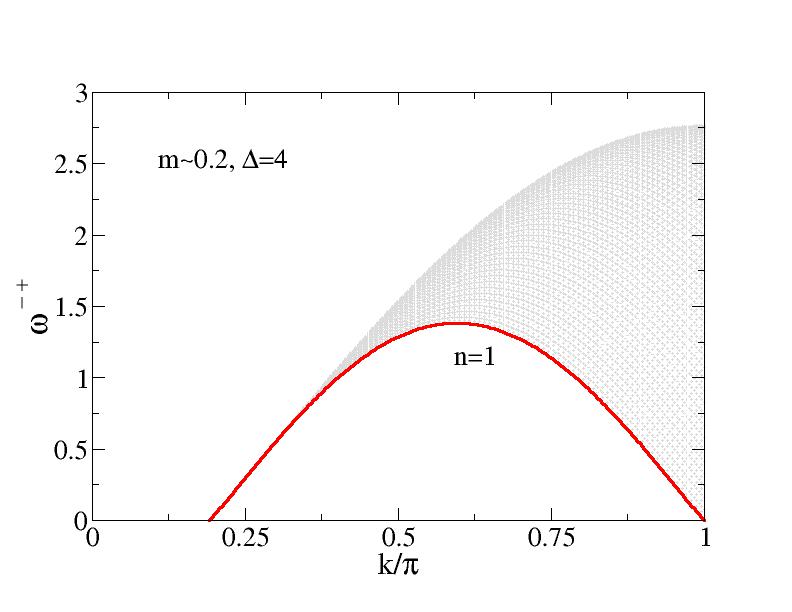}}
\hspace{0.50cm}
\subfigure{\includegraphics[width=8.5cm]{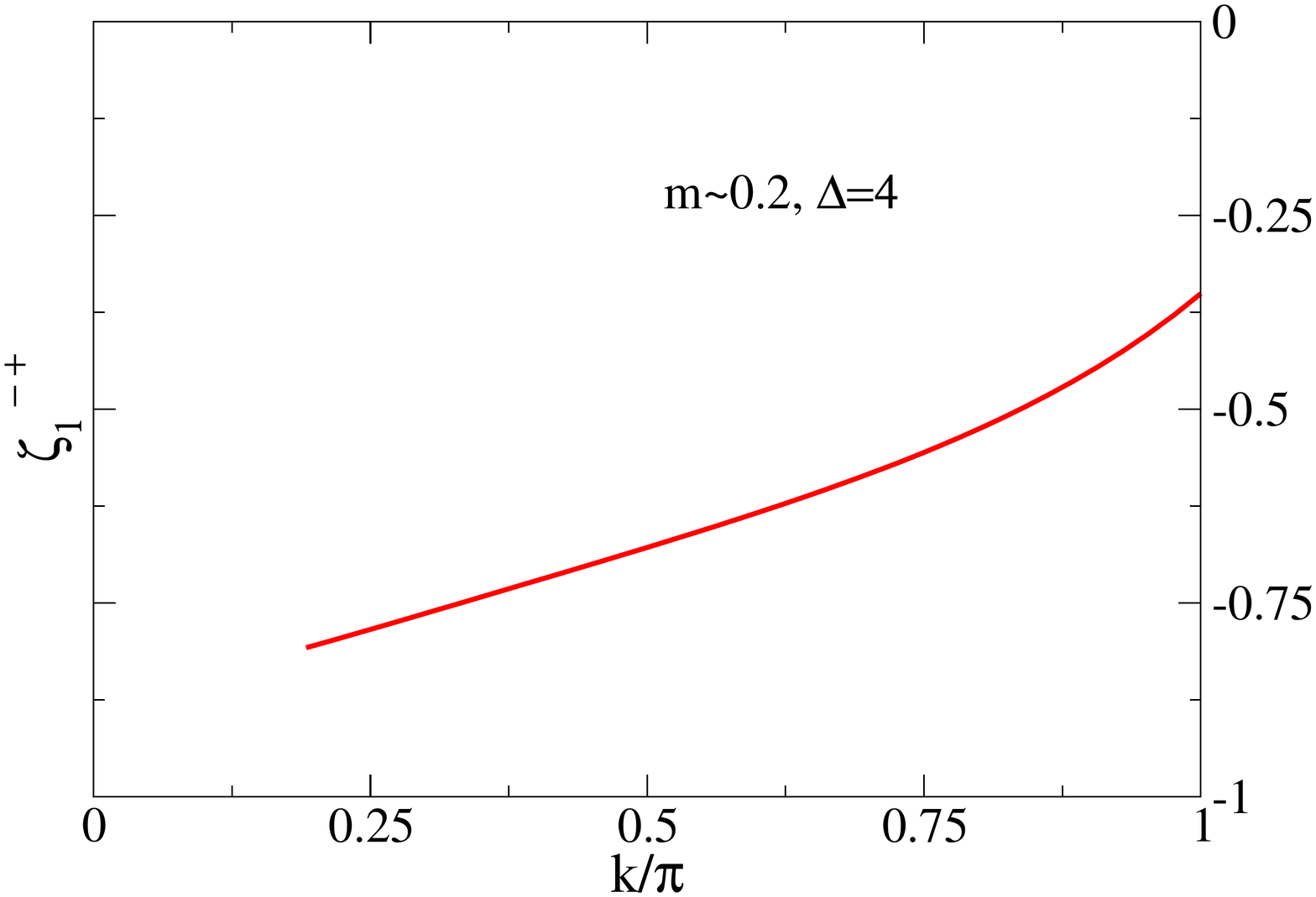}}
\subfigure{\includegraphics[width=8.5cm]{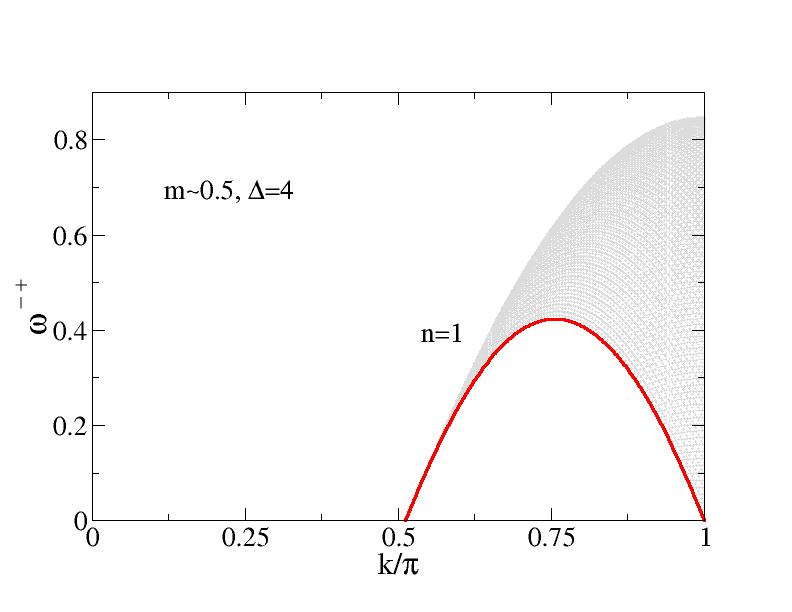}}
\hspace{0.50cm}
\subfigure{\includegraphics[width=8.5cm]{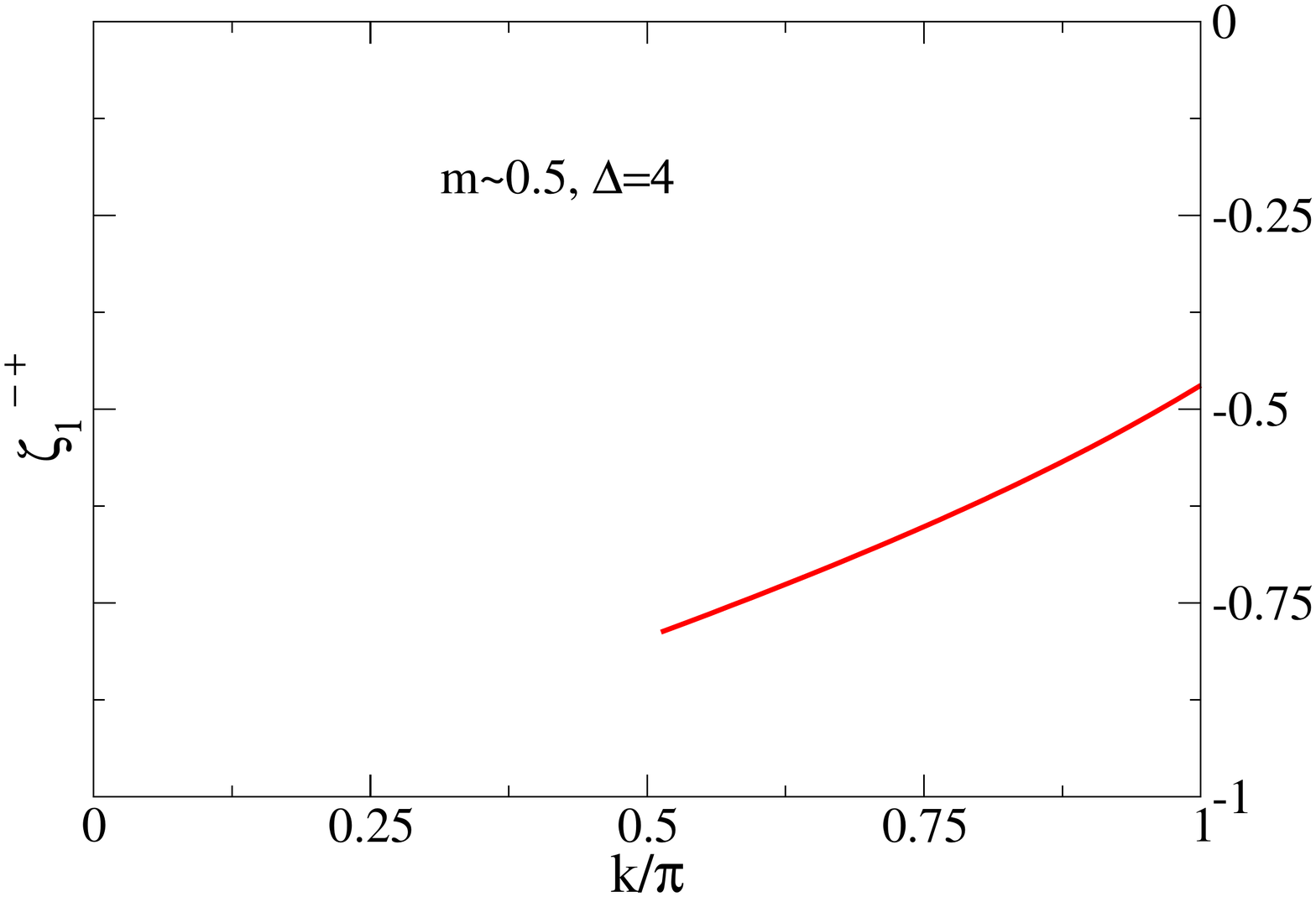}}
\subfigure{\includegraphics[width=8.5cm]{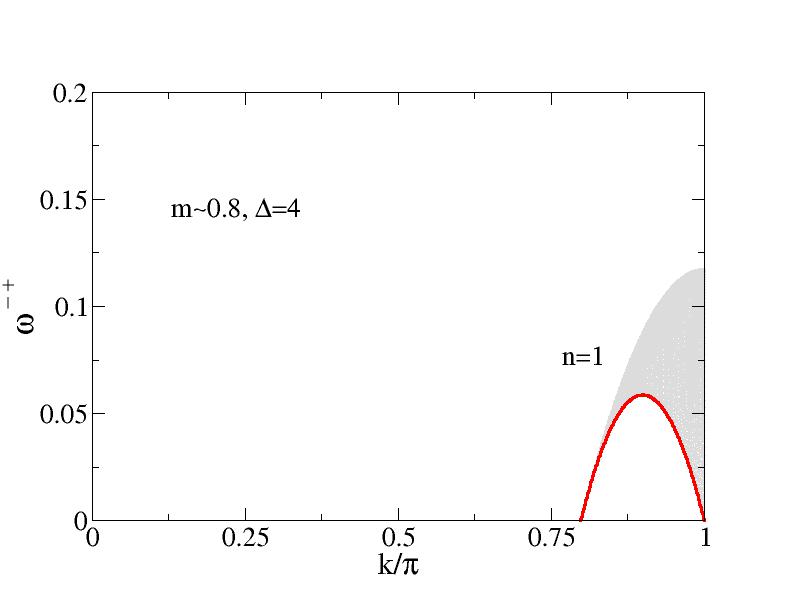}}
\hspace{0.50cm}
\subfigure{\includegraphics[width=8.5cm]{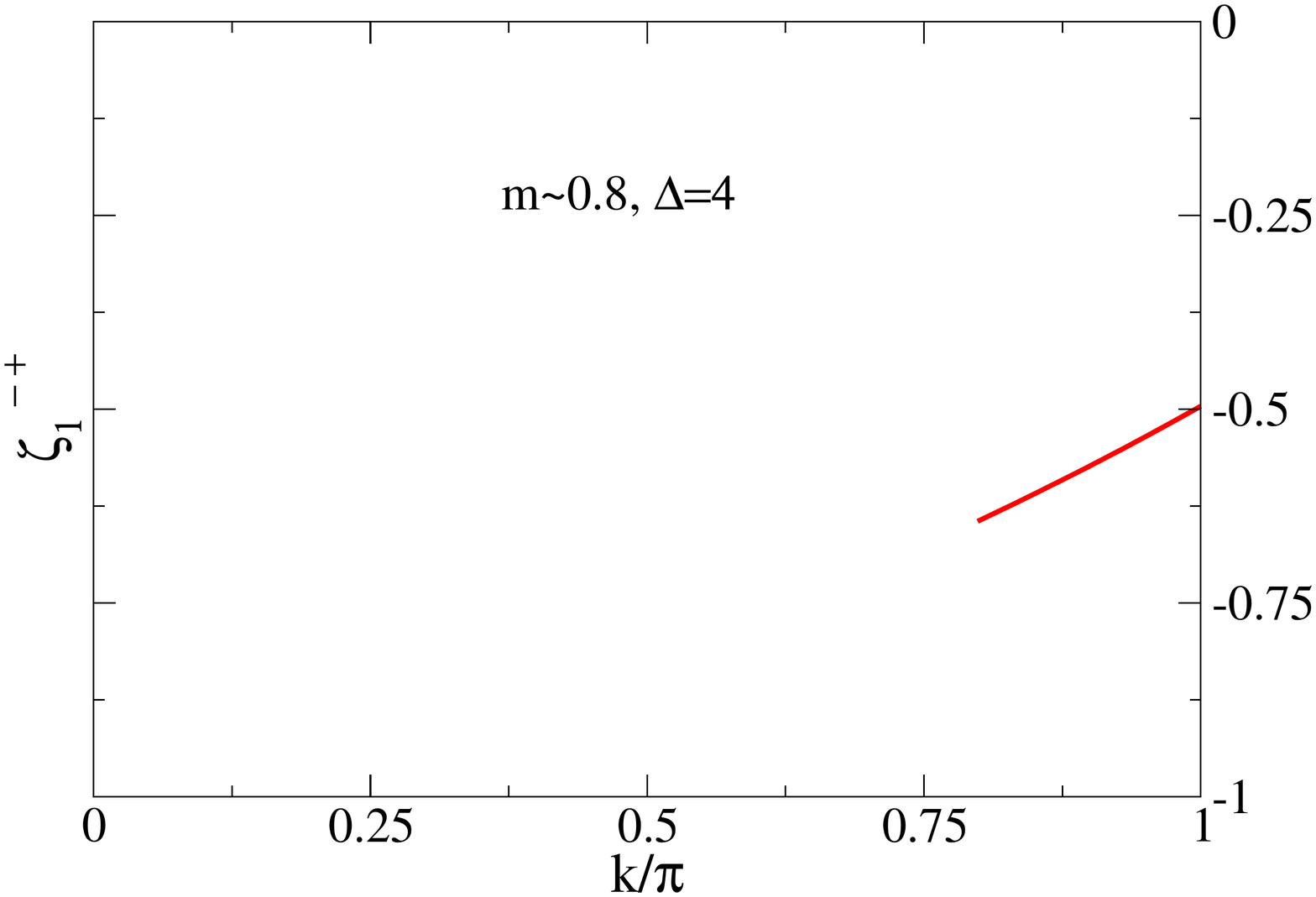}}
\caption{The $(k,\omega)$-plane continuum where there is more
spectral weight in $S^{-+} (k,\omega)$ at $\Delta = 4$ 
for $m=0.1920\approx 0.2$, $m=0.5125\approx 0.5$, and $m=0.7985\approx 0.8$ (left) and
the negative $k$ dependent exponents that
control the line shape $S^{-+} (k,\omega)\propto (\omega - \omega^{+-}_1 (k))^{\zeta^{-+}_1 (k)}$
in the marked $k$ intervals near the lower threshold of that continuum (right).}
\label{figure6NPB}
\end{center}
\end{figure}

The expressions of the two-parametric spectra given in the following refer to a $k$ extended zone scheme. However, in 
Figs. \ref{figure4NPB},\ref{figure5NPB},\ref{figure6NPB},\ref{figure7NPB},\ref{figure8NPB},\ref{figure9NPB}
they have been brought to the first Brillouin zone for positive momentum values $k>0$.
In the case of $S^{+-} (k,\omega)$, the spectrum associated with the lower $(k,\omega)$-plane $n=1$ continuum 
shown in Figs. \ref{figure4NPB},\ref{figure5NPB} is the superposition of the following two spectra,
\begin{eqnarray}
\omega^{+-}_1 (k) & = & \omega^{+-}_{1\,A} (k) + \omega^{+-}_{1\,B} (k)
\hspace{0.40cm}{\rm where}
\nonumber \\
\omega^{+-}_{1\,A} (k) & = & \varepsilon_{1} (q) - \varepsilon_{1} (q') 
\hspace{0.40cm}{\rm with}\hspace{0.40cm} k = \iota\pi + q - q'\hspace{0.40cm}{\rm for}
\nonumber \\
\vert q\vert & \in & [k_{F\downarrow},k_{F\uparrow}] \, ,
\hspace{0.40cm}q' \in [-k_{F\downarrow},k_{F\downarrow}]\hspace{0.40cm}{\rm and}\hspace{0.40cm}\iota = \pm 1
\nonumber \\
\omega^{+-}_{1\,B} (k) & = & \varepsilon_{1} (q) + \varepsilon_{1} (q') 
\hspace{0.40cm}{\rm with}\hspace{0.40cm} k = \iota\pi + q + q'\hspace{0.40cm}{\rm for}
\nonumber \\
q & \in & [k_{F\downarrow},k_{F\uparrow}] \, ,
\hspace{0.40cm}q' \in [-k_{F\uparrow},-k_{F\downarrow}]\hspace{0.40cm}{\rm and}\hspace{0.40cm}\iota = \pm 1 \, .
\label{dkEdPPMn1}
\end{eqnarray}

The spectrum of that dynamical structure factor component associated
with the gapped middle $(k,\omega)$-plane $n=2$ continuum 
shown in such figures reads,
\begin{eqnarray}
\omega^{+-}_{2} (k) & = & - \varepsilon_{1} (q) + \varepsilon_{2} (q') 
\hspace{0.40cm}{\rm with}\hspace{0.40cm} k = \iota k_{F\downarrow} - q + q'\hspace{0.40cm}{\rm for}
\nonumber \\
q & \in & [-k_{F\downarrow},k_{F\downarrow}] \, , \hspace{0.40cm}
q' \in [0,(k_{F\uparrow}-k_{F\downarrow})] \hspace{0.40cm}{\rm and}\hspace{0.40cm}\iota = 1 
\nonumber \\
q & \in & [-k_{F\downarrow},k_{F\downarrow}] \, , \hspace{0.40cm}
q' \in [-(k_{F\uparrow}-k_{F\downarrow}),0]\hspace{0.40cm}{\rm and}\hspace{0.40cm}\iota = - 1 \, .
\label{dkEdPPMn2}
\end{eqnarray}

The $S^{+-} (k,\omega)$ spectrum associated
with the gapped upper $(k,\omega)$-plane $n=3$ continuum also
shown in Figs. \ref{figure4NPB},\ref{figure5NPB} is the superposition of the
following two spectra,
\begin{eqnarray}
\omega^{+-}_{3} (k) & = & \omega^{+-}_{3\,A} (k) +  \omega^{+-}_{3\,B} (k) 
\hspace{0.40cm}{\rm where}
\nonumber \\
\omega^{+-}_{3\,A} (k) & = & - \varepsilon_{1} (q) + \varepsilon_{3} (q') 
\hspace{0.40cm}{\rm with}\hspace{0.40cm} k = \iota k_{F\uparrow} - q + q' \hspace{0.40cm}{\rm for}
\nonumber \\
q & \in & [-k_{F\downarrow},k_{F\downarrow}] \, , \hspace{0.40cm}
q' \in [-(k_{F\uparrow}-k_{F\downarrow}),0]\hspace{0.40cm}{\rm and}\hspace{0.40cm}\iota = 1 
\nonumber \\
q & \in & [-k_{F\downarrow},k_{F\downarrow}] \, , \hspace{0.40cm}
q' \in [0,(k_{F\uparrow}-k_{F\downarrow})] \hspace{0.40cm}{\rm and}\hspace{0.40cm} \iota = - 1
\nonumber \\
\omega^{+-}_{3\,B} (k) & = &- \varepsilon_{1} (q) - \varepsilon_{1} (q') + \varepsilon_{3} (0) 
\hspace{0.40cm}{\rm with} k = \iota\pi - q - q' \hspace{0.40cm}{\rm for}
\nonumber \\
q & \in & [0,k_{F\downarrow}] \, , \hspace{0.40cm} q' \in [-k_{F\downarrow},0] 
\hspace{0.40cm}{\rm and}\hspace{0.40cm}\iota = \pm 1 \, . 
\label{dkEdPPMn3}
\end{eqnarray}

As justified in Sec. \ref{SECV}, in the case of $S^{-+} (k,\omega)$ only the $(k,\omega)$-plane $n=1$ 
continuum has a significant amount of spectral weight. It is shown in Figs. \ref{figure6NPB},\ref{figure7NPB}.
The corresponding spectrum reads,
\begin{eqnarray}
\omega^{-+}_1 (k) & = & - \varepsilon_{1} (q) - \varepsilon_{1} (q')
\hspace{0.40cm} {\rm where}\hspace{0.40cm} k = \iota\pi - q - q' \hspace{0.40cm}{\rm for}
\nonumber \\
q & \in & [-k_{F\downarrow},k_{F\downarrow}] \, , \hspace{0.40cm}
q' \in [-k_{F\downarrow},k_{F\downarrow}]\hspace{0.40cm}{\rm and}\hspace{0.40cm}\iota = \pm 1 \, .
\label{dkEdPxxMP}
\end{eqnarray}

The $S^{zz} (k,\omega)$ spectrum associated with the lower $(k,\omega)$-plane $n=1$ 
continuum shown in Figs. \ref{figure8NPB},\ref{figure9NPB} is given by,
\begin{eqnarray}
\omega^{zz}_1 (k) & = & \varepsilon_{1} (q) - \varepsilon_{1} (q') 
\hspace{0.40cm}{\rm with}\hspace{0.40cm} k = q - q'\hspace{0.40cm}{\rm for}
\nonumber \\
\vert q\vert & \in & [k_{F\downarrow},k_{F\uparrow}] \, ,
\hspace{0.40cm} q' \in [-k_{F\downarrow},k_{F\downarrow}] \, .
\label{sepectzzn1}
\end{eqnarray}

The gapped $n=2$ continuum shown in these figures is the superposition of the
following two spectra,
\begin{eqnarray}
\omega^{zz}_{2} (k) & = & \omega^{zz}_{2\,A} (k) +  \omega^{zz}_{2\,B} (k) 
\hspace{0.40cm}{\rm where}
\nonumber \\
\omega^{zz}_{2\,A} (k) & = & - \varepsilon_{1} (q) + \varepsilon_{2} (q') 
\hspace{0.40cm}{\rm with}\hspace{0.40cm} k = \iota k_{F\uparrow} - q + q' \hspace{0.40cm}{\rm for} 
\nonumber \\
q & \in & [-k_{F\downarrow},k_{F\downarrow}] \, , \hspace{0.40cm}
q' \in [-(k_{F\uparrow}-k_{F\downarrow}),0] \hspace{0.40cm}{\rm and}\hspace{0.40cm}\iota = 1
\nonumber \\
q & \in & [-k_{F\downarrow},k_{F\downarrow}] \, , \hspace{0.40cm}
q' \in [0,(k_{F\uparrow}-k_{F\downarrow})]\hspace{0.40cm}{\rm and}\hspace{0.40cm}\iota = - 1
\nonumber \\
\omega^{zz}_{2\,B} (k) & = & - \varepsilon_{1} (q) - \varepsilon_{1} (q') + \varepsilon_{2} (0) 
\nonumber \\
{\rm with} && k = \iota\pi - q - q' \hspace{0.40cm}{\rm for}
\nonumber \\
q & \in & [0,k_{F\downarrow}] \, , \hspace{0.40cm} q' \in [-k_{F\downarrow},0] 
\hspace{0.40cm}{\rm and}\hspace{0.40cm}\iota = \pm 1 \, .
\label{sepectzzn2}
\end{eqnarray}

\section{The power-law line shape near the sharp peaks controlled by the $1$-$1$ and $1$-$n$ scattering
of $1$-particles and $n$-string-particles}
\label{SECV}

\begin{figure}
\begin{center}
\subfigure{\includegraphics[width=8.5cm]{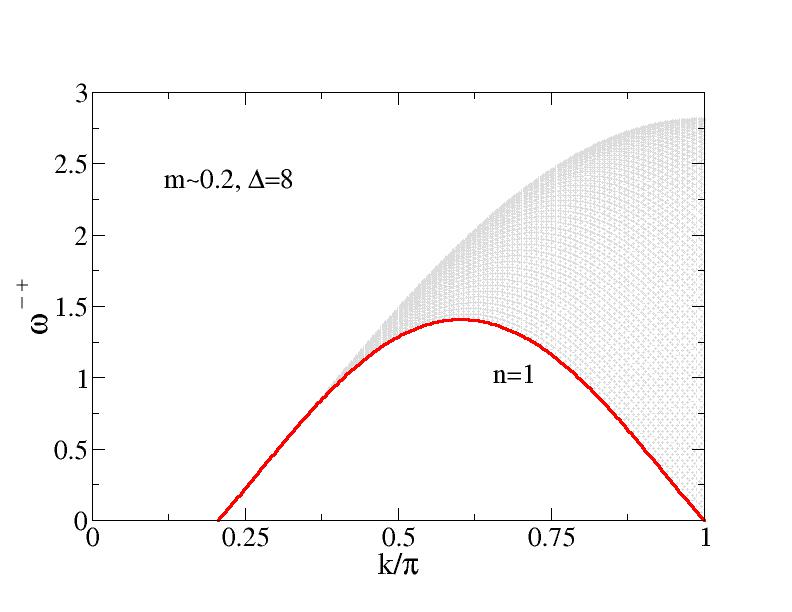}}
\hspace{0.50cm}
\subfigure{\includegraphics[width=8.5cm]{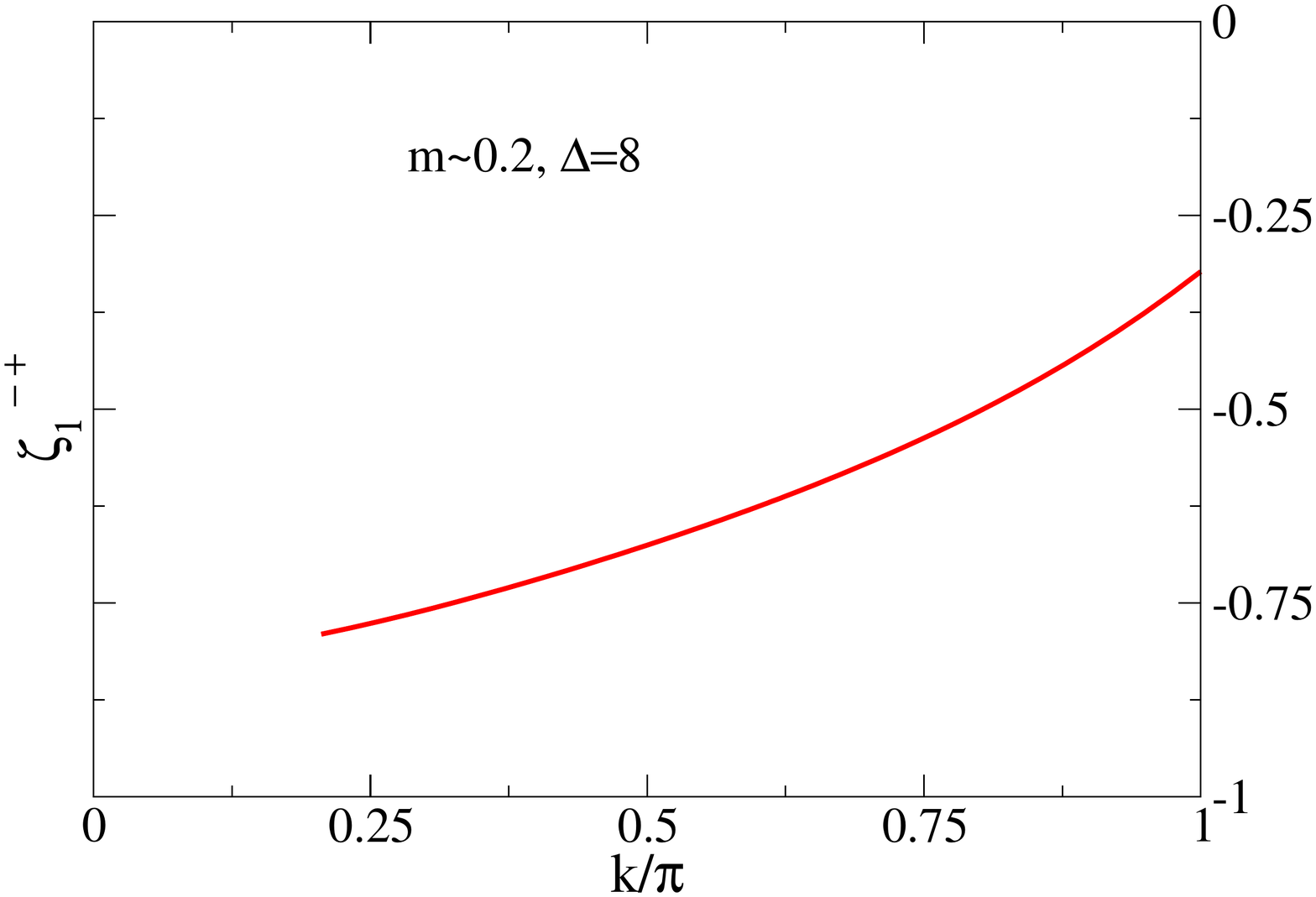}}
\subfigure{\includegraphics[width=8.5cm]{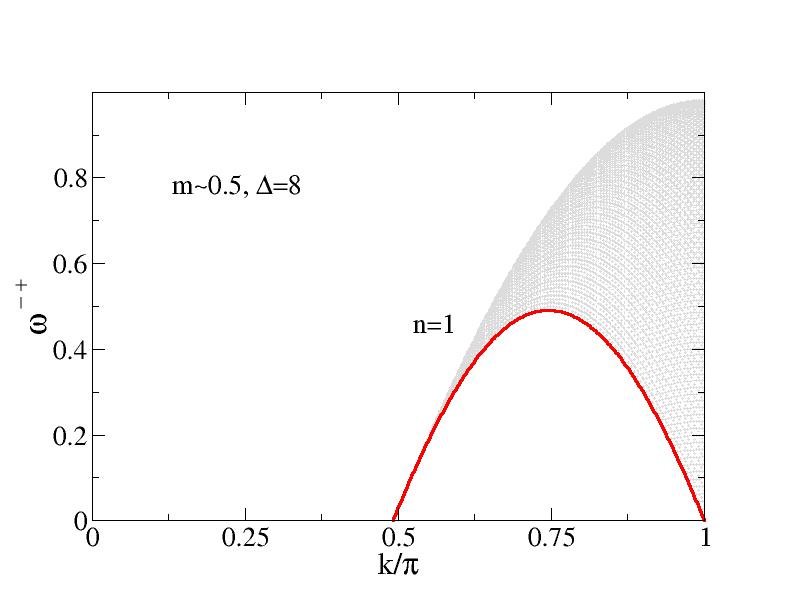}}
\hspace{0.50cm}
\subfigure{\includegraphics[width=8.5cm]{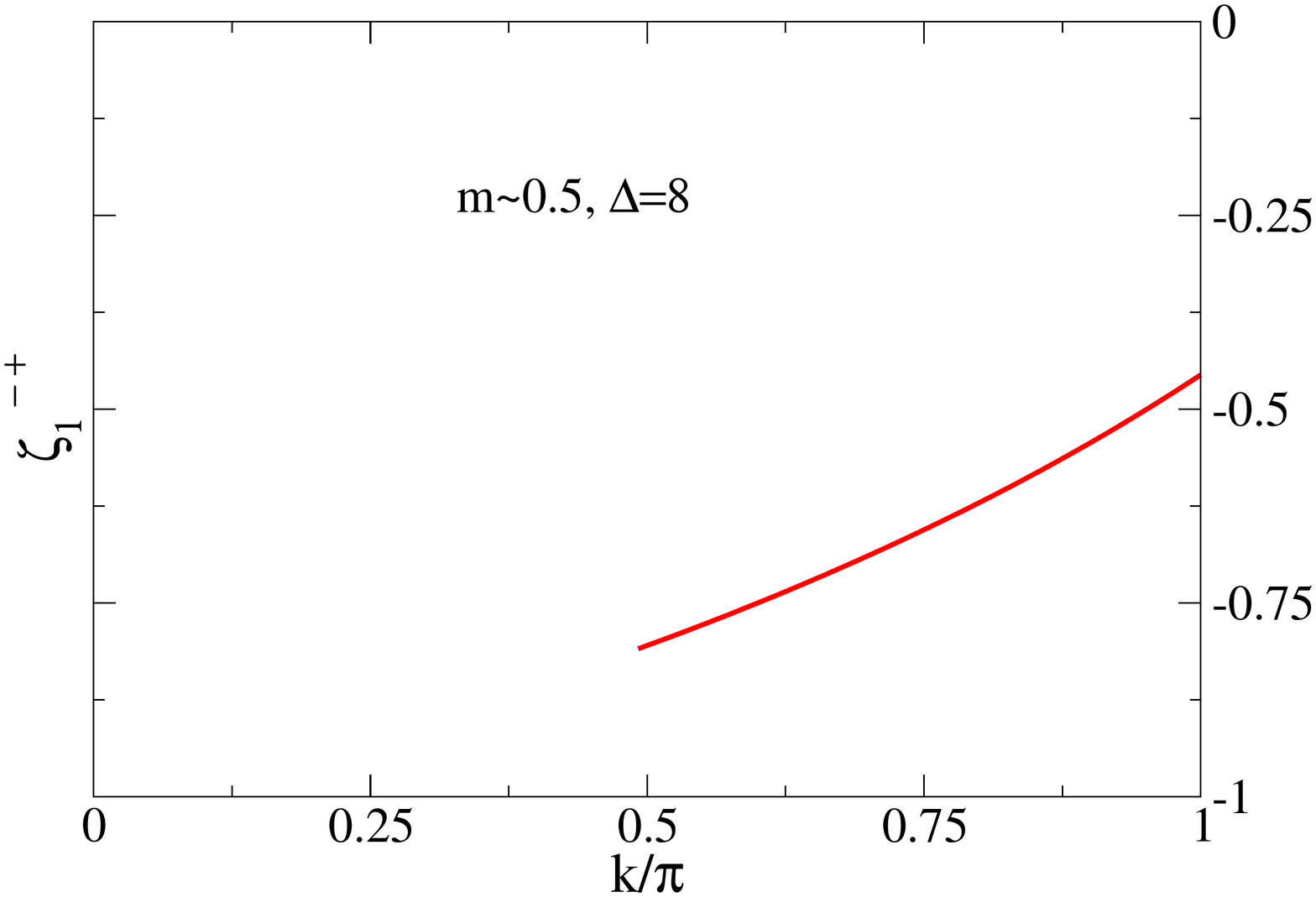}}
\subfigure{\includegraphics[width=8.5cm]{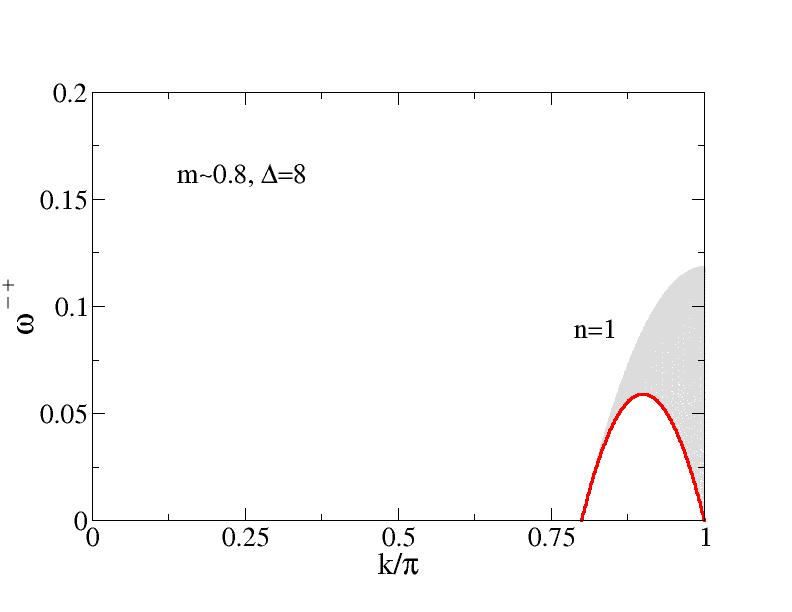}}
\hspace{0.50cm}
\subfigure{\includegraphics[width=8.5cm]{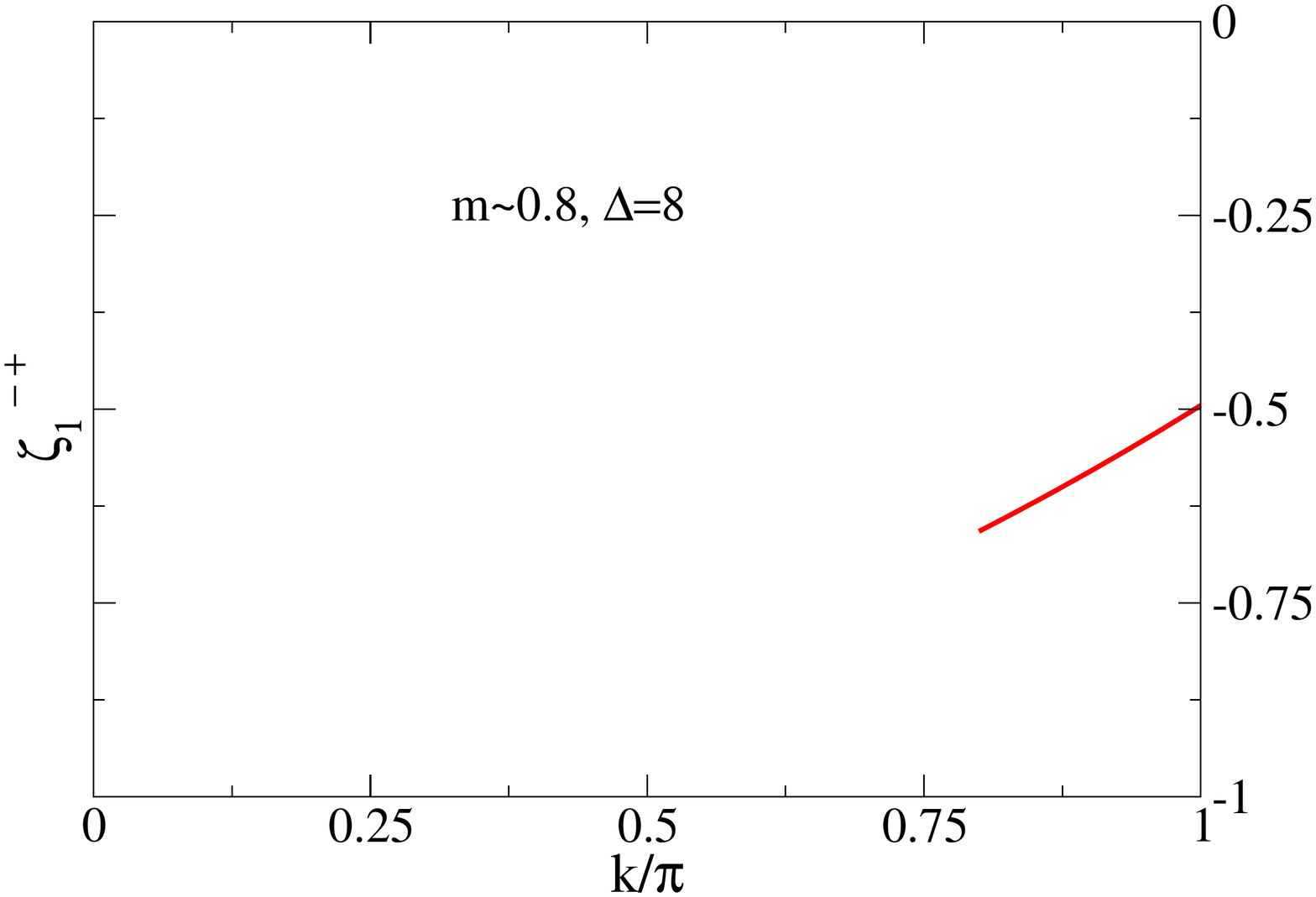}}
\caption{The same as in Fig. \ref{figure6NPB} for spin densities
$m=0.2056\approx 0.2$, $m=0.4918\approx 0.5$, and $m=0.7997\approx 0.8$ 
and anisotropy $\Delta = 8$.}
\label{figure7NPB}
\end{center}
\end{figure}

\begin{figure}
\begin{center}
\subfigure{\includegraphics[width=8.5cm]{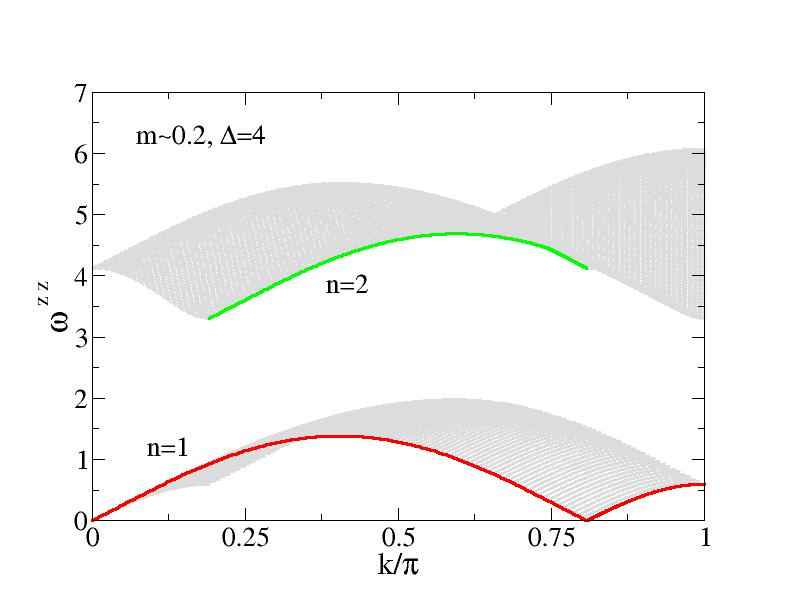}}
\hspace{0.50cm}
\subfigure{\includegraphics[width=8.5cm]{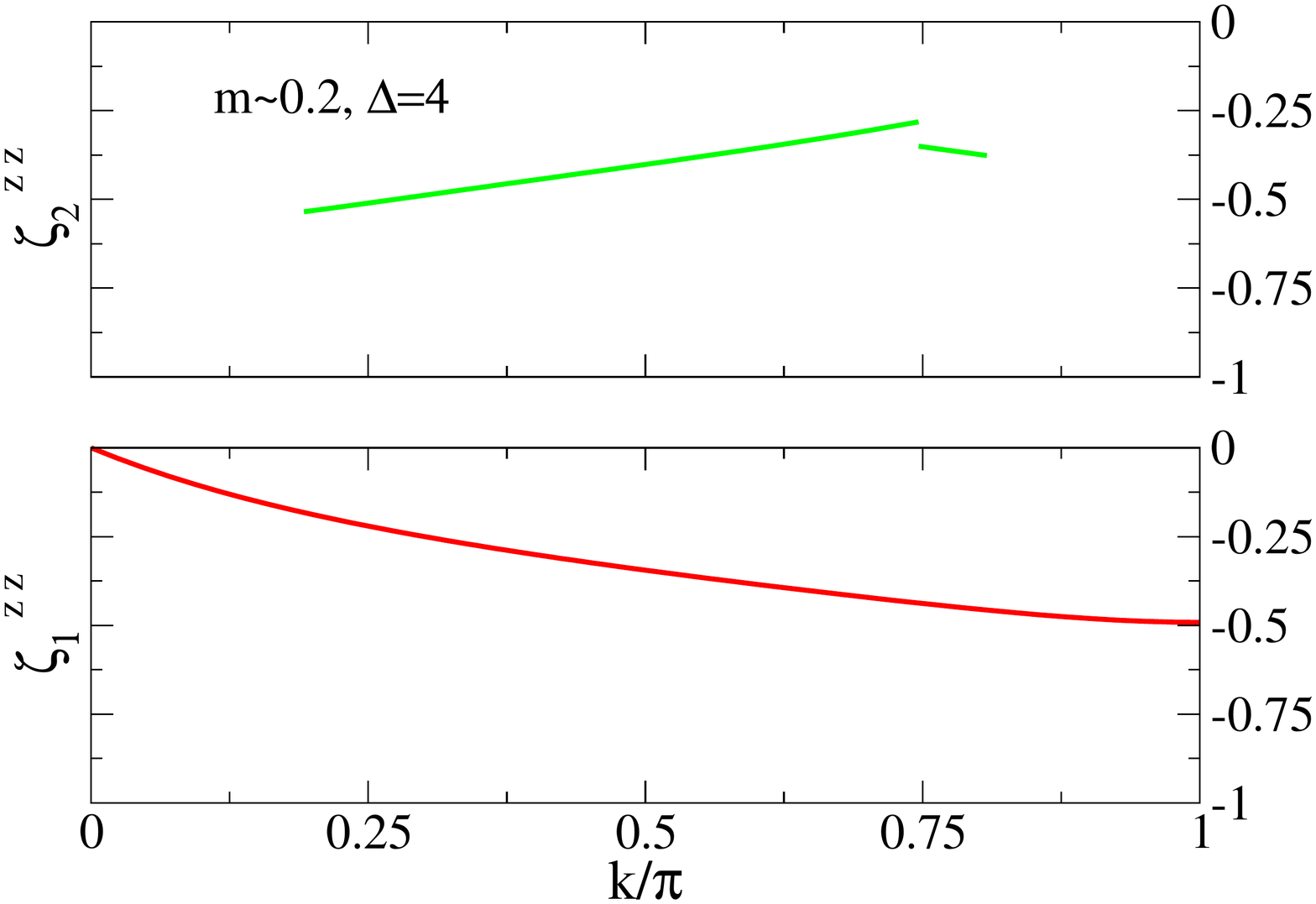}}
\subfigure{\includegraphics[width=8.5cm]{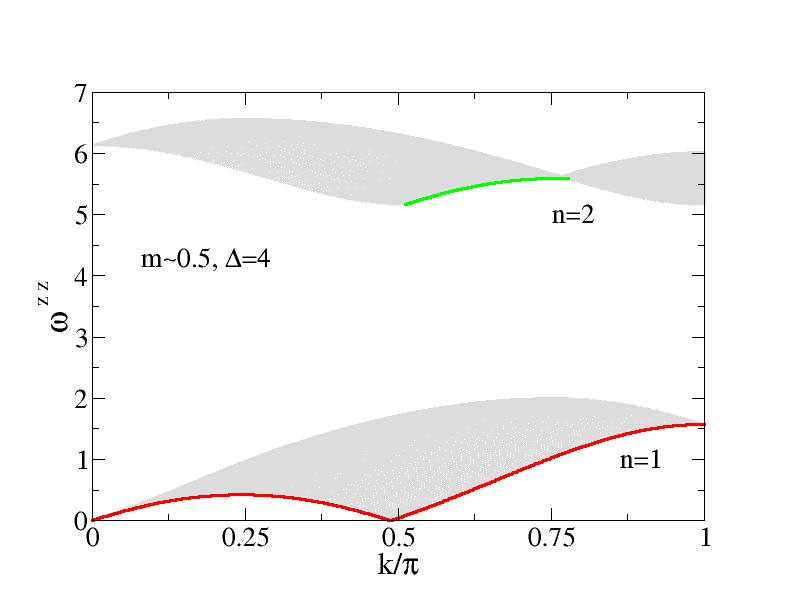}}
\hspace{0.50cm}
\subfigure{\includegraphics[width=8.5cm]{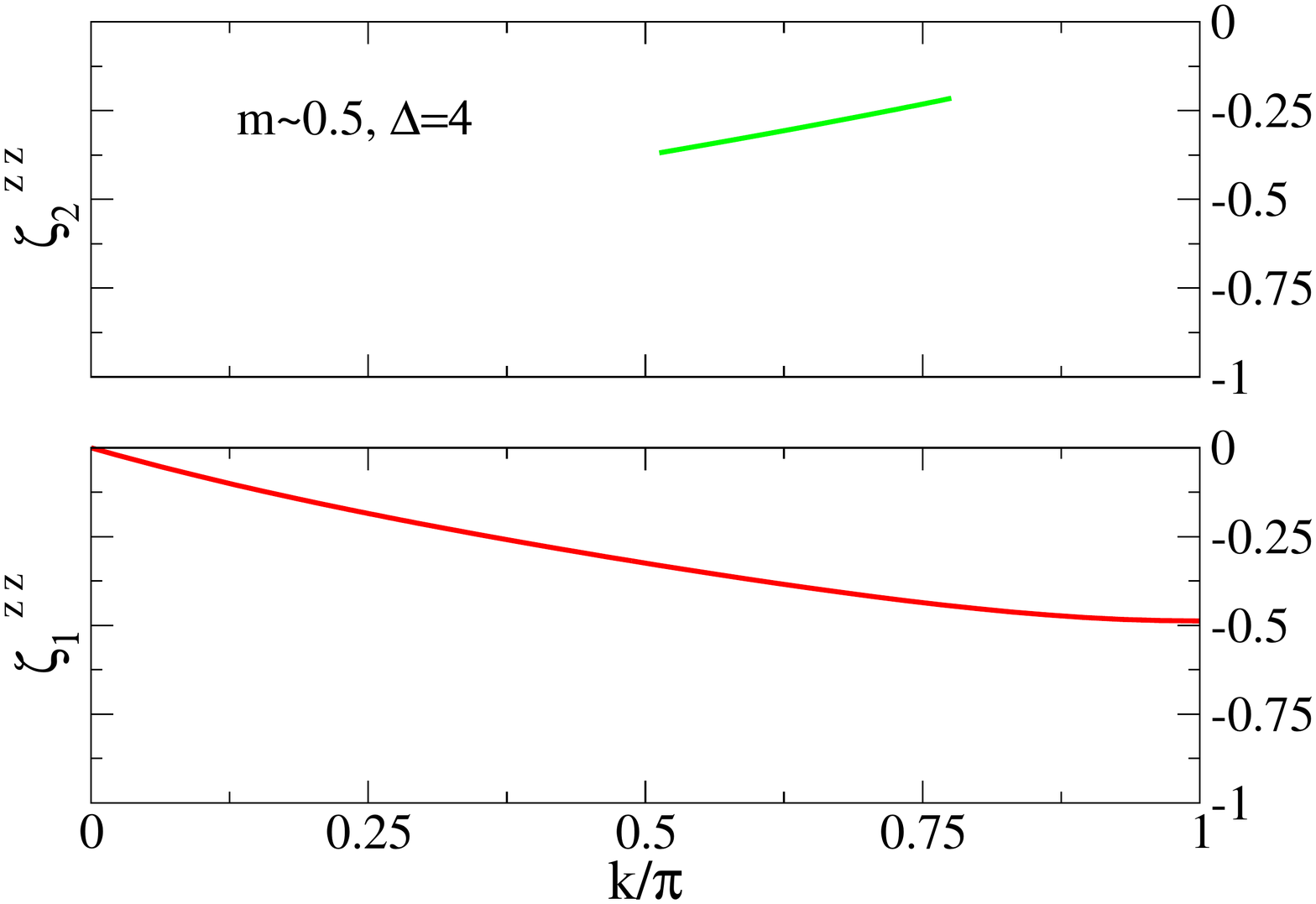}}
\subfigure{\includegraphics[width=8.5cm]{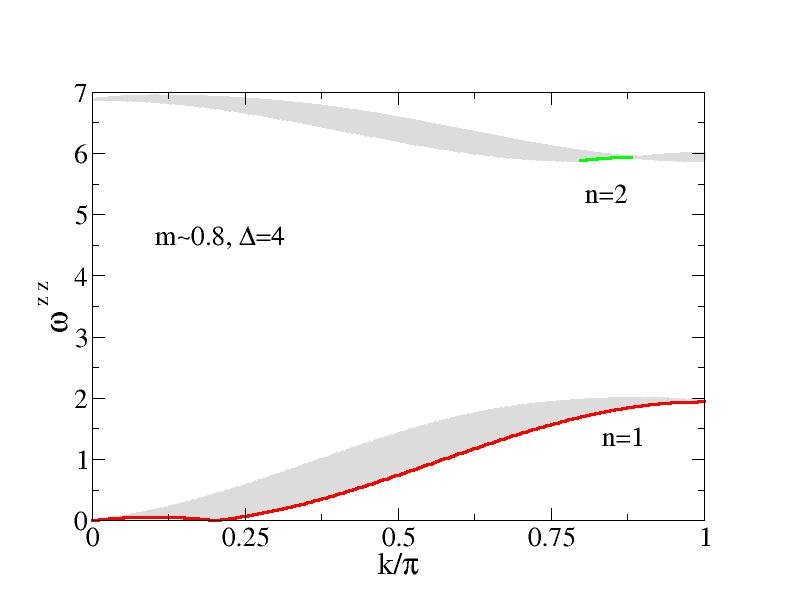}}
\hspace{0.50cm}
\subfigure{\includegraphics[width=8.5cm]{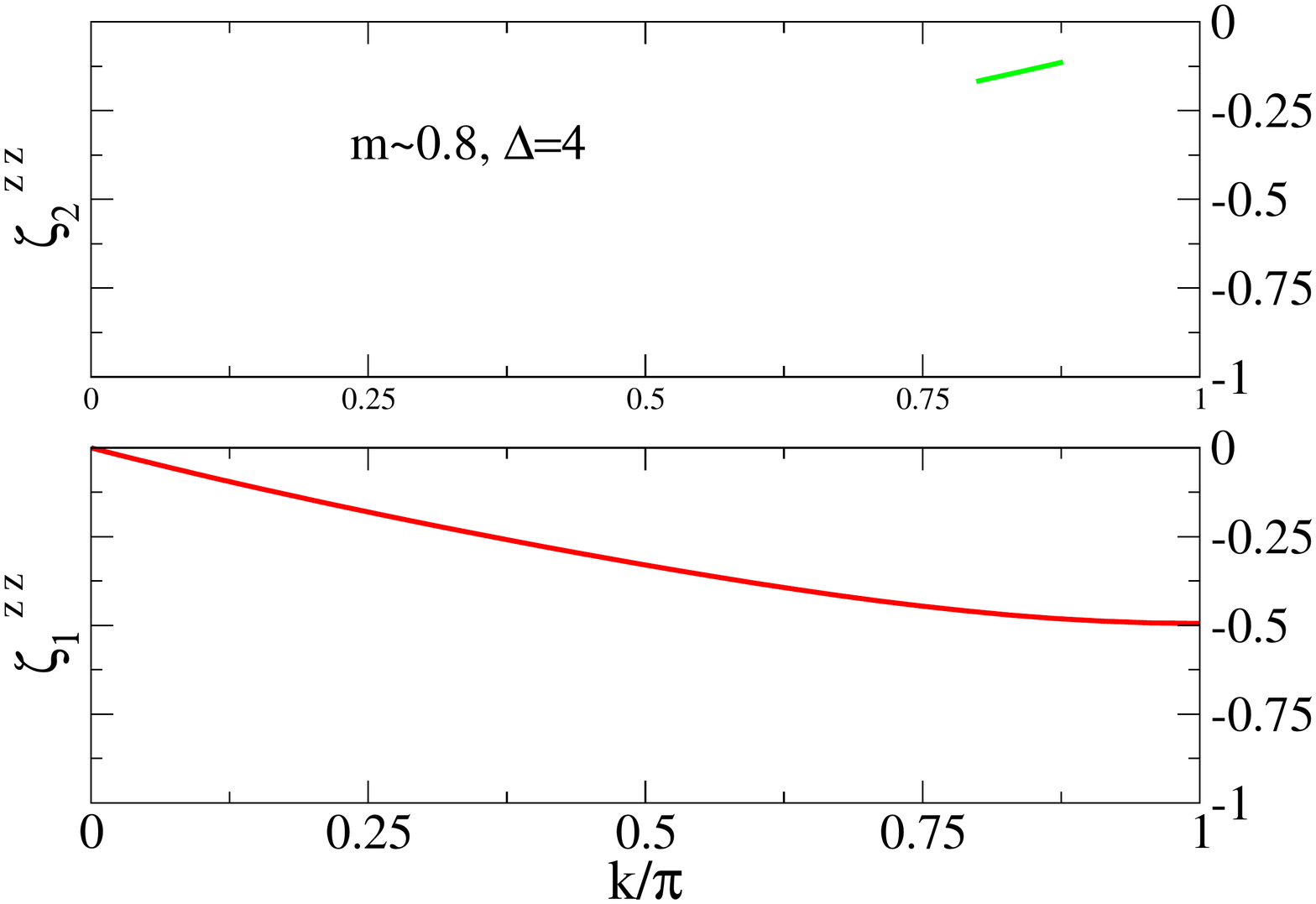}}
\caption{The $(k,\omega)$-plane continua where there is more
spectral weight in $S^{zz} (k,\omega)$ at $\Delta = 4$ for 
$m=0.1920\approx 0.2$, $m=0.5125\approx 0.5$, and $m=0.7985\approx 0.8$ (left) and
the negative $k$ dependent exponents that
control the line shape $S^{zz} (k,\omega)\propto (\omega - \omega^{+-}_n (k))^{\zeta^{zz}_n (k)}$
in the $k$ intervals near the lower thresholds of the $n=1$ and $n=2$
continua marked in the spectra (right.) In the $k$ intervals of such exponents $S^{zz} (k,\omega)$ 
displays sharp peaks.}
\label{figure8NPB}
\end{center}
\end{figure}

In this section, the $(k,\omega)$-plane power-law line shape near the sharp peaks in
$S^{+-} (k,\omega)$, $S^{-+} (k,\omega)$, and $S^{zz} (k,\omega)$ is studied
in the thermodynamic limit. Such sharp peaks are located 
in specific $k$ intervals of the lower thresholds of the set of $(k,\omega)$-plane continua shown in 
Figs. \ref{figure4NPB},\ref{figure5NPB},\ref{figure6NPB},\ref{figure7NPB},\ref{figure8NPB},\ref{figure9NPB}. 
(Below we also consider sharp peaks in the $k$ interval of a branch line running inside the $n=1$
$(k,\omega)$-plane continuum of $S^{+-} (k,\omega)$.)

The results of Sec. \ref{SECIII} render the extension to anisotropy $\Delta >1$ of the 
dynamical theory suitable to the isotropic point \cite{Carmelo_20,Carmelo_15A} straightforward and direct. 
That some dynamical properties are different from those of the isotropic point is naturally captured by 
the $\eta$-dependent expressions of the dynamical structure factor components for $\eta>0$ provided below
in Sec. \ref{SECVB}.

The general dynamical theory used for the isotropic point in Refs. \onlinecite{Carmelo_20,Carmelo_15A} and for anisotropy
$\Delta >1$ in the following, was introduced in Ref. \onlinecite{Carmelo_05} for the 1D Hubbard model
with onsite repulsion $U$ and transfer integral $t$. It is a generalization to the whole $u=U/4t>0$ range 
of the approach used for the $u\rightarrow\infty$ limit in Refs.
\onlinecite{Karlo_96,Karlo_97}. Momentum-dependent exponents in the expressions
of dynamical correlation functions have also been obtained in Refs. \onlinecite{Sorella_96,Sorella_98}.
That dynamical theory is easily generalized to several integrable problems \cite{Carmelo_18}, including here to the 
$\Delta >1$ anisotropic spin-$1/2$ $XXZ$ chain in a longitudinal magnetic field. 
For integrable problems, that general theory is equivalent 
to the mobile quantum impurity model scheme of Refs. \onlinecite{Imambekov_09} and \onlinecite{Imambekov_12}, 
accounting for exactly the same microscopic elementary excitation processes \cite{Carmelo_18,Carmelo_06}.

Although the general dynamical theory applies in principle to the whole $(k,\omega)$ plane, simple analytical 
expressions for the dynamical correlation functions are achieved near their sharp peaks. 
Power-law line shapes associated with positive momentum dependent exponents are also
valid at and near $(k,\omega)$-plane branch lines as defined in Ref. \onlinecite{Carmelo_15A},
provided there is no spectral weight or very little such a weight below them. However, the studies of this paper focus on the line
shape near sharp peaks. Complementarily to the results given below in Sec. \ref{SECVB},
in both these simplest cases, the relation of the matrix elements square 
$\vert\langle \nu\vert\hat{S}^{a}_k\vert GS\rangle\vert^2$ in Eq. (\ref{SDSF}) to the expressions of 
$S^{+-} (k,\omega)$, $S^{-+} (k,\omega)$, and $S^{zz} (k,\omega)$
near the corresponding power-law line shapes is an issue discussed in Appendix \ref{C}. 

\subsection{The $S$ matrices and phase shifts associated with the $n$-particle scattering}
\label{SECVA}

The dynamical theory whose expressions for the present model are given below
in Sec. \ref{SECVB} fully relies on the $S$ matrices associated with the $n$-particles
scattering. Such a scattering is the physically important issue discussed here.
As mentioned previously, the singlet $S^z=S_q=0$ nature of both the unbound pairs of physical spins $1/2$ 
described by $n=1$ real single Bethe rapidities and $n=2,3,...$ bound pairs described by Bethe 
$n$-strings ensures that the present dynamical theory applies. Indeed, the corresponding $1$-particle and $n>1$ $n$-string-particle 
$S$ matrices are then found to have for $\eta>0$ and magnetic fields $h\in [h_{c1},h_{c2}]$ the following same exact scalar form
as for the isotropic case,
\begin{equation}
S_{n} (q_j) = \prod_{n'=1}^{\infty}\,\prod_{j'=1}^{L_{n'}}\,S_{n ,n'} (q_j, q_{j'})
\hspace{0.40cm}{\rm where}\hspace{0.40cm}
S_{n,n'} (q_j, q_{j'}) =
e^{i\,\delta N_{n'}(q_{j'})\,2\pi\Phi_{n,n'}(q_j,q_{j'})} \, .
\label{Smatrix}
\end{equation}
The quantities $2\pi\Phi_{n,n'}(q_j,q_{j'})$ in this equation are $1$-particle and $n>1$ $n$-string-particle 
phase shifts. 

The technically difficult task of extending the dynamical theory to the whole $(k,\omega)$ plane
would involve $S$ matrices for all $n\geq 1$ and $n'\geq 1$ values in the
quantity $S_{n,n'} (q_j, q_{j'})$ given in Eq. (\ref{Smatrix}). This though leads
to formal expressions containing state summations that are both analytically 
and numerically very difficult to be carried out.

However, for the technically simpler problem referring to our study of the power-law line shape 
near the sharp peaks of the dynamical structure factor components, only the $1$-partitcle $S$ matrix
$S_{1} (q)$ at the $1$-band Fermi points $q = \pm k_{F\downarrow}$ and corresponding
$\eta >0$ phase shifts $2\pi\Phi_{1,1}(\pm k_{F\downarrow},q)$ and $2\pi\Phi_{1,n}(\pm k_{F\downarrow},q)$ for $n>1$
defined by Eqs. (\ref{Phi-barPhi})-(\ref{Phis-all-qq}) of Appendix \ref{C} 
play an active role. Indeed, ground states are not populated by $n$-string particles. Hence only the
ground-state preexisting $1$-particles play the role of scatterers. It follows that the 
corresponding $1$-particle $S$ matrix, within which the created 
$1$-particles, $1$-holes, and $n$-string-particles under transitions to excited
states play the role of scattering centers,
determines the momentum $k$ dependence of the exponents that control
the power-law line shapes near the sharp peaks in $S^{+-} (k,\omega)$, 
$S^{-+} (k,\omega)$, and $S^{zz} (k,\omega)$.

Physically, (i) $-2\pi\Phi_{1,1} (\pm k_{F\downarrow},q)$ and (ii) 
$2\pi\Phi_{1,1} (\pm k_{F\downarrow},q)$
are the phase shift acquired by a $1$-particle of momentum $\pm k_{F\downarrow}$ 
upon creation (i) of one $1$-hole at a $1$-band momentum whose maximum interval is
$q\in [-k_{F\downarrow},k_{F\downarrow}]$ and (ii) of one $1$-particle at a $1$-band momentum 
whose maximum intervals are $q\in [-k_{F\uparrow},-k_{F\downarrow}]$ and $q\in [k_{F\downarrow},k_{F\uparrow}]$.
On the other hand, $2\pi\Phi_{1,n} (\pm k_{F\downarrow},q)$ is for
$n=2,3$ the phase shift acquired by such a $1$-particle of momentum $\pm k_{F\downarrow}$ upon creation of one 
$n$-string-particle at a $n$-band momentum whose maximum interval is
$q\in [-(k_{F\uparrow}-k_{F\downarrow}),(k_{F\uparrow}-k_{F\downarrow})]$.

Hence for the present quantum problem only the dominant $1-1$ and $1-n$ scattering channels
where $n=2,3$ contribute to the dynamical properties in what nearly the whole spectral weight is concerned.
Importantly, the power-law line shape near the sharp peaks under study is {\it only}
controlled by such dominant scattering processes. Within them, the $1$-particles at and very near the $1$-band Fermi 
points $\iota\,k_{F\downarrow} = \pm k_{F\downarrow}$ are the scatterers and the $1$-particles 
or $1$-holes and the $2$-string-particle or $3$-string-particle created under the transitions from 
the ground state to the excited states are the scattering centers. 

Actually and as discussed in Appendix \ref{C}, the $1$-band momentum of the $1$-particle scatterers refers to the 
{\it reference-state} $\iota =\pm 1$ Fermi points $q_{F1,\iota} = q_{F1,\iota}^0+ {\pi\over L}\,\delta N_{1,\iota}^F$, Eq. (\ref{RSFP}) 
of that Appendix. Here $q_{F1,\iota}^0$ are the initial ground-state $\iota =\pm 1$ Fermi points that in 
the thermodynamic limit can be written as $q_{F1,\iota}^0=\iota\,k_{F\downarrow}$
and $\delta N_{1,\iota}^F$ is the deviation under the ground-state - excited state transitions
in the number of $1$-particles at such $\iota = \pm 1$ Fermi points. 
As was given in Sec. \ref{SECII}, the quantum numbers $I_j^1$ in $q_j = {2\pi\over L}I_j^1$
are integers or half-odd integers for $L_1$ odd and even, respectively. Hence, $1$-band momentum shifts $\pm\pi/L$ occur 
when the deviations $\delta L_1$ under such transitions are odd integers. As a result, $\delta N_{1,\iota}^F$ may be
an integer or a half-odd integer number. 

For each ground-state - excited state transition, the relevant $1$-particle $S$ matrix is then $S_{1} (q_{F1,\iota})$, which for the present
quantum problem involves some of the phase shifts $2\pi\Phi_{1,n}(q_{F1,\iota},q)$ where $n=1,2,3$. However, in the thermodynamic
limit one can replace the reference-state $\iota =\pm 1$ Fermi points $q_{F1,\iota}$ in the argument of such a $S$ matrix and corresponding 
phase shifts by the ground-state $\iota = \pm 1$ Fermi points $q_{F1,\iota}^0=\iota\,k_{F\downarrow}$. The neglected contributions are irrelevant
and vanish in that limit. The apparently neglected deviation $\delta N_{1,\iota}^F$
in $q_{F1,\iota} = q_{F1,\iota}^0+ {\pi\over L}\,\delta N_{1,\iota}^F$ actually emerges in a
higher order contribution that does not vanish in the thermodynamic limit.

Indeed, the momentum $k$ dependence of the exponents considered below in Sec. \ref{SECVB}
is determined by both that deviation $\delta N_{1,\iota}^F$ and 
the $1$-particle $S$ matrix $S_{1} (q)$ at the $1$-band Fermi points $q=\iota k_{F\downarrow} = \pm k_{F\downarrow}$.
This occurs through an important functional of the following general form,
\begin{eqnarray}
\Phi_{\iota} & = & \iota\,\delta N_{1,\iota}^F  - {i\over 2\pi}\ln S_{1} (\iota k_{F\downarrow}) =
\iota{\delta N_1^F\over 2} + \delta J_1^F  - {i\over 2\pi}\ln S_{1} (\iota k_{F\downarrow})
\hspace{0.40cm}{\rm where}
\nonumber \\
S_{1} (\iota k_{F\downarrow}) & = & 
\prod_{n=1}^{3}\,\prod_{j=1}^{L_{n}}\,e^{i\,\delta N_{n}(q_{j})\,2\pi\Phi_{1,n}(\iota k_{F\downarrow},q_{j})} 
= e^{i\sum_{n=1}^{3}\sum_{j=1}^{L_{n}}\delta N_{n}(q_{j})2\pi\Phi_{1,n}(\iota k_{F\downarrow},q_{j})} \, .
\label{functional}
\end{eqnarray}
Here the number deviation $\delta N_1^F$ and the current-like deviation $\delta J_1^F$ in which
$\delta N_{1,\iota}^F$ can be decomposed read $\delta N_{1}^F = \sum_{\iota = \pm 1}\delta N_{1,\iota}^F$ and
$\delta J_{1}^F =  {1\over 2}\sum_{\iota = \pm 1}\iota\,\delta N_{1,\iota}^F$, respectively.
The deviations $\delta N_{n}(q_{j})$ refer to the $n$-band momentum distributions.

For each specific $(k,\omega)$-plane line of sharp peaks in the dynamical structure factor components,
the functional $\Phi_{\iota}$, Eq. (\ref{functional}), refers to a form of the $S$ matrix $S_{1} (\iota k_{F\downarrow})$ in that equation
specific to the one-parametric spectrum ${\bar{\omega}}^{ab}_{n} (k)$ in Eq. (\ref{MPSs})
corresponding to a branch line, as defined in Ref. \onlinecite{Carmelo_15A}. 
For instance, for both the branch-line $k$ intervals that coincide with lower threshold of the $n=1$ continua of
$S^{+-} (k,\omega)$, $S^{-+} (k,\omega)$, and $S^{zz} (k,\omega)$ and for a specific  branch line considered
below that for some $m$ values runs on the $n=1$ continuum of $S^{+-} (k,\omega)$,
the functional $\Phi_{\iota} (k)$ is a function of the excitation momentum $k$ that has the following simple form,
\begin{eqnarray}
\Phi_{\iota} (k) & = & \iota\,\delta {\cal{N}}^F_{1,\iota} \pm \Phi_{1,1}(\iota k_{F\downarrow},q)
\hspace{0.40cm}{\rm where}\hspace{0.40cm}k = k_1 \pm q \, .
\label{Ph1n1}
\end{eqnarray}
Here $+$ or $-$ refers to creation of one $1$-particle at a $1$-band momentum $q$ whose maximum interval 
is $\vert q\vert\in [k_{F\downarrow},k_{F\uparrow}]$ or creation of 
one $1$-hole at a $1$-band momentum $q$ whose maximum interval is $q\in [-k_{F\downarrow},k_{F\downarrow}]$,
respectively. The values of the fixed momentum value $k_1$ specific to the branch-line $k$ intervals under consideration 
are provided in Eqs. (\ref{OkMPRs}), (\ref{OkPMstar}), (\ref{OkPMRs}), and (\ref{OkPMRsL}) of Appendix \ref{D}, and
the renormalized $\iota =\pm 1$ number deviation $\delta {\cal{N}}^F_{1,\iota}$ is given by,
\begin{equation}
\delta {\cal{N}}^F_{1,\iota} = {\delta N_1^F\over 2\xi_{1\,1}} + \iota\,\xi_{1\,1}\delta J_1^F 
\hspace{0.40cm}{\rm where}\hspace{0.40cm}
\xi_{1\,1} = 1 + \Phi_{1,1}(k_{F\downarrow},k_{F\downarrow}) - \Phi_{1,1}(k_{F\downarrow},-k_{F\downarrow}) \, .
\label{dXxi11}
\end{equation}
The $1$-band $q$ intervals for which the corresponding exponent is negative refer either to
the above maximum intervals or to their well defined subintervals.

For the $n=2,3$ continua lower thresholds of $S^{+-} (k,\omega)$
and $n=2$ continuum lower threshold of $S^{zz} (k,\omega)$ the functional $\Phi_{\iota} (k) $ 
has for the momentum $k$ intervals for which the corresponding exponent is negative 
one of the following three simple forms,
\begin{eqnarray}
\Phi_{\iota} (k) & = &  \iota\,\delta {\cal{N}}^F_{1,\iota} + \iota\,{\xi^0_n\over 2} - \Phi_{1,1}(\iota k_{F\downarrow},q)
\hspace{0.40cm}{\rm where}\hspace{0.40cm}k = k_n - q
\nonumber \\
\Phi_{\iota} (k) & = &  \iota\,\delta {\cal{N}}^F_{1,\iota} + \xi^{\iota,\pm}_n
- \Phi_{1,1}(\iota k_{F\downarrow},q)\hspace{0.40cm}{\rm where}\hspace{0.40cm}k = k_n - q
\nonumber \\
\Phi_{\iota} (k) & = &  \iota\,\delta {\cal{N}}^F_{1,\iota} + \Phi_{1,n}(\iota k_{F\downarrow},q)
\hspace{0.40cm}{\rm where}\hspace{0.40cm}k = k_n + q
\hspace{0.40cm}{\rm for}\hspace{0.40cm}n=2,3 \, .
\label{Ph1n23several}
\end{eqnarray}

In the first two expressions of this equation, $q$ is the $1$-band momentum
for creation of one $1$-hole whose maximum interval is $q\in [-k_{F\downarrow},k_{F\downarrow}]$.
In the third expression, $q$ is for $n=2,3$ the $n$-band momentum 
for creation of one $n$-string-particle whose maximum interval is 
$q \in [- (k_{F\uparrow}-k_{F\downarrow}), (k_{F\uparrow}-k_{F\downarrow})]$.
In these expressions, the values of the fixed momentum values $k_2$ and $k_3$ specific to the branch-line $k$ 
intervals under consideration are provided in
Eqs. (\ref{Ds2})-(\ref{Ds2p}), (\ref{Ds2pL})-(\ref{Ds2L}) and
Eqs. (\ref{Ds3pP})-(\ref{Ds3P}), respectively, of Appendix \ref{D}. The renormalized 
number deviation $\delta {\cal{N}}^F_{1,\iota}$ is given in 
Eq. (\ref{dXxi11}) and the parameters $\xi^0_n$ and $\xi^{\iota,\pm}_n$ for $\iota =\pm 1$ read,
\begin{eqnarray}
\xi^0_n & = &  2\Phi_{1,n}(k_{F\downarrow},0)\hspace{0.40cm}{\rm for}\hspace{0.40cm}n=2,3
\nonumber \\
\xi^{\iota,\pm}_n & = & \Phi_{1,n}(\iota k_{F\downarrow},\pm (k_{F\uparrow}-k_{F\downarrow}))
\hspace{0.40cm}{\rm for}\hspace{0.40cm}n=2,3\hspace{0.40cm}{\rm and}\hspace{0.40cm}\iota = \pm 1 \, .
\label{xpmnx0n}
\end{eqnarray}

In the four types of expressions for the functional $\Phi_{\iota} (k)$ given in Eqs. (\ref{Ph1n1}) and (\ref{Ph1n23several}),
which are specific to $(k,\omega)$-plane line shapes near branch lines of sharp peaks whose spectrum ${\bar{\omega}}^{ab}_{n} (k)$ in Eq. (\ref{MPSs})
is one-parametric, all except one of the few deviations $\delta N_{n}(q_{j})$ that have finite values in the 
$S$ matrix $S_{1} (\iota k_{F\downarrow})$ general expression, Eq. (\ref{functional}), 
refer to fixed $n$-band momentum values $q_j=q$. Those are either at the $q = \pm k_{F\downarrow}$ Fermi points
for $n=1$ or at $q = 0$ or $q = \pm (k_{F\uparrow}-k_{F\downarrow})$ for $n=2$ or $n=3$. 

A question is to which $n$-band refers the only finite deviation $\delta N_{n}(q_{j})$ whose momentum $q_{j}=q$ can have arbitrary values
within one of the maximum intervals provided above? As given in Eq. (\ref{Ph1n1}), for the $n=1$ $(k,\omega)$-plane 
continua one has always that such a deviation refers to the $1$-band.
As revealed by inspection of the form of the expressions provided in Eq. (\ref{Ph1n23several}), for the $n=2$ and $n=3$ $(k,\omega)$-plane 
continua such a deviation can either refer to the $1$-band or to the $2$-band and $3$-band, respectively.

\subsection{The expressions of the dynamical structure factor components near sharp peaks}
\label{SECVB}

In all integrable problems to which the dynamical theory addressed and used here applies, one finds
a direct relationship between the negativity and the length of the momentum $k$ interval
of the momentum dependent exponents that control the power-law line shape of the dynamical correlation function
components near the lower threshold of a given $(k,\omega)$-plane continuum
and the amount of spectral weight over the latter. This has been confirmed for instance 
in the case of the present spin-$1/2$ chain at the isotropic point \cite{Carmelo_20}.
\begin{figure}
\begin{center}
\subfigure{\includegraphics[width=8.5cm]{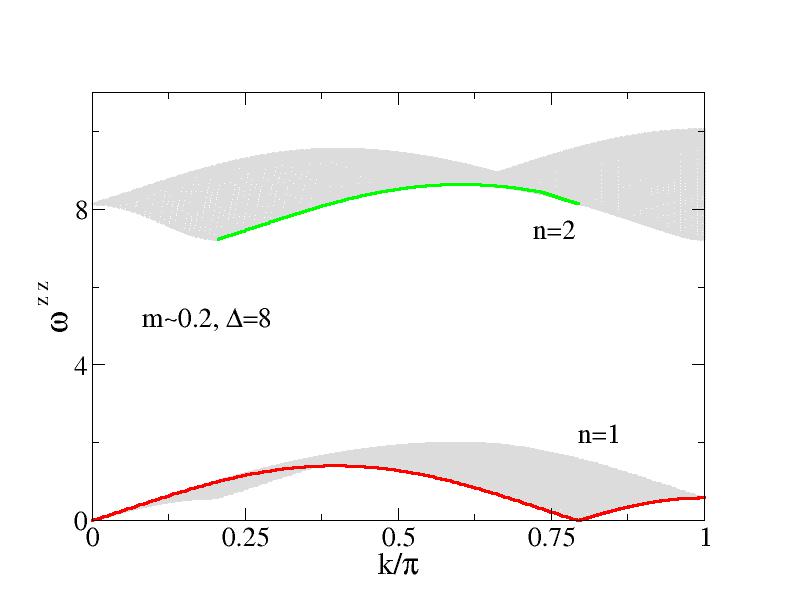}}
\hspace{0.50cm}
\subfigure{\includegraphics[width=8.5cm]{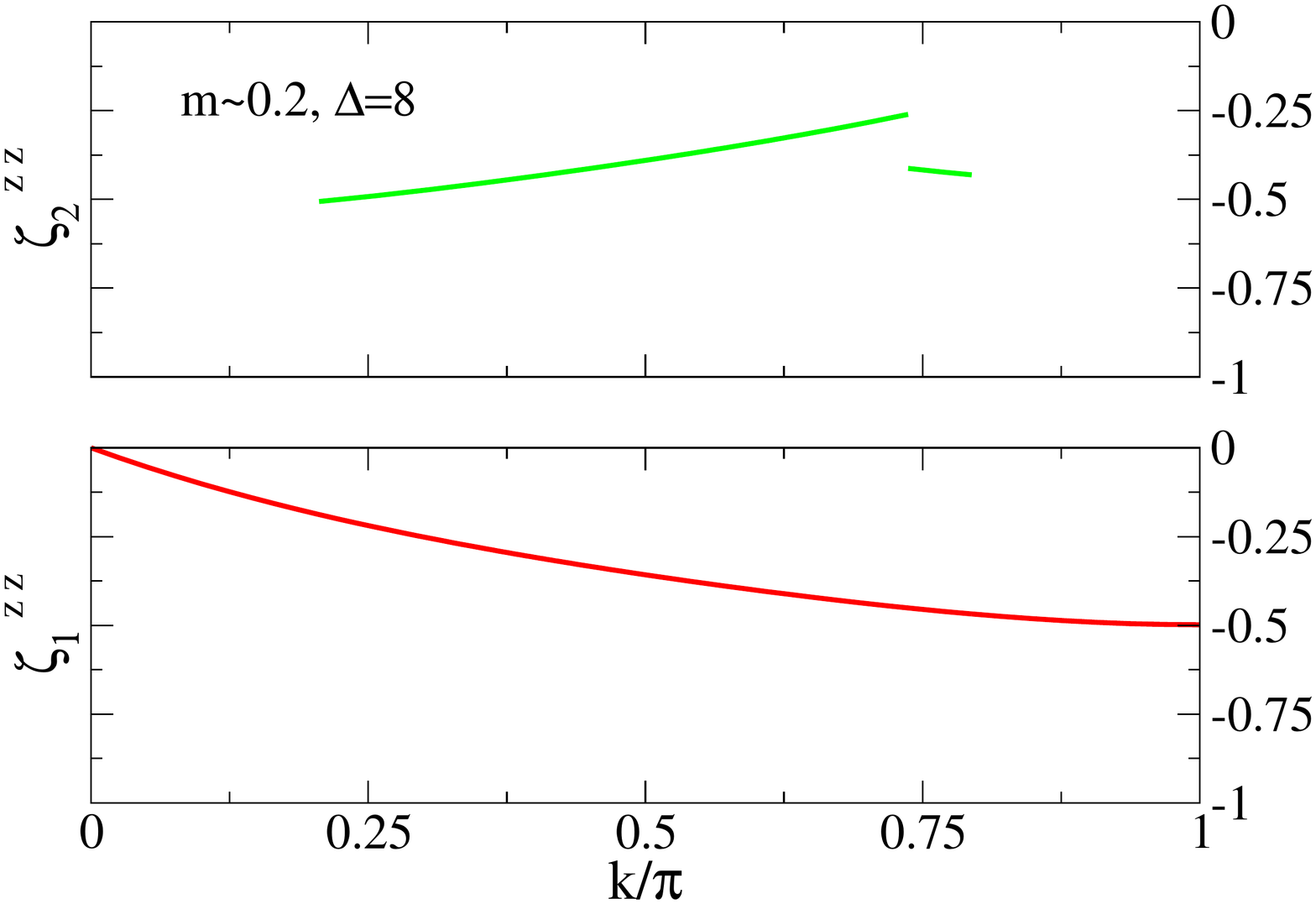}}
\subfigure{\includegraphics[width=8.5cm]{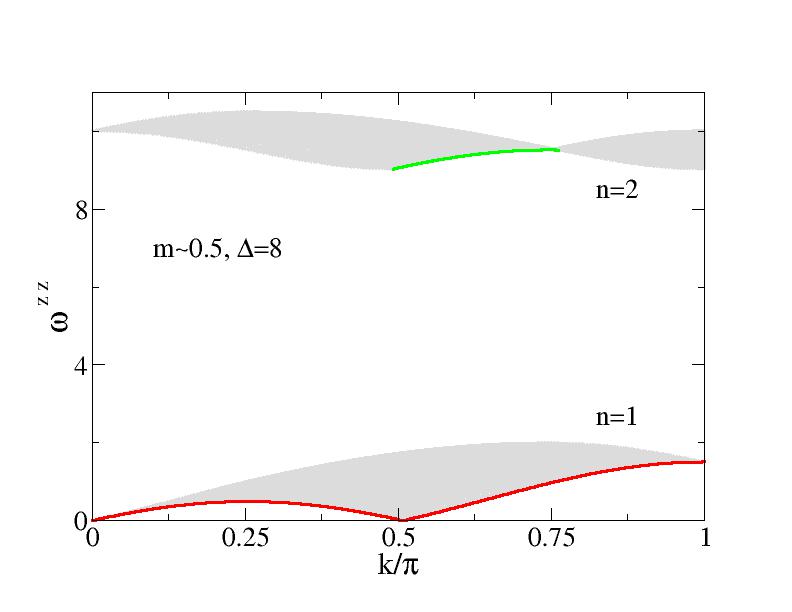}}
\hspace{0.50cm}
\subfigure{\includegraphics[width=8.5cm]{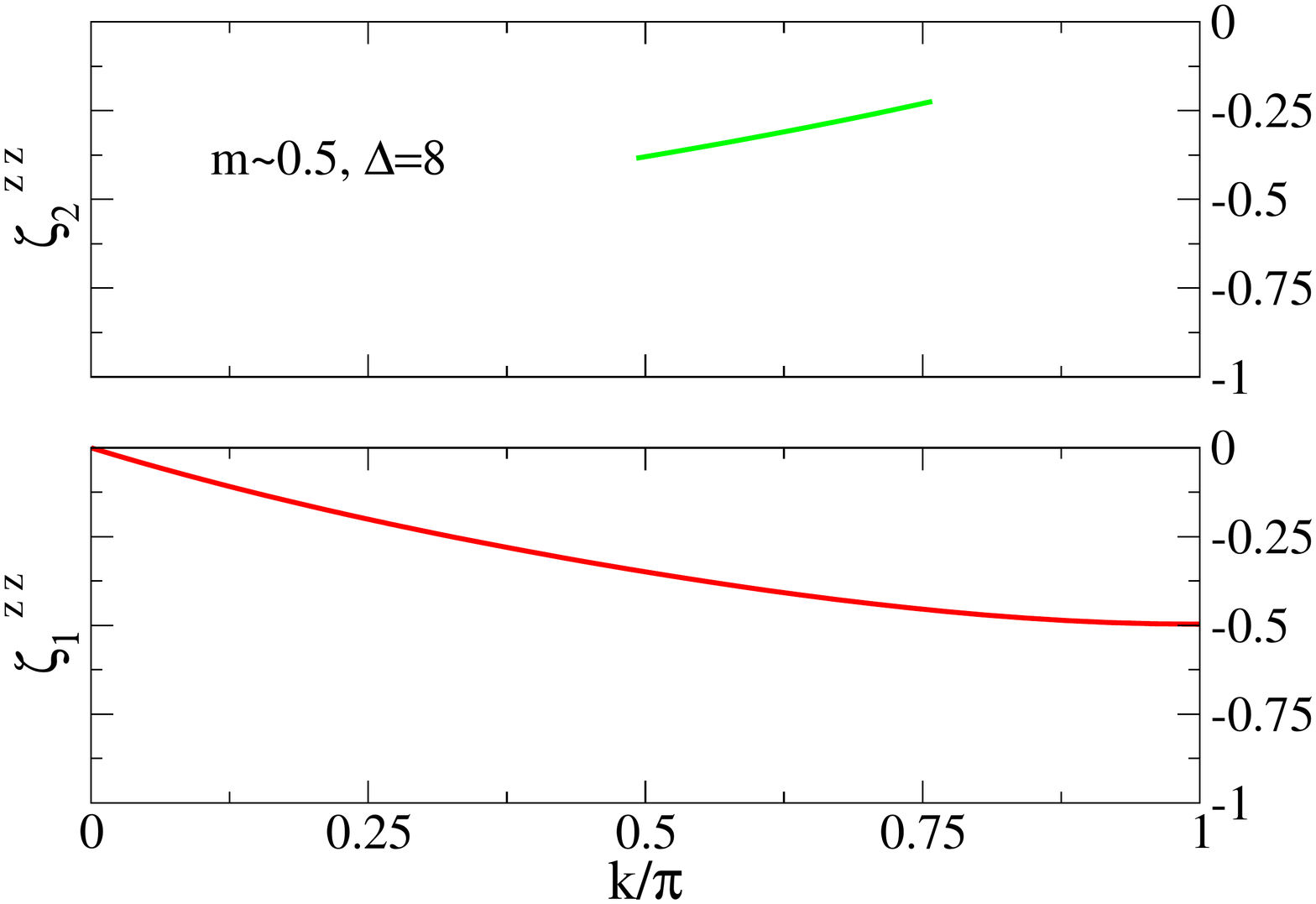}}
\subfigure{\includegraphics[width=8.5cm]{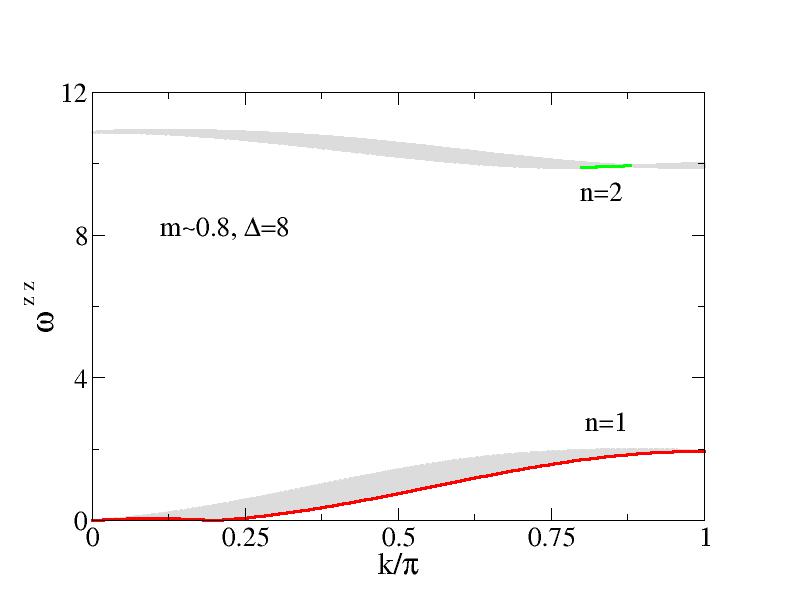}}
\hspace{0.50cm}
\subfigure{\includegraphics[width=8.5cm]{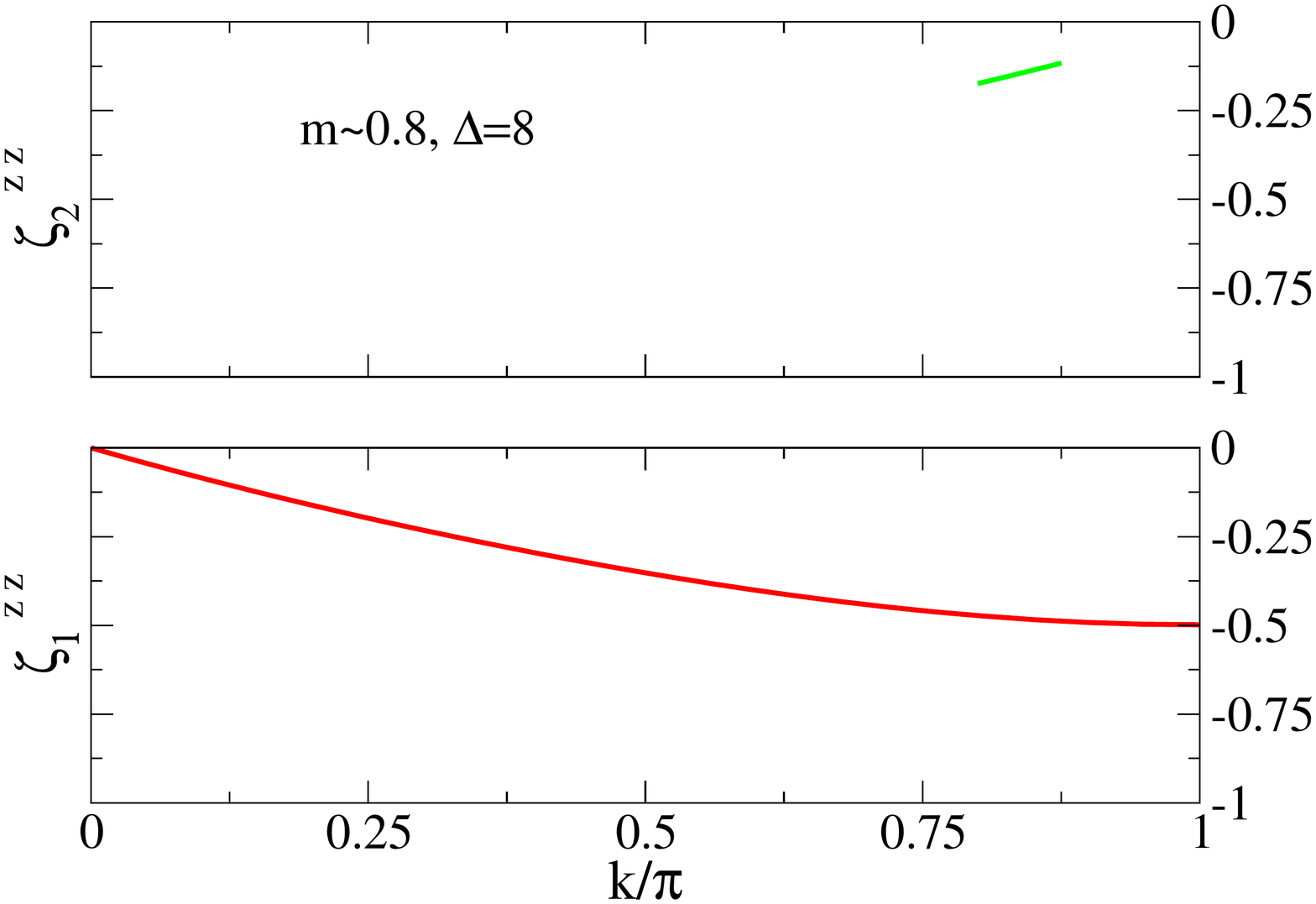}}
\caption{The same as in Fig. \ref{figure8NPB} for spin densities
$m=0.2056\approx 0.2$, $m=0.4918\approx 0.5$, and $m=0.7997\approx 0.8$ 
and anisotropy $\Delta = 8$.}
\label{figure9NPB}
\end{center}
\end{figure}

We have systematically used the present dynamical theory whose dynamical correlation
functions expressions for the present model are given below to identify {\it all} branch lines 
containing sharp peaks in $S^{+-} (k,\omega)$, $S^{-+} (k,\omega)$, and $S^{zz} (k,\omega)$ for $\eta >0$.
One then finds that the corresponding $(k,\omega)$-plane continua with a significant
amount of spectral weight are those shown in
Figs. \ref{figure4NPB},\ref{figure5NPB},\ref{figure6NPB},\ref{figure7NPB},\ref{figure8NPB},\ref{figure9NPB}.
{\it All } $(k,\omega)$-plane one-parametric lines of such sharp peaks emerge {\it only} in the 
two-parametric spectra, Eqs. (\ref{dkEdPPMn1})-(\ref{sepectzzn2}), corresponding to
such continua. Those refer to the classes of selected excited states that contribute
to a significant amount of spectral weight. Note though that the aim of these figures is to provide 
the $(k,\omega)$-plane location in the continua of the sharp peaks whose line shape is studied in the following, 
which refers to the marked lines in them. It is not to provide information on the relative intensities of the 
spectral-weight distribution over such continua. 

Specifically, the found sharp peaks whose line shape is studied in the following
are located in $k$ intervals of the lower thresholds of the three $n=1,2,3$ continua, one $n=1$ continuum,
and two $n=1,2$ continua. They are shown in Figs. \ref{figure4NPB},\ref{figure5NPB}
for $S^{+-} (k,\omega)$, Figs. \ref{figure6NPB},\ref{figure7NPB} for $S^{-+} (k,\omega)$,
and Figs. \ref{figure8NPB},\ref{figure9NPB} for $S^{zz} (k,\omega)$, respectively. 
We denote by ${\bar{\omega}}^{ab}_{n} (k)$ where $n=1,2,3$ for $ab=+-$, $n=1$ for $ab=-+$, 
and $n=1,2$ for $ab=zz$ the corresponding one-parametric spectra of the lower thresholds of the continua
containing nearly the whole spectral weight in the components $S^{ab} (k,\omega)$.
The corresponding two-parametric spectra $\omega^{ab}_{n} (k)$ associated with such continua 
are given for $m>0$ in Eqs. (\ref{dkEdPPMn1})-(\ref{sepectzzn2}).

The expressions of the $n=1,2,3$ one-parametric spectra ${\bar{\omega}}^{+-}_n (k)$ are provided in Eqs. (\ref{OkMPRs}),
(\ref{Ds2})-(\ref{Ds2p}), and (\ref{Ds3pP})-(\ref{Ds3P}), respectively, of Appendix \ref{D}. That
of ${\bar{\omega}}^{-+}_1 (k)$ is given in Eq. (\ref{OkPMRs}) of that Appendix.
The expressions of the $n=1,2$ spectra ${\bar{\omega}}^{zz}_n (k)$ are given in Eqs. (\ref{OkPMRsL}) and
(\ref{Ds2pL})-(\ref{Ds2L}), respectively, of the same Appendix. (The expressions provided in Eqs. (\ref{DsL3})-(\ref{Ds3L}) 
of Appendix \ref{D} refer to ${\bar{\omega}}^{zz}_3 (k)$ whose $(k,\omega)$-plane $n=3$ continuum does not contain 
sharp peaks and thus has no significant amount of spectral weight over it.)

As justified in Appendix \ref{C} in terms of the matrix elements square 
$\vert\langle \nu\vert\hat{S}^{a}_k\vert GS\rangle\vert^2$ in Eq. (\ref{SDSF}),
the line shape of the spin dynamical structure factor components $S^{ab} (k,\omega)$ 
where $ab = +-,-+,zz$ at and just above the $(k,\omega)$-plane $n=1,2,3$ continua lower thresholds 
$k$ intervals under consideration at which there are sharp peaks has in the thermodynamic limit the following general power-law form,
\begin{equation}
S^{ab} (k,\omega) = C_{ab}^n (k)
\left({\omega - {\bar{\omega}}^{ab}_{n} (k)\over 4\pi\,B_1^{ab}\,v_1}\right)^{\zeta_{n}^{ab} (k)}  
\hspace{0.40cm}{\rm for}\hspace{0.40cm} (\omega - {\bar{\omega}}^{ab}_{n} (k)) \geq 0  
\hspace{0.40cm} {\rm where}\hspace{0.40cm} ab = +-,-+,zz \, .
\label{MPSs}
\end{equation}
This expression is valid for small values of the energy deviation $(\omega - {\bar{\omega}}^{ab}_{n} (k)>0$
at fixed excitation momentum $k$. In it $v_1 = v_1 (k_{F\downarrow})$ where $v_1 (q)$ is the $1$-band group velocity, 
Eq. (\ref{equA4Bv}) of Appendix \ref{B} for $n=1$ and $B_1^{ab}$ is a $\eta$ and $m$ dependent constant 
such that $0<B_1^{ab}\leq 1$. The exponent in Eq. (\ref{MPSs}) has the following general expression,
\begin{eqnarray}
\zeta_{n}^{ab} (k) = - 1  + \sum_{\iota = \pm 1} \Phi_{\iota}^2 (k) \, ,
\label{zetaabk}
\end{eqnarray}
where $\Phi_{\iota} (k)$ is the corresponding functional in Eq. (\ref{Ph1n1}) for $n=1$ and
in Eq. (\ref{Ph1n23several}) for $n=2,3$ and the index $n=1,2,3$ refers to one of the continua shown in 
Figs. \ref{figure4NPB},\ref{figure5NPB},\ref{figure6NPB},\ref{figure7NPB},\ref{figure8NPB},\ref{figure9NPB}.
The multiplicative factor function $C_{ab}^n (k)$ in Eq. (\ref{MPSs}) also involves the functional $\Phi_{\iota} (k)$.
It is found in Appendix \ref{C} to have the following general form,
\begin{eqnarray}
C_{ab}^n (k) =  {1\over \vert\zeta_{n}^{ab} (k)\vert}\prod_{\iota =\pm 1}
{e^{-f_0^{ab} + f_2^{ab}\left(2{\tilde{\Phi}}_{\iota}\right)^2 - f_4^{ab}\left(2{\tilde{\Phi}}_{\iota}\right)^4} \over \Gamma (\Phi_{\iota}^2 (q))} \, .
 \label{Cabn}
\end{eqnarray}
As given in Eq. (\ref{functional2}) of Appendix \ref{C}, here ${\tilde{\Phi}}_{\iota} = \Phi_{\iota} -\iota\,\delta N_{1,\iota}^F$
is the scattering part of the general functional, Eq. (\ref{functional}), {\it i. e.} ${\tilde{\Phi}}_{\iota} = - {i\over 2\pi}\ln S_{1} (\iota k_{F\downarrow})$. 
The $C_{ab}^n (k)$ expression also involves $1/\vert\zeta_{n}^{ab} (k)\vert$ and three $\eta$ 
dependent constants $0<f_l^{ab}<1$ where $l=0,2,4$. Both $B_1^{ab}$ in Eq. (\ref{MPSs}) and these three constants stem 
from the general expression of the matrix elements square, Eq. (\ref{ME}) of 
Appendix \ref{C}. Their precise values and $\eta$ dependences for each 
dynamical structure factor component are difficult to access.

As discussed in that Appendix, the $S^{ab} (k,\omega)$ expression, Eq. (\ref{MPSs}), is valid both
for exponent values $-1 < \zeta_{n}^{ab} (k)<0$ and $0 < \zeta_{n}^{ab} (k)<1$, provided there
is no spectral weight or very little such a weight below the corresponding branch line. Our study though 
focus on the case when $-1 < \zeta_{n}^{ab} (k)<0$ whose corresponding power-law line shape
refers to a sharp-peak singularity. However, the $S^{ab} (k,\omega)$ expression, Eq. (\ref{MPSs}), is not 
valid when $\zeta_{n}^{ab} (k) = 0$. Then the line shape near the sharp peak is rather logarithmic. 
Its corresponding general expression is provided in Eq. (\ref{MPSs2}) of Appendix \ref{C}.

The one-parametric spectrum of a branch line that for small spin densities coincides with
the $n=1$ lower threshold of $S^{+-} (k,\omega)$ for $k\in [2k_{F\downarrow},\pi]$ and for intermediate and large spin densities 
does not coincide with it, is marked in Figs. \ref{figure4NPB},\ref{figure5NPB}. 
We denote that one-parametric spectrum by ${\tilde{\omega}}^{+-}_1 (k)$, which is given in Eq. (\ref{OkPMstar})
of Appendix \ref{D}. At and just above that branch line,
the spin dynamical structure factor component $S^{+-} (k,\omega)$ also has 
for small deviations $(\omega -  {\tilde{\omega}}^{+-}_{1} (k)) \geq 0$ the form,
\begin{equation}
S^{+-} (k,\omega) = {\tilde{C}}_{+-}^{1} (k)
\left({\omega - {\tilde{\omega}}^{+-}_{1} (k)\over 4\pi\,{\tilde{B}}_1^{+-}\,v_1}\right)^{{\tilde{\zeta}}_{1}^{+-} (k)} \, .
\label{Sotherline}
\end{equation}
Its exponent expression is provided in Eq. (\ref{expstar}) of Appendix \ref{D}. (The general expression 
of ${\tilde{C}}_{+-}^{1} (k)$ is similar to that of $C_{ab}^n (k)$ given in Eq. (\ref{Cabn}).)
This line shape applies to $k$ interval for which there is very little spectral weight below the branch line.

The relation of the excitation momentum $k$ to the $n$-band momentum $q$ in Eqs. (\ref{Ph1n1}) and (\ref{Ph1n23several}) 
suitable to each $n=1,2,3$ lower threshold $k$ interval of the components $S^{ab} (k,\omega)$ 
as well as the corresponding specific values of the deviations 
$\delta N_1^F$ and $\delta J_1^F$ in Eq. (\ref{dXxi11}) were accounted for in the exponent expressions
provided in Appendix \ref{D}. The expressions of the exponents $\zeta_{n}^{+-} (k)$ for $n=1,2,3$ are given in 
Eqs. (\ref{expsPM1}), (\ref{expG+-}), (\ref{expG3+-}), respectively, of that Appendix. Those of the exponents 
$\zeta_{n}^{-+} (k)$ for $n=1,2,3$ are provided in Eqs. (\ref{expsMP1}), (\ref{expG-+}), (\ref{expG3-+}), respectively, of the
same Appendix. Those of the exponents $\zeta_{n}^{zz} (k)$ for $n=1,2,3$ are given in Eqs. (\ref{expszz1}), 
(\ref{exps2pL}), (\ref{exps3pL}), respectively, of Appendix \ref{D}.

All such exponents as well as other exponents associated with line shapes on $(k,\omega)$-plane continua not
shown in Figs. \ref{figure4NPB},\ref{figure5NPB},\ref{figure6NPB},\ref{figure7NPB},\ref{figure8NPB},\ref{figure9NPB}
have been used in our search of sharp peaks located in corresponding lower thresholds. 
However, in these figures {\it only} the exponents that are negative in some $k$ interval are plotted.
The corresponding lower thresholds intervals are marked in the spectra shown in them.
The $k$ interval of the branch line associated with the line shape in Eq. (\ref{Sotherline})
in which the corresponding exponent ${\tilde{\zeta}}_{1}^{+-} (k)$ is negative is also marked in 
the $n=1$ continuum of Figs. \ref{figure4NPB},\ref{figure5NPB}. 

Specifically, the three $n=1,2,3$ exponents $\zeta_{n}^{+-} (k)$ 
are plotted in Figs. \ref{figure4NPB},\ref{figure5NPB} as a function of $k$ for the $k$ intervals for which they are negative. 
The exponents $\zeta_{2}^{-+} (k)$ and $\zeta_{3}^{-+} (k)$ are positive and large for all their $k$ intervals. 
Therefore, in Figs. \ref{figure6NPB},\ref{figure7NPB} only the exponent $\zeta_{1}^{-+} (k)$ is plotted. This reveals that the corresponding
$n=2,3$ continua have nearly no spectral weight and therefore are not shown in these figures.
Also the exponent $\zeta_{3}^{zz} (k)$ is positive and large for its $k$ interval.
This reveals again that there is no significant amount of spectral weight in the corresponding $n=3$ continuum of $S^{zz} (k,\omega)$
that is not shown in Figs. \ref{figure8NPB},\ref{figure9NPB}. The exponents $\zeta_{1}^{zz} (k)$ and $\zeta_{2}^{zz} (k)$ 
are plotted in these figures. 

In the $k$ intervals of Figs. \ref{figure4NPB},\ref{figure5NPB},\ref{figure6NPB},\ref{figure7NPB},\ref{figure8NPB},\ref{figure9NPB}
for which the momentum dependent exponents are negative there are power-law sharp peaks of form, 
Eq. (\ref{MPSs}), in the corresponding component $S^{ab} (k,\omega)$. 
Such sharp peaks are located at and just above the lower thresholds of the $(k,\omega)$-plane three $n=1,2,3$ continua 
for $ab=+-$, $n=1$ continuum for $ab=-+$, and two $n=1,2$ continua for $ab=zz$.
The sharp peaks, Eq. (\ref{Sotherline}), are located in the branch line that does not coincide with the lower 
threshold of the $n=1$ continuum of $S^{+-} (k,\omega)$ for intermediate and large $m$ values.

Inspection of the spectra and exponents plotted in 
Figs. \ref{figure4NPB},\ref{figure5NPB},\ref{figure6NPB},\ref{figure7NPB},\ref{figure8NPB},\ref{figure9NPB}, 
reveals that the effects of increasing anisotropy from $\Delta =4$ to $\Delta =8$ are in the cases
of $S^{+-} (k,\omega)$ and $S^{zz} (k,\omega)$ a decrease of the continua energy
bandwidths and an increase of the energies of the $n=2$ and $n=3$ continua.
The $k$ dependent exponent values do not show significant variations upon increasing $\Delta$
from $\Delta =4$ to $\Delta =8$. In the case of $S^{+-} (k,\omega)$, that $\Delta$'s
increase has nearly no effect on the corresponding $n=1$ continuum energy bandwidth.

At the isotropic point, $\Delta =1$, there are no sharp peaks in the gapped lower
threshold of the $S^{+-} (k,\omega)$'s $n=3$ continuum \cite{Carmelo_20}. This indicates that 
the presence of anisotropy tends to increase the spectral weight of that gapped $n=3$ continuum associated
with excited states populated by one $3$-string particle.

The dynamical structure factor expression, Eq. (\ref{MPSs}), is exact at and just above the
lower thresholds of the $(k,\omega)$-plane $n=1$ continua shown in 
Figs. \ref{figure4NPB},\ref{figure5NPB},\ref{figure6NPB},\ref{figure7NPB},\ref{figure8NPB},\ref{figure9NPB}
under which there is no spectral weight. There is no or a negligible amount of spectral weight just below the gapped lower thresholds of 
the $(k,\omega)$-plane $n=2$ continua in Figs. \ref{figure4NPB},\ref{figure5NPB},\ref{figure8NPB},\ref{figure9NPB} and
$n=3$ continua in Figs. \ref{figure4NPB},\ref{figure5NPB}. This ensures that the line shape behavior in
Eq. (\ref{MPSs}) is an excellent approximation for excitation energies at and just above 
such gapped lower thresholds in their $k$ intervals for which $\zeta_{n}^{ab} (k)<0$.

In $(k,\omega)$-plane $n=2,3$ continua gapped lower thresholds $k$ intervals for which there is a negligible amount of spectral weight just
below them, the weak coupling to that weight leads to very small higher order contributions to the expression of the
exponent $\zeta_{n}^{ab} (k)$, Eqs. (\ref{zetaabk}), that can be neglected in the present thermodynamic limit. 

\section{Concluding remarks}
\label{SECVI}

One of the aims of this paper is the clarification of the nature in terms of physical spins $1/2$ configurations 
of scientifically important and interesting elementary magnetic configurations. 
We have exactly shown that such elementary magnetic configurations, that can occur either
unbound and described by $n=1$ real single Bethe rapidities or bound and described by 
$n>1$ $n$-strings, are singlet $S^z=S_q=0$ pairs of physical spins $1/2$. 
For the present $\eta >0$ spin-$1/2$ $XXZ$ chain 
in a longitudinal magnetic field, spin $S$ {\it is not} a good quantum number, so that
such physical spins pairs are labelled by $q$-spin $S_q$. 

A direct consequence of the neutral $q$-spin nature of such $S^z=S_q=0$ pairs of physical spins $1/2$ 
is that the corresponding $2\Pi = N-2S_q = \sum_{n=1}^{\infty}2n\,N_n$ paired physical spins $1/2$ 
of an energy eigenstate do not couple to a vector potential and thus do not carry spin current \cite{Carmelo_21}. 
Only the $M = 2S_q$ unpaired physical spins $1/2$ of that state carry such a current. This property 
can be confirmed by adding a uniform vector potential \cite{Shastry_90} to the Hamiltonian, Eq. (\ref{Hphi}), which
remains solvable by the Bethe ansatz \cite{Zotos_99,Herbrych_11}.
It has direct consequences for the spin transport properties 
of the spin-$1/2$ $XXZ$ chain \cite{Jepsen_20,Steinigeweg_11,Znidaric_11}.

The form of the one-dimensional scalar $S$ matrices, Eqs. (\ref{Smatrix}) and (\ref{functional}), associated with the 
scattering of $n$-particles is another consequence of the singlet $S^z = S_q =0$ nature of these pairs in terms of physical 
spins $1/2$ configurations. This is an important result, since such scattering controls the power-law line shape near the sharp peaks 
in the spin dynamical structure factor components.

We have used a dynamical theory that relies on such $S$ matrices to derive analytical
expressions valid in the thermodynamic limit for the power-law line shape of
$S^{+-} (k,\omega)$, $S^{-+} (k,\omega)$, and  $S^{zz} (k,\omega)$ in the vicinity of sharp peaks.Those are
located in specific $(k,\omega)$-plane momentum $k$ intervals.
Our results refer to anisotropies $\Delta >1$. For magnetic fields in the interval $h\in [h_{c1},h_{c2}]$,
the main difference relative to the $\Delta =1$ isotropic point
\cite{Carmelo_20}, is that for $\Delta >1$ excited states containing one $3$-string-particle lead to
a significant yet small amount of spectral weight in $S^{+-} (k,\omega)$. 

The quantum problem studied in this paper is scientifically interesting in its own right.
On the other hand, the elementary magnetic configurations and corresponding $n$-strings studied in it
have been recently realized and identified in the spin chains of SrCo$_2$V$_2$O$_8$ \cite{Bera_20,Wang_18} and
BaCo$_2$V$_2$O$_8$ \cite{Wang_19} by experiments. Our general theoretical results refer to lines 
of sharp peaks located in $k$ intervals much beyond the few momentum values $k=0$, $k=\pi/2$, and $k=\pi$ of the sharp 
modes that could be experimentally observed \cite{Bera_20,Wang_18,Wang_19}.

There is a unjustified belief that for the isotropic case the elementary magnetic configurations studied in this paper 
are spin-$1$ magnons \cite{Karbach_02,Karbach_00}. This though contradicts exact results that show they rather being
singlet $S^z=S=0$ pairs of physical spins $1/2$ \cite{Carmelo_15,Carmelo_17}. Such a misleading interpretation
was likely imported from the isotropic $\Delta =1$ case by the authors of Refs. \onlinecite{Bera_20,Wang_18,Wang_19}.
For $\Delta >1$ they have identified the $n>1$ elementary magnetic configurations bound within Bethe $n$-strings 
with spin-$1$ magnons. Our result that they are singlet $S^z=S_q=0$ pairs of physical spins $1/2$
corrects that interpretation, which otherwise has not affected
the validity of the results of Refs. \onlinecite{Bera_20,Wang_18,Wang_19}.

There is full qualitative consistency and agreement between finite-size results for $\Delta = 2$ relying on the algebraic Bethe ansatz 
formalism \cite{Yang_19} and those for the $\Delta =4$ and $\Delta =8$ anisotropy values used in the present study
of the present dynamical theory, which refers to the thermodynamic limit.
This concerns both the form of the energy spectra and corresponding dynamical correlation functions
line shapes. As confirmed elsewhere \cite{Carmelo_21}, such a full consistency and agreement becomes 
quantitative under the use of the present dynamical theory at anisotropy $\Delta =2$.

The general expressions of the dynamical structure factor components obtained in this paper
will also be used to show that all the experimentally observed sharp peaks are particular cases of those
considered in the present study \cite{Carmelo_21}. This will include extracting from the line shape of general form, Eq. (\ref{MPSs}), 
the $h$ dependencies in the thermodynamic limit of the energies of the experimentally observed sharp peaks, to
be compared with those observed in the spin chains of SrCo$_2$V$_2$O$_8$
\cite{Bera_20,Wang_18}. This study will be extended to the spin chains of BaCo$_2$V$_2$O$_8$ \cite{Wang_19} in a
future publication \cite{Carmelo_22}. Also the consequences to the physics of such
spin chains of the found $q$-spin neutral nature of the elementary magnetic configurations under
study in this paper will be addressed \cite{Carmelo_21,Carmelo_22}.\\ \\
{\bf CRediT authorship contribution statement}\\

The two authors contributed equally to the formulation of the research goals and aims and to the discussion 
and physical interpretation of the results. J. M. P. C. extended to anisotropy larger than one results for the 
isotropic point concerning both the nature in terms of physical spins $1/2$ of important elementary magnetic 
configurations and a suitable dynamical theory and prepared the original manuscript. P. D. S. has designed the 
computer programs to solve the integral equations that define the phase shifts and spectra of that theory and implemented them.\\ \\
{\bf Declaration of competing interest}\\

The authors declare that they have no known competing financial interests or personal relationships that could have appeared to influence the work reported in this paper.

\acknowledgements
We thank Toma\v{z} Prosen and David K. Campbell for fruitful discussions with the authors and Tilen \v{C}ade\v{z} for useful remarks.
J. M. P. C. would like to thank the Boston University's Condensed Matter Theory Visitors Program for support and
Boston University for hospitality during the initial period of this research. He acknowledges the support from
FCT through the Grants Grant UID/FIS/04650/2013, PTDC/FIS-MAC/29291/2017, and SFRH/BSAB/142925/2018.
P. D. S. acknowledges the support from FCT through the Grant UID/CTM/04540/2019.


\appendix

\section{Confirmation of the $\eta$ independence of the quantum numbers represented by $l_{\rm r}$ in $\vert l_{\rm r},S_q,S^z,\eta\rangle$}
\label{A}

In order to relate to the notations used in Ref. \onlinecite{Takahashi_71}
for Bethe-ansatz quantities of the isotropic model ($\eta =0$), the real part of the rapidities $\varphi_{n,i}$
of Ref. \onlinecite{Gaudin_71} is replaced in this Appendix by the following related quantity,
\begin{equation}
\Lambda_j^n = - {\varphi_{n,j}\over \eta} \in \left[-{\pi\over\eta},{\pi\over\eta}\right]
\hspace{0.40cm}{\rm where}\hspace{0.40cm}
n = 1,...,\infty\hspace{0.40cm}{\rm and}\hspace{0.40cm}j=1,...,L_n \, .
\label{LambdaInterv}
\end{equation}

In the notation $\vert l_{\rm r},S_q,S^z,\eta\rangle$ used in this paper for the $2^N$ energy eigenstates, Eq. (\ref{Ophi}),
all quantum numbers other than $S_q$, $S^z$, and $\eta$ needed to specify such a state are represented by $l_{\rm r}$. 
For each of the $n$-bands of a given energy eigenstate for which $N_n>0$, this 
refers to one of the ${L_n\choose N_n}$ independent occupancy configurations of the $N_n$ $n$-particles
over the $j = 1,...,L_n$ available discrete quantum number values. Such quantum numbers $I_j^n$ refer to the
$n$-band momentum values $q_j = {2\pi\over L}\,I_j^n$ in units of ${2\pi\over L}$. They are given by,
\begin{eqnarray}
q_j & = & {2\pi\over L}\,I_j^n\hspace{0.40cm}{\rm for}\hspace{0.40cm}j = 1,...,L_n 
\hspace{0.40cm}{\rm where}
\nonumber \\
I_j^n & = & 0,\pm 1,...,I_{\pm}^n \hspace{0.40cm}{\rm for}\hspace{0.40cm}L_n\hspace{0.40cm}{\rm odd}
\nonumber \\
& = & \pm 1/2,\pm 3/2,...,I_{\pm}^n \hspace{0.40cm}{\rm for}\hspace{0.40cm}L_n\hspace{0.40cm}{\rm even} \, .
\label{qjDeltal1}
\end{eqnarray}
These quantum numbers are such that $I_j^{n+1}-I_j^n = 1$ and only admit Pauli-like occupancies one or zero. 

Our goal is to show that the $n=1,2,...,\infty$ numbers $L_n$ are independent of $\eta$. This implies
that the values of the quantum numbers represented by $l_{\rm r}$ in $\vert l_{\rm r},S_q,S^z,\eta\rangle = 
\hat{U}_{\eta}^+\vert l_{\rm r},S=S_{\eta},S^z,0\rangle$ indeed are also independent of $\eta$ and thus
remain invariant under the unitary operators $\hat{U}_{\eta}^+$, as $S_q$ and $S^z$ do.
The numbers $L_{n}$ were called $\nu_m$ with
$m=n = 1,...,\infty$ in Ref. \onlinecite{Gaudin_71}, yet they were not derived previously for $\eta>0$. 

In the thermodynamic limit, $\Lambda_j^n$ is the real part of the Bethe-ansatz complex rapidity 
given in Eq. (\ref{LambdaIm}). The set of Bethe-ansatz equations considered in Ref. \onlinecite{Gaudin_71} 
can be written in functional form as,
\begin{equation}
2\arctan\left(\coth\left({n\eta\over 2}\right)\tan \left({\eta\Lambda^n (q_j)\over 2}\right)\right) = q_j
+ {1\over L}\sum_{(n',j')\neq (n,j)}N_{n'} (q_{j'})\Theta^{\eta}_{n\,n'} (\Lambda^n (q_j) - \Lambda^{n'} (q_{j'})) \, ,
\label{BAEqDeltal1}
\end{equation}
where $n=1,...,\infty$. Here the summations refer to $\sum_{n'=1}^{\infty}\sum_{j=1}^{L_{n'}}$ with the 
restriction $(n',j')\neq (n,j)$ and the energy eigenstates distributions $N_{n} (q_j)$ read $N_{n} (q_j)=0$ 
and $N_{n} (q_j)=1$ for unoccupied and occupied discrete momentum values $q_j$, respectively. Those
are given by in Eq. (\ref{qjDeltal1}). Their occupancy configurations generate the energy eigenstates 
described by the Bethe-ansatz solution. Such states refer to
fixed values for the set of numbers $\{N_n\}$ of occupied $q_j$ values in each $n$ band. The
distributions $N_{n} (q_j)$ then obey the sum rule,
\begin{equation}
\sum_{j=1}^{L_n} N_{n} (q_j) = N_n \hspace{0.40cm}{\rm for}\hspace{0.40cm}n=1,...,\infty \, .
\label{sumNnq}
\end{equation}

Moreover, the function $\Theta^{\eta}_{n\,n'}(x)$ in Eq. (\ref{BAEqDeltal1}) is given by,
\begin{eqnarray}
\Theta^{\eta}_{n\,n'}(x) & = & \delta_{n,n'}\Bigl\{2\arctan\left(\coth\left(n\eta\right)\tan \left({\eta\,x\over 2}\right)\right)
+ \sum_{l=1}^{n -1}4\arctan\left(\coth\left(l\eta\right)\tan \left({\eta\,x\over 2}\right)\right)\Bigr\} 
\nonumber \\
& + & (1-\delta_{n,n'})\Bigl\{2\arctan\left(\coth\left({(n+n')\eta\over 2}\right)\tan \left({\eta\,x\over 2}\right)\right) 
+ 2\arctan\left(\coth\left({(\vert n-n'\vert)\eta\over 2}\right)\tan \left({\eta\,x\over 2}\right)\right) 
\nonumber \\
& + & \sum_{l=1}^{{n+n'-\vert\,n-n'\vert\over 2} -1}
4\arctan\left(\coth\left({(\vert n-n'\vert + 2l)\eta\over 2}\right)
\tan \left({\eta\,x\over 2}\right)\right)\Bigr\} \, ,
\label{ThetaDeltal1}
\end{eqnarray}
where $n, n' = 1,...,\infty$ and $\delta_{n,n'}$ is the Kronecker symbol.

As given in Eq. (\ref{LambdaInterv}), $\Lambda_j^n \in \left[-{\pi\over\eta},{\pi\over\eta}\right]$, 
so that the number $L_n$ is derived here by finding, by means of the Bethe-ansatz equations, Eq. (\ref{BAEqDeltal1}), 
the two limiting momentum values $q_j={2\pi\over L} I_{\pm}^n = q_n^{\pm}$ that correspond to $\Lambda_j^n = \pm {\pi\over\eta}$. The
numbers $L_n$ are then given by,
\begin{equation}
L_n = {(q_n^+ - q_n^-)\over (2\pi/L)} + 1\hspace{0.40cm}{\rm for}\hspace{0.40cm}n=1,...,\infty \, .
\label{LnDeltal1}
\end{equation}

One then finds,
\begin{equation}
q_n^{\pm} = {\pi\over L}\Bigl\{\pm N
- {1\over \pi}\sum_{(n',j')\neq (n,j)}N_{n'} (q_{j'})\Theta^{\eta}_{n\,n'} \left(\pm{\pi\over 2} - \Lambda^{n'} (q_{j'})\right)\Bigr\} \, .
\label{qnpmDeltal1}
\end{equation}
After some algebra, one achieves the following general expression,
\begin{equation}
q_n^{\pm} = \pm {\pi\over L}\Bigl\{L_n - 1 \mp \delta L_n^{\eta}\Bigr\} \, ,
\label{qnpmDeltal1Sphi}
\end{equation}
where $L_n$ and $\delta L_n^{\eta}$ are given in Eqs. (\ref{Ln}) and (\ref{deltaLn}), respectively.

While $L_n$ is independent of $\eta$, the term $\delta L_n^{\eta}$ depends on that parameter,
its limiting values being,
\begin{eqnarray}
\delta L_n^{\eta} & = & n\,\eta\sum_{n'=1}^{\infty}\sum_{j=1}^{L_{n'}}n'\,N_{n'} (q_j)\,\left({\eta\,\Lambda^{n'} (q_{j})\over\pi}\right)
\hspace{0.40cm}{\rm for}\hspace{0.40cm}\eta\ll 1
\nonumber \\
\delta L_n^{\eta} & = & \sum_{n'=1}^{\infty}\sum_{j=1}^{L_{n'}}N_{n'} (q_j)
(n + n' - \vert n-n'\vert - \delta_{n',n}) 
\left({\eta\,\Lambda^{n'} (q_{j})\over\pi}\right)
\hspace{0.40cm}{\rm for}\hspace{0.40cm}\eta\gg 1 \, ,
\label{deltaLnphiLimits}
\end{eqnarray}
where we recall that ${\eta\,\Lambda^{n'} (q_{j})\over\pi}\in [-1,1]$ for all 
$\eta \geq 0$.

From the use of the first limiting expression in Eq. (\ref{deltaLnphiLimits}),
one finds that $\delta L_n^{\eta}=0$ at $\eta =0$,
consistently with the Bethe-ansatz solution at the isotropic point \cite{Takahashi_71}.
For $\eta >0$, the numbers $\delta L_n^{\eta}$ vanish when
all distributions $N_{n} (q_j)$ are symmetrical, $N_{n} (q_j)=N_{n} (-q_j)$.
Such numbers are typically of order $1/L$ for excited energy eigenstates
contributing to the dynamical properties studied in this paper.

\section{Ground-state rapidity functions and energy dispersions for the excited states}
\label{B}

\subsection{Ground-state rapidity functions and energy dispersions for the excited states at general $m$ values}
\label{B1}

Here we use the rapidity-function notation $\varphi_{n,j} = \varphi_{n} (q_j)$, whose relation to the notation
suitable to the isotropic case, $\Lambda_j^n = \Lambda^n (q_j)$,
is given in Eq. (\ref{LambdaInterv}) of Appendix \ref{A}. For $n=1$ and $n>1$ the rapidity function $\varphi_{n} (q_j)$ is real and the real part of
a complex rapidity, respectively.  

Three types of excited energy eigenstates span the subspaces associated with the spectra
whose $n=1,2,3$ continua, respectively, are shown in 
Figs. \ref{figure4NPB},\ref{figure5NPB},\ref{figure6NPB},\ref{figure7NPB},\ref{figure8NPB},\ref{figure9NPB}.
The $n=1$ continua stem from excited energy eigenstates
with $N_n = 0$ for $n>1$. They have $1$-band numbers $N_1 = N_{F\downarrow}$,
$N_1^h = N_{F\uparrow} - N_{F\downarrow}$, and $L_1 = N_1 + N_1^h = N_{F\uparrow}$.
Excited energy eigenstates with $N_n = 1$ for a single $n>1$ $n$-band are associated with the $n=2$ and $n=3$ continua
in Figs. \ref{figure4NPB},\ref{figure5NPB},\ref{figure8NPB},\ref{figure9NPB}.
Such excited states have numbers 
$N_1 = N_{F\downarrow} - n$, $N_1^h = N_{F\uparrow} - N_{F\downarrow} + 2 (n-1)$, 
$L_1 = N_1 + N_1^h = N_{F\uparrow} + n - 2$, $N_n =1$, $N_n^h = N_{F\uparrow} - N_{F\downarrow}$,
and $L_n = N_n + N_n^h = N_{F\uparrow} - N_{F\downarrow} + 1$. In all these expressions,
either $n=2$ or $n=3$.

As discussed in Sec. \ref{SECIIA}, in the thermodynamic limit one may use continuous momentum variables $q$ that
replace the discrete $n$-band momentum values $q_j$. The corresponding $n$-band intervals then read
$q\in [-k_{F\uparrow},k_{F\uparrow}]$ for the $1$-band and $q \in [-(k_{F\uparrow} - k_{F\downarrow}),(k_{F\uparrow} - k_{F\downarrow})]$
for the $n=2$ and $n=3$ $n$-bands. Here $q=\pm k_{F\uparrow}$ and $q = \pm (k_{F\uparrow} - k_{F\downarrow})$ play 
the role of $1$-band and $n$-band zone limits, respectively.

The ground-state $n=1$ rapidity distribution function $\varphi_{1} (q)$ where $q \in [-k_{F\uparrow},k_{F\uparrow}]$ 
can be defined in terms of its $1$-band inverse function $q = q_1 (\varphi)$ where $\varphi \in [-\pi,\pi]$. 
From the use of the Bethe-ansatz equation, Eq. (\ref{BAEqDeltal1}) of Appendix \ref{A} for $n=1$
with $\Lambda_j^1 = \Lambda^1 (q_j) = - \varphi_{1} (q_j)/\eta$, one finds that for the ground state
 $q = q_1 (\varphi)$ is defined by the equation,
\begin{equation}
q_1 (\varphi) = 2\arctan\left(\coth\left({\eta\over 2}\right)\tan\left({\varphi\over 2}\right)\right) 
- \frac{1}{\pi} \int_{-B}^{B}d\varphi^{\prime}\,2\pi\sigma_1 (\varphi^{\prime})\, \arctan 
\left(\coth (\eta)\tan\left({\varphi-\varphi^{\prime}\over 2}\right)\right)
\hspace{0.40cm}{\rm for}\hspace{0.40cm} \varphi \in [-\pi,\pi] \, .
\label{equA7}
\end{equation}
Here the parameter $B$ and distribution $2\pi\sigma_1 (\varphi)$ are defined below.

The ground-state $n>1$ rapidity function $\varphi_{n} (q)$ where
$q\in [-(k_{F\uparrow} - k_{F\downarrow}),(k_{F\uparrow} - k_{F\downarrow})]$
is also defined in terms of its $n$-band inverse function 
$q = q_n (\varphi)$ where $\varphi \in [-\pi,\pi]$. From the use of the Bethe-ansatz equations, Eq. (\ref{BAEqDeltal1}) 
of Appendix \ref{A} for $n>1$ with $\Lambda_j^n = \Lambda^n (q_j) = - \varphi_{n} (q_j)/\eta$, one finds that for the ground state
$q_n (\varphi)$ is defined by the following equation,
\begin{eqnarray}
q_n (\varphi) & = & 2\arctan\left(\coth\left({n\,\eta\over 2}\right)\tan\left({\varphi\over 2}\right)\right)
\nonumber \\
& - & \frac{1}{\pi}\sum_{\iota=\pm 1}\int_{-B}^{B}d\varphi^{\prime}\,2\pi\sigma_1 (\varphi^{\prime})
\arctan\left(\coth \left({(n + \iota)\,\eta\over 2}\right)\tan\left({\varphi-\varphi^{\prime}\over 2}\right)\right)
\hspace{0.40cm}{\rm for}\hspace{0.40cm}  \varphi \in [-\pi,\pi] \, ,
\label{qtwoprime}
\end{eqnarray}
where $n>1$.

The parameter $B$ appearing in the above equations is such that,
\begin{equation}
B = \varphi_{1} (k_{F\downarrow}) \hspace{0.40cm}{\rm with}
\hspace{0.40cm}
\lim_{m\rightarrow 0} B = \pi\hspace{0.40cm}{\rm and}\hspace{0.40cm}
\lim_{m\rightarrow 1} B = 0 \, .
\label{QB-r0rs}
\end{equation}
By considering that $\varphi = B$ in Eq. (\ref{equA7}), one finds that $B$ can be defined by the equation,
\begin{equation}
k_{F\downarrow} = 2\arctan\left(\coth\left({\eta\over 2}\right)\tan\left({B\over 2}\right)\right) 
- \frac{1}{\pi} \int_{-B}^{B}d\varphi^{\prime}\,2\pi\sigma_1 (\varphi^{\prime})
\arctan \left(\coth (\eta)\tan\left({B-\varphi^{\prime}\over 2}\right)\right) \, .
\label{BF}
\end{equation}

The distribution,
\begin{equation}
2\pi\sigma_1 (\varphi) = {\partial q_1 (\varphi)\over\partial\varphi} \, ,
\label{sig1}
\end{equation}
also appearing in the above equations is the solution of the following integral equation,
\begin{equation}
2\pi\sigma_1 (\varphi) = {\sinh (\eta)\over \cosh (\eta) - \cos (\varphi)} 
+ \int_{-B}^{B}d\varphi^{\prime}\,G_1 (\varphi - \varphi^{\prime})\,2\pi\sigma_1 (\varphi^{\prime}) \, .
\label{equA10}
\end{equation}
The kernel $G_1 (\varphi)$ in this equation is given by,
\begin{equation}
G_1 (\varphi) = - {1\over{2\pi}}{\sinh (2\eta)\over \cosh (2\eta) - \cos (\varphi)}  \, .
\label{Gne1}
\end{equation} 

The distribution $2\pi\sigma_1 (\varphi)$ obeys the sum rule,
\begin{equation}
\int_{-B}^{B}d\varphi\,2\pi\sigma_1 (\varphi) = \int_{-B}^{B}d\varphi\,{\partial q_1 (\varphi)\over\partial\varphi}
= 2k_{F\downarrow} \, .
\label{equA10B}
\end{equation}

The corresponding $n>1$ distribution $2\pi\sigma_n (\varphi) = \partial q_n (\varphi)/\partial\varphi$ 
is given by,
\begin{equation}
2\pi\sigma_n (\varphi) = {\partial q_n (\varphi)\over\partial\varphi}
= {\sinh (n\,\eta)\over \cosh (n\,\eta) - \cos (\varphi)} 
+ \int_{-B}^{B}d\varphi^{\prime}\,G_n (\varphi - \varphi^{\prime})\,2\pi\sigma_1 (\varphi^{\prime}) \, ,
\label{equA10n}
\end{equation}
where the $n>1$ function $G_n (\varphi)$ reads,
\begin{equation}
G_n (\varphi) = - {1\over{2\pi}}\sum_{\iota=\pm 1}{\sinh ((n+\iota)\,\eta)\over \cosh ((n+\iota)\,\eta) - \cos (\varphi)}  \, .
\label{Gnen}
\end{equation} 

The $n>1$ distribution $2\pi\sigma_n (\varphi)$ obeys the sum rule,
\begin{equation}
\int_{-\pi}^{\pi}d\varphi\,2\pi\sigma_n (\varphi) = \int_{-\pi}^{\pi}d\varphi\,{\partial q_n (\varphi)\over\partial\varphi} = 2(k_{F\uparrow}-k_{F\downarrow}) \, .
\label{equA10Bn}
\end{equation}

The limiting values of the ground-state rapidity functions $\varphi_{1} (q)$ and $\varphi_{n} (q)$
for $n>1$ are,
\begin{eqnarray}
\varphi_{1} (0) & = & 0 \, ; \hspace{0.40cm}\varphi_{1} (\pm k_{F\downarrow}) = \pm B
\hspace{0.40cm}{\rm and}\hspace{0.40cm}
\varphi_{1} (\pm k_{F\uparrow}) = \pm\pi
\nonumber \\
\varphi_{n} (0) & = & 0\hspace{0.40cm}{\rm and}\hspace{0.40cm}
\varphi_{n}  (\pm (k_{F\uparrow} - k_{F\downarrow})) = \pm\pi \, .
\label{qtwoprimelimits}
\end{eqnarray}

The $1$-particle energy dispersion $\varepsilon_1 (q)$ plotted in Fig. \ref{figure1NPB}
and the $n>1$ $n$-string-particle energy dispersions $\varepsilon_{n} (q)$ plotted
in Figs. \ref{figure2NPB} and \ref{figure3NPB} for $n=2$ and $n=3$, respectively,
that appear in the excited states spectra of the dynamical structure factor components play an important role in
the studies of this paper. The energy dispersion $\varepsilon_1 (q)$ is defined as follows,
\begin{eqnarray}
\varepsilon_{1} (q) & = & {\bar{\varepsilon}_{1}} (\varphi_1 (q)) 
\hspace{0.40cm}{\rm for}\hspace{0.40cm}q \in [-k_{F\uparrow},k_{F\uparrow}] 
\hspace{0.40cm}{\rm where}
\nonumber \\
{\bar{\varepsilon}_{1}} (\varphi) & = & {\bar{\varepsilon}_{1}^0} (\varphi) + {1\over 2}\,g\mu_B\,h
\hspace{0.40cm}{\rm for}\hspace{0.40cm}h\in [0,h_{c1}]
\nonumber \\
& = & {\bar{\varepsilon}_{1}^0} (\varphi) - {\bar{\varepsilon}_{1}^0} (B)
\hspace{0.40cm}{\rm for}\hspace{0.40cm}h\in [h_{c1},h_{c2}] \, ,
\label{equA4}
\end{eqnarray}
and
\begin{equation}
{\bar{\varepsilon}_{1}^0} (\varphi) = - J{\sinh^2 (\eta)\over \cosh (\eta) - \cos (\varphi)} 
- {1\over\pi}\int_{-B}^{B}d\varphi^{\prime}\,2J\gamma_{1} (\varphi^{\prime})
\arctan\left(\coth (\eta)\tan\left({\varphi-\varphi^{\prime}\over 2}\right)\right) \, .
\label{equA4B}
\end{equation}
Here $\varphi \in [-\pi,\pi]$ and the distribution $2J\gamma_{1} (\varphi)$ is defined below.
The expression of the magnetic field $h$ and those of the critical magnetic fields $h_{c1}$ and $h_{c2}$ associated with quantum
phase transitions appearing in Eq. (\ref{equA4}) are given in Eqs. (\ref{magcurve}) and (\ref{hc1hc2}), respectively.

The energy dispersions $\varepsilon_n (q)$ are for $n>1$ defined as follows,
\begin{eqnarray}
\varepsilon_{n} (q) & = & {\bar{\varepsilon}_{n}} (\varphi_n (q)) 
\hspace{0.40cm}{\rm for}\hspace{0.40cm}
q \in [-(k_{F\uparrow} - k_{F\downarrow}),(k_{F\uparrow} - k_{F\downarrow})]
\hspace{0.40cm}{\rm where}
\nonumber \\
{\bar{\varepsilon}_{n}} (\varphi) & = & {\bar{\varepsilon}_{n}^0} (\varphi) + (n-1)\,g\mu_B\,h
\hspace{0.40cm}{\rm for}\hspace{0.40cm}h\in [0,h_{c1}]
\nonumber \\
{\bar{\varepsilon}_{n}} (\varphi) & = & {\bar{\varepsilon}_{n}^0} (\varphi) - n\,{\bar{\varepsilon}_{1}^0} (B)
\hspace{0.40cm}{\rm for}\hspace{0.40cm}h\in [h_{c1},h_{c2}] \, ,
\label{equA4n}
\end{eqnarray}
and
\begin{eqnarray}
{\bar{\varepsilon}_{n}^0} (\varphi) & = & - {J\over n}\left({\sinh^2 (n\,\eta)\over \cosh (n\,\eta) - \cos (\varphi)} - C_n (\eta)\right)
\nonumber \\
& - & {1\over\pi}\sum_{\iota=\pm 1}\int_{-B}^{B}d\varphi^{\prime}\,2J\gamma_{1} (\varphi^{\prime})
\arctan\left(\coth \left({(n + \iota)\,\eta\over 2}\right)\tan\left({\varphi-\varphi^{\prime}\over 2}\right)\right) \, .
\label{equA4nb}
\end{eqnarray}
Here $\varphi \in [-\pi,\pi]$, the constant $C_n (\eta)$ is given in Eq. (\ref{Cneta}), and the distribution 
$2J\gamma_{1} (\varphi)$ is defined below.

The energy dispersions $\varepsilon_{1} (q)$ and $\varepsilon_{n} (q)$ for $n>1$ 
that appear in the expressions of the physical excited energy eigenstates spectra and the corresponding
rapidity energy dispersions ${\bar{\varepsilon}_{1}} (\varphi)$ and ${\bar{\varepsilon}_{n}} (\varphi)$, respectively,
are continuous functions of the magnetic field $h$ for the whole range $h\in [0,h_{c2}]$. The auxiliary energy dispersions
though have the following discontinuities at $h=h_{c1}$,
\begin{eqnarray}
\lim_{m\rightarrow 0}\varepsilon_{1}^0 (q) & = & \varepsilon_{1}^0 (q)\vert_{m=0,h=h_{c1}} - {1\over 2}\,g\mu_B\,h_{c1}
\nonumber \\
\lim_{m\rightarrow 0}{\bar{\varepsilon}_{1}^0} (\varphi) & = & {\bar{\varepsilon}_{1}^0} (\varphi)\vert_{m=0,h=h_{c1}} - {1\over 2}\,g\mu_B\,h_{c1}
\nonumber \\
\lim_{m\rightarrow 0}\varepsilon_{n}^0 (q) & = & \varepsilon_{n}^0 (q)\vert_{m=0,h=h_{c1}} - g\mu_B\,h_{c1}
\nonumber \\
\lim_{m\rightarrow 0}{\bar{\varepsilon}_{n}^0} (\varphi) & = & {\bar{\varepsilon}_{n}^0} (\varphi)\vert_{m=0,h=h_{c1}} - g\mu_B\,h_{c1} \, ,
\label{discontinuities}
\end{eqnarray}
where $n>1$.

The energy dispersions ${\bar{\varepsilon}_{n}^0} (\varphi)$ defined by Eqs. (\ref{equA4}) and (\ref{equA4B}) 
for $n=1$ and in Eqs. (\ref{equA4n}) and (\ref{equA4nb}) for $n>1$ can for $n\geq 1$ be expressed as,
\begin{eqnarray}
{\bar{\varepsilon}_{n}^0} (\varphi) & = & \int_{0}^{\varphi }d\varphi^{\prime}2J\gamma_{n} (\varphi^{\prime}) + A_n^{0}
\hspace{0.40cm}{\rm where}
\nonumber \\
A_1^{0} & = & - J(1 + \cosh (\eta)) 
+ {1\over\pi}\int_{-B}^{B}d\varphi^{\prime}\,2J\gamma_{1} (\varphi^{\prime})
\arctan\left(\coth (\eta)\tan\left({\varphi^{\prime}\over 2}\right)\right)
\nonumber \\
A_n^{0} & = & -J {\sinh (\eta)\over\sinh (n\,\eta)}\left(1 + \cosh (n\,\eta)\right) 
\nonumber \\
& + & {1\over\pi}\sum_{\iota=\pm 1}\int_{-B}^{B}d\varphi^{\prime}\,2J\gamma_{1} (\varphi^{\prime})
\arctan\left(\coth \left({(n + \iota)\,\eta\over 2}\right)\tan\left({\varphi^{\prime}\over 2}\right)\right) 
\hspace{0.40cm}{\rm for}\hspace{0.40cm}n>1 \, .
\label{equA4n10}
\end{eqnarray}

One has that,
\begin{equation}
{\partial{\bar{\varepsilon}_{n}^0} (\varphi) \over \partial\varphi}
= 2J\gamma_{n} (\varphi) \hspace{0.40cm}{\rm for}\hspace{0.40cm}n\geq 1 \, ,
\label{defvarepn}
\end{equation}
where the distribution $2J\gamma_{n} (\varphi)$ obeys the following equation,
\begin{equation}
2J\gamma_{n} (\varphi) = {J\over n}{\sinh^2 (n\,\eta)\sin (\varphi)\over (\cosh (n\,\eta) - \cos (\varphi))^2} 
+ \int_{-B}^{B}d\varphi^{\prime}\,G_n (\varphi - \varphi^{\prime})\,2J\gamma_{1} (\varphi^{\prime})  \, .
\label{equA6}
\end{equation}
Here $G_n (\varphi - \varphi^{\prime})$ is given in Eqs. (\ref{Gne1}) and (\ref{Gnen})
for $n=1$ and $n>1$, respectively. For $n=1$, Eq. (\ref{equA6}) is the integral equation,
\begin{equation}
2J\gamma_{1} (\varphi) = J{\sinh^2 (\eta)\sin (\varphi)\over (\cosh (\eta) - \cos (\varphi))^2} 
+ \int_{-B}^{B}d\varphi^{\prime}\,G_1 (\varphi - \varphi^{\prime})\,2J\gamma_{1} (\varphi^{\prime})  \, ,
\label{equA6n1}
\end{equation}
whose solution is the distribution $2J\gamma_{1} (\varphi)$ that appears in 
Eqs. (\ref{equA4}), (\ref{equA4n}), and (\ref{equA6}). 

The following equalities apply,
\begin{eqnarray}
\varepsilon_{1} (0) & = & {\bar{\varepsilon}_{1}} (0) 
\, ; \hspace{0.40cm} \varepsilon_{1} (k_{F\downarrow}) = {\bar{\varepsilon}_{1}} (B) = 0
\, ; \hspace{0.40cm} \varepsilon_{1} (k_{F\uparrow}) = {\bar{\varepsilon}_{1}} (\pi)
\nonumber \\
\varepsilon_{n} (0) & = & {\bar{\varepsilon}_{n}} (0) \, ; \hspace{0.40cm} 
\varepsilon_{n} (k_{F\uparrow} - k_{F\downarrow}) = {\bar{\varepsilon}_{n}} (\pi)
\hspace{0.40cm}{\rm for}\hspace{0.40cm}n>1 \, .
\label{equalities}
\end{eqnarray}

The $n$-band energy bandwidths read,
\begin{eqnarray}
W_1 & = & W_1^p + W_1^h\hspace{0.40cm} {\rm where}
\nonumber \\
W_1^p & = & \varepsilon_{1} (k_{F\downarrow}) - \varepsilon_{1} (0) = {\bar{\varepsilon}_{1}} (B) - {\bar{\varepsilon}_{1}} (0)
\hspace{0.40cm} {\rm and}
\nonumber \\
W_1^h & = & \varepsilon_{1} (k_{F\uparrow}) - \varepsilon_{1} (k_{F\downarrow}) 
= {\bar{\varepsilon}_{1}} (\pi) - {\bar{\varepsilon}_{1}} (B) \, ,
\label{W1}
\end{eqnarray}
for $n=1$ and,
\begin{equation}
W_n = W_n^h =  \varepsilon_{n} (k_{F\uparrow}-k_{F\downarrow}) - \varepsilon_{n} (0) 
 = {\bar{\varepsilon}_{n}} (\pi) - {\bar{\varepsilon}_{n}} (0) \, ,
\label{Wn}
\end{equation}
for $n>1$. 

The $n$-particles $n$-band group velocities are given by,
\begin{equation}
v_n (q) = {\partial\varepsilon_n (q)\over\partial q}\hspace{0.40cm}{\rm for}\hspace{0.40cm}n\geq 1 \, ,
\label{equA4Bv}
\end{equation}
where $\varepsilon_n (q)$ are the $n=1$ and $n>1$ energy dispersions
defined by Eqs. (\ref{equA4})-(\ref{equA4B}) and  (\ref{equA4n})-(\ref{equA4nb}), respectively.

\subsection{Ground-state rapidity functions and energy dispersions for the excited states at $m=0$}
\label{B2}

An important case is that of vanishing spin density, $m=0$, at which $B=\pi$ and the
present model is a gapped spin insulator for magnetic fields $0\leq h < h_{c1}$.
The combined use of Fourier series and the convolution theorem permits the derivation of closed-form expressions for several quantities.
The expressions in Eqs. (\ref{equA7}) and (\ref{qtwoprime}) are then found to read,
\begin{eqnarray}
q_1 (\varphi) & = & {\varphi\over 2} + \sum_{l=1}^{\infty}{1\over l}{\sin (l\,\varphi)\over\cosh (l\,\eta)} 
\hspace{0.40cm}{\rm for}\hspace{0.40cm}\varphi\in [-\pi,\pi] 
\nonumber \\
q_n (\varphi) & = & 0\hspace{0.40cm}{\rm for}\hspace{0.40cm}n>1
\hspace{0.40cm}{\rm and}\hspace{0.40cm}\varphi\in [-\pi,\pi] \, .
\label{equA7m0}
\end{eqnarray}
After performing the infinite summation, the function $q_1 (\varphi)$ is defined
in terms of its inverse function, the rapidity function $\varphi_1 (q)$, as follows,
\begin{equation}
\varphi_1 (q) = \pi {F (q,u_{\eta})\over K (u_{\eta})} 
\hspace{0.40cm}{\rm and}\hspace{0.40cm}
{\rm thus}\hspace{0.40cm}\varphi = \pi {F (q_1 (\varphi),u_{\eta})\over K (u_{\eta})} \, .
\label{varphi1}
\end{equation}
Here $F (q,u_{\eta})$ and $K (u_{\eta})$ are the elliptic integral of the first kind
and the complete elliptic integral of the first kind. The former is given by,
\begin{equation}
F (q,u_{\eta}) = \int_0^{q}d\theta {1\over \sqrt{1 - u_{\eta}^2\sin^2\theta}} \, .
\label{Felliptic}
\end{equation}  
The complete elliptic integral of the first kind $K (u_{\eta})= F (\pi/2,u_{\eta})$ and the $\eta$ 
dependent function $u_{\eta}$ are defined in Eq. (\ref{uphi}).

The function $F (q,u_{\eta}) = F (q_1 (\varphi),u_{\eta})$ is
such that $F (0,u_{\eta})=0$, $F (q,0)=q$,
and $F (q,u_{\eta}) = q$ for $q\rightarrow 0$. This gives the correct limiting values
$q_1 (0)=0$ and $q_1 (\pi) = \pi/2$. The expressions
in Eq. (\ref{varphi1}) define the functions $q_1 (\varphi)$ 
and $\varphi_1 (q)$ for the intervals $q_1 (\varphi)\in [0,\pi/2]$ and
$\varphi_1 (q) \in [0,\pi]$, respectively. That $q_1 (\varphi) = - q_1 (-\varphi)$ 
and $\varphi_1 (q) = - \varphi_1 (-q)$ ensures that 
such an equation defines these functions for their
full intervals $q_1 (\varphi)\in [-\pi/2,\pi/2]$
and $\varphi_1 (q) \in [-\pi,\pi]$, respectively.

The functions $K (u_{\eta})=F (\pi/2,u_{\eta})$ 
and $K' (u_{\eta}) = K \left(\sqrt{1 - u_{\eta}^2}\right)$
have limiting values $K (0) = K' (1) = \pi/2$ and $K (1) = K' (0) = \infty$.
Related useful limiting behaviors for $\eta\ll 1$ and $\eta\gg 1$ are,
\begin{eqnarray}
K (u_{\eta}) & = & {\pi^2\over 2\eta} \, ; \hspace{0.40cm} K' (u_{\eta})  = {\pi\over 2} 
\, ; \hspace{0.40cm} u_{\eta} = 1\hspace{0.40cm}{\rm for}\hspace{0.40cm}\eta\ll 1
\nonumber \\
K (u_{\eta}) & = & {\pi\over 2}\left(1 + 4\,e^{-\eta}\right)
\, ; \hspace{0.40cm}
K' (u_{\eta})  = {\eta\over 2}\left(1 + 4\,e^{-\eta}\right)
\, ; \hspace{0.40cm}
u_{\eta} = 4\,e^{-\eta/2} 
\hspace{0.40cm}{\rm for}\hspace{0.40cm}\eta\gg 1\, .
\label{limitsKK}
\end{eqnarray}

The distribution $2\pi\sigma_1 (\varphi)$, Eq. (\ref{equA10}), is at $m=0$
also expressed as an infinite summation that can again be solved in closed form 
in terms of elliptic integrals of the first kind as follows,
\begin{equation}
2\pi\sigma_1 (\varphi) = {\partial q_1 (\varphi)\over \partial \varphi} =
{1\over 2} + \sum_{l=1}^{\infty}{\cos (l\,\varphi)\over\cosh (l\,\eta)} 
= {1\over\pi} K (u_{\eta})\,\sqrt{1 - u_{\eta}^2\sin^2 q_1 (\varphi)} \, .
\label{equA10m0}
\end{equation}

At $m=0$ the $1$-band energy dispersion, Eqs. (\ref{equA4}) and (\ref{equA4B}), 
can then be expressed as,
\begin{eqnarray}
\varepsilon_{1} (q) & = & {\bar{\varepsilon}_{1}} (\varphi_1 (q)) 
\hspace{0.40cm}{\rm for}\hspace{0.40cm}q \in [-\pi/2,\pi/2] 
\hspace{0.40cm}{\rm where}
\nonumber \\
{\bar{\varepsilon}_{1}} (\varphi) & = & {\bar{\varepsilon}_{1}^0} (\varphi) + {1\over 2}\,g\mu_B\,h
\hspace{0.40cm}{\rm for}\hspace{0.40cm}h \in [0,h_{c1}] \, ,
\label{band1m0}
\end{eqnarray}
and
\begin{equation}
{\bar{\varepsilon}_{1}^0} (\varphi) = - J\sinh (\eta)\left({1\over 2} + \sum_{l=1}^{\infty}{\cos (l\,\varphi)\over\cosh (l\,\eta)}\right)
= - {J\over\pi}\sinh (\eta)\,K (u_{\eta})\,\sqrt{1 - u_{\eta}^2\sin^2 q_1 (\varphi)} 
\hspace{0.40cm}{\rm for}\hspace{0.40cm}\varphi\in [-\pi,\pi] \, ,
\label{band1m0b}
\end{equation}
where the critical magnetic field $h_{c1}$ is defined in Eq. (\ref{hc1hc2}). 
This directly gives an explicit simple dependence of $\varepsilon_{1} (q)$ on the 
$1$-band momentum $q$ as follows,
\begin{equation}
\varepsilon_{1} (q) = - {J\over\pi}\sinh (\eta)\,K (u_{\eta})\,\sqrt{1 - u_{\eta}^2\sin^2 q} 
+ {1\over 2}\,g\mu_B\,h \hspace{0.40cm}
{\rm for}\hspace{0.40cm}q \in [-\pi/2,\pi/2]\hspace{0.40cm}{\rm and}\hspace{0.40cm}h \in [0,h_{c1}] \, .
\label{vareband1m0}
\end{equation}

At $m=0$ the $n$-band energy dispersion, Eqs. (\ref{equA4n}) and (\ref{equA4nb}), is for $n>1$ found to be given by,
\begin{eqnarray}
\varepsilon_{n} (q) & = & {\bar{\varepsilon}_{n}} (\varphi_n (q)) 
\hspace{0.40cm}{\rm for}\hspace{0.40cm}q \in [0^-,0^+]\hspace{0.40cm}{\rm where}
\nonumber \\
{\bar{\varepsilon}_{n}} (\varphi) & = & {\bar{\varepsilon}_{n}^0} (\varphi) + (n-1)\,g\mu_B\,h
\hspace{0.40cm}{\rm for}\hspace{0.40cm}h \in [0,h_{c1}] \, ,
\label{bandnm0}
\end{eqnarray}
and
\begin{equation}
{\bar{\varepsilon}_{n}^0} (\varphi) = - {J\over n}\left(1 - {n\,\sinh (\eta)\over\sinh (n\,\eta)}\right)
\left({\sinh^2 (n\,\eta)\over \cosh (n\,\eta) - \cos (\varphi)} - 1 - \cosh (n\,\eta)\right)
\hspace{0.40cm}{\rm for}\hspace{0.40cm}\varphi\in [-\pi,\pi] \, .
\label{bandnm0b}
\end{equation}
At the limiting $\varphi$ values $\varphi = 0$ and $\varphi = \pm\pi$ the $n>1$ dispersion ${\bar{\varepsilon}_{n}^0} (\varphi)$ reads,
\begin{equation}
{\bar{\varepsilon}_{n}^0} (0) = 0\hspace{0.40cm}{\rm and}\hspace{0.40cm}
{\bar{\varepsilon}_{n}^0} (\pm\pi) = {2J\over n}\left(1 - {n\,\sinh (\eta)\over\sinh (n\,\eta)}\right) \, .
\label{bandnm00pi}
\end{equation}

At $m=0$ one has that $W_1^h =0$ and thus $W_1 = W_1^p$. The 
finite energy dispersions bandwidths read,
\begin{eqnarray}
W_1^p & = & \varepsilon_{1} (\pi/2) - \varepsilon_{1} (0) = {\bar{\varepsilon}_{1}} (\pi) - {\bar{\varepsilon}_{1}} (0)
= {J\over\pi}\sinh (\eta) K (u_{\eta})\,\left(1 - \sqrt{1 - u_{\eta}^2}\right) 
\nonumber \\
W_n^h & = & \varepsilon_{n} (0^{\pm}) = {\bar{\varepsilon}_{n}} (\pm\pi) 
= {2J\over n}\left(1 - {n\,\sinh (\eta)\over\sinh (n\,\eta)}\right)
\hspace{0.40cm}{\rm for}\hspace{0.40cm}n > 1 \, .
\label{bandwidth1m0}
\end{eqnarray}

These energy bandwidths have for $n=1$ and $n>1$ the following limiting behaviors,
\begin{eqnarray}
W_1^p & = & {\pi\over 2}J\hspace{0.40cm}{\rm and}\hspace{0.40cm}W_n^h = 0
\hspace{0.40cm}{\rm for}\hspace{0.40cm}\eta\rightarrow 0
\nonumber \\
W_1^p & = & 2J\hspace{0.40cm}{\rm and}\hspace{0.40cm}W_n^h = {2J\over n}
\hspace{0.40cm}{\rm for}\hspace{0.40cm}\eta\rightarrow\infty \, .
\label{Limitsbandwidthsm0}
\end{eqnarray}
Those of $W_1^p$ are obtained from the use of the corresponding behaviors $K (u_{\eta}) = {\pi^2\over 2}{1\over\eta}$ 
and $u_{\eta} = 4\,e^{-\eta/2}$ for $\eta\gg 1$, Eq. (\ref{limitsKK}).
Those of $W_n^h$ for $n>1$ are obtained from the use of that quantity
general expression provided in Eq. (\ref{bandwidth1m0}). 

Consistently, that $K (u_{\eta}) = {\pi^2\over 2}{1\over\eta}$ for $\eta\ll 1$
in Eq. (\ref{vareband1m0}) gives the following correct isotropic-point expression for 
the $1$-particle energy dispersion $\varepsilon_{1} (q)$ at $h=0$ \cite{Carmelo_20,Carmelo_15A},
\begin{equation}
\lim_{\eta\rightarrow 0}\varepsilon_{1} (q) = - J{\pi\over 2}\cos q\hspace{0.40cm}{\rm for}\hspace{0.40cm}h=0 \, .
\label{epsilonm0eta0}
\end{equation}

The behaviors found for $\eta\rightarrow 0$ are indeed consistent with the corresponding known 
values $W_1^p= {\pi\over 2}J$ and $W_n^h = 0$ where $n>0$ for the isotropic case \cite{Carmelo_20,Carmelo_15A}.
In that case, both the momentum bandwidths and energy bandwidths of
the $n$-bands vanish for $n>1$ in the spin density $m\rightarrow 0$ limit.
For $\eta >1$ this applies to the momentum bandwidths, yet the 
$n>1$ energy bandwidths $W_n = W_n^h$ are finite in that limit, 
as given in Eq. (\ref{bandwidth1m0}).

\subsection{Ground-state rapidity functions and energy dispersions for the excited states in the $m\rightarrow 1$ limit}
\label{B3}

In the $m\rightarrow 1$ limit one has that $B=0$ and all expressions have the same general
$n$-dependent form for $n\geq 1$. Specifically, the expressions, Eqs. (\ref{equA7}) and (\ref{qtwoprime}), are given by,
\begin{equation}
q_n (\varphi) = 2\arctan\left(\coth\left({n\,\eta\over 2}\right)\tan\left({\varphi\over 2}\right)\right)
\hspace{0.40cm}{\rm for}\hspace{0.40cm} \varphi \in [-\pi,\pi] \, .
\label{qtwoprimem1}
\end{equation}
Inversion of this function gives the following
closed-form expression for the $n$-band rapidity functions,
\begin{equation}
\varphi_n (q) = 2\arctan\left(\tanh\left({n\,\eta\over 2}\right)\tan\left({q\over 2}\right)\right)
\hspace{0.40cm}{\rm for}\hspace{0.40cm} q \in [-\pi,\pi] \, .
\label{varphinqm1}
\end{equation}

The distributions, Eqs. (\ref{equA10}) and (\ref{equA10n}), read,
\begin{equation}
2\pi\sigma_n (\varphi) = {\sinh (n\,\eta)\over \cosh (n\,\eta) - \cos (\varphi)} 
\hspace{0.40cm}{\rm for}\hspace{0.40cm} \varphi \in [-\pi,\pi] \, .
\label{equA10nm1}
\end{equation}

The energy dispersions in Eqs. (\ref{equA4}) and (\ref{equA4n}) can be written as,
\begin{eqnarray}
\varepsilon_{n} (q) & = & {\bar{\varepsilon}_{n}} (\varphi_n (q)) 
\hspace{0.40cm}{\rm for}\hspace{0.40cm}
q \in [-\pi,\pi]\hspace{0.40cm}{\rm where}
\nonumber \\
{\bar{\varepsilon}_{n}} (\varphi) & = & {\bar{\varepsilon}_{n}^0} (\varphi) + n\,g\mu_B\,h_{c2} 
\nonumber \\
{\bar{\varepsilon}_{n}}^0 (\varphi) & = & - {J\over n}\Bigl({\sinh^2 (n\,\eta)\over \cosh (n\,\eta) - \cos (\varphi)} 
- C_n (\eta)\Bigr) \, .
\label{equA4nm1}
\end{eqnarray}
Here $C_n (\eta)$ is the constant, Eq. (\ref{Cneta}), and the critical magnetic field $h_{c2}$ 
to the fully-polarized ferromagnetic quantum phase reads $J (\Delta +1 )/g\mu_B$, Eq. (\ref{hc1hc2}). 

After some algebra, one finds that in the $m\rightarrow 1$ limit the $q$-dependent energy dispersions in 
Eqs. (\ref{equA4}) and  (\ref{equA4n}) have the following simple closed-form expressions,
\begin{eqnarray}
\varepsilon_{1} (q) & = & J (1 - \cos q)\hspace{0.40cm}{\rm and}
\nonumber \\
\varepsilon_{n} (q) & = & {J\over n}\Bigl(n^2 (\Delta +1) +  C_n (\eta) - \cosh (n\,\eta) - \cos q\Bigr)
\hspace{0.40cm}{\rm for}\hspace{0.40cm} q \in [-\pi,\pi] \, ,
\label{qdepvarepsilon}
\end{eqnarray}
respectively, where $\Delta = \cosh \eta$. 

Finally, in that limit one has that the $n$-band energy dispersions bandwidths are given by,
\begin{equation}
W_n = W_n^h = \varepsilon_{n} (\pi) - \varepsilon_{n} (0) = {\bar{\varepsilon}_{n}} (\pi) - {\bar{\varepsilon}_{n}} (0)
= {2J\over n} \hspace{0.40cm}{\rm for}\hspace{0.40cm}n\geq 1 \, .
\label{bandwidth1m1}
\end{equation}

\section{Matrix elements that control the line shape near the sharp peaks and related phase shifts}
\label{C}

The $q$-spin neutral nature found in this paper for anisotropy $\Delta >1$ of the unbound $S^z = S_q = 0$ 
pairs of physical spins $1/2$ described by $n=1$ 
real single Bethe rapidities and the $n>1$ bound $S^z = S_q = 0$ pairs of physical spins $1/2$ described by $n$-strings
implies that the $S$ matrices associated with their scattering have the same form,
Eqs. (\ref{Smatrix}) and (\ref{functional}), as for the isotropic point, $\Delta =1$. It follows that the
corresponding dynamical theory has the same general structure as that used in
the studies of Refs. \onlinecite{Carmelo_20,Carmelo_15A} for the isotropic spin-$1/2$ chain. 

Here the relation of the line shape near the sharp peaks located in continua lower thresholds, Eq. (\ref{MPSs}),
to the matrix elements in Eq. (\ref{SDSF}) is shortly addressed for $\Delta >1$. 
The present analysis also applies to the line shape, Eq. (\ref{Sotherline}), of a sharp peak located in a $+-$ branch line
that for intermediate and large spin density values does not coincide with the $n=1$ continuum lower threshold.
The following general dynamical theory expressions are similar to those of the isotropic point, $\Delta =1$.
The difference is the dependence on anisotropy $\Delta = \cosh\eta>1$ of some of the 
quantities in them. 

The dynamical structure factor components are within the dynamical theory used in the studies
of this paper expressed as a sum of $1$-particle spectral functions $B_{1} (k',\omega')$
(denoted by $B_{Q} (k',\omega')$ in Ref. \onlinecite{Carmelo_15A}.) Each of them is associated with 
a reference energy eigenstate and corresponding compactly occupied $1$-band sea. Specifically, the $1$-band
occupancy of such a state changes 
at the $\iota =+1$ right and $\iota =-1$ left {\it reference-state} Fermi points,
\begin{equation}
q_{F1,\iota} = q_{F1,\iota}^0+ {\pi\over L}\,\delta N_{1,\iota}^F \, .
\label{RSFP}
\end{equation}
Here $q_{F1,\iota}^0$ stands for the initial ground-state Fermi points that in 
the thermodynamic limit are given by $q_{F1,\iota}^0=\iota\,k_{F\downarrow}$
and $\delta N_{1,\iota}^F$ are the number deviations at the $\iota = \pm 1$ $1$-band Fermi points 
in the functional $\Phi_{\iota}$ expression, Eq. (\ref{functional}), that define the reference state.

Each $1$-particle spectral function $B_1 (k',\omega')$ contributing to a dynamical correlation function
$S^{ab} (k,\omega)$, Eq. (\ref{MPSs}), involves sums that run over $m_{\iota}=1,2,3,...$ elementary particle-hole processes of 
$\iota = \pm 1$ elementary momentum values $\pm {2\pi\over L}$ around the corresponding reference-state Fermi points $q_{F1,\iota}$.
Such processes generate a tower of excited states upon that reference-state. The
$1$-particle spectral function reads, 
\begin{eqnarray}
B_1 (k',\omega') & = & \sum_{m_{+1};m_{-1}}\,A^{(0,0)}\,a (m_{+1},\,m_{-1})
\nonumber \\
& \times & \delta \Bigl(\omega' - {\bar{\omega}}^{ab}_n (k) - {2\pi\over L}\,v_1\sum_{\iota =\pm1} (m_{\iota}+\Phi_{\iota}^2/4)\Bigr)
\delta \Bigl(k' - k - {2\pi\over L}\,\sum_{\iota =\pm1}\iota\,(m_{\iota}+\Phi_{\iota}^2/4)\Bigr) \, .
\label{BQ-gen}
\end{eqnarray}
Here $v_1 = v_1 (k_{F\downarrow})$ where $v_1 (q)$ is the $1$-band group velocity, Eq. (\ref{equA4Bv}) of Appendix \ref{B} for $n=1$,
and ${\bar{\omega}}^{ab}_n (k)$  is the branch-line one-parametric spectrum in the expression of $S^{ab} (k,\omega)$, Eq. (\ref{MPSs}).
It is given in Appendix \ref{D} for $n=1,2,3$ and $ab=+-$, $n=1$ and $ab = -+$, and $n=1,2$ and $ab = zz$.

The {\it lowest peak weight} $A^{(0,0)}$ and the weights $A^{(0,0)}\,a (m_{+1},\,m_{-1})$ in Eq. (\ref{BQ-gen})
refer to the matrix elements square $\vert\langle \nu\vert\hat{S}^{a}_{k}\vert GS\rangle\vert^2$
in Eq. (\ref{SDSF}) between the ground state and the $m_{+1}=m_{-1}=0$ reference excited state
and the $\iota =\pm 1$ corresponding $m_{\iota}>0$ tower excited states, respectively.
For the subspaces spanned by the ground states and their excited energy eigenstates that 
contribute to the power-law spectral weight line shape near the sharp peaks in the dynamical correlation function
components $S^{ab} (k,\omega)$, Eq. (\ref{MPSs}), the $\iota =\pm 1$ functionals $\Phi_{\iota}$ 
have the general expressions given in Eqs. (\ref{Ph1n1}) and (\ref{Ph1n23several}). They stem from
the form of the $1-particle$ $S$ matrix $S_{1} (\iota k_{F\downarrow})$ in Eq. (\ref{functional}) specific to branch lines
as defined in Ref. \onlinecite{Carmelo_15A}.

The relative weights $a (m_{+1},\,m_{-1})$ in Eq. (\ref{BQ-gen})
can be expressed in terms of the gamma function as \cite{Carmelo_15A},
\begin{equation}
a (m_{+1},m_{-1}) = \prod_{\iota =\pm 1}
a_{\iota}(m_{\iota})\hspace{0.40cm}{\rm where}\hspace{0.40cm}
a_{\iota}(m_{\iota}) = \frac{\Gamma (m_{\iota} +
\Phi_{\iota}^2)}{\Gamma (m_{\iota}+1)\,
\Gamma (\Phi_{\iota}^2)} \, .
\label{amm}
\end{equation}

For excited states whose $\iota =\pm 1$ functional values belong to the 
interval $\Phi_{\iota}\in [-1,1]$, the matrix-element weights have 
in the present thermodynamic limit the following asymptotic behavior,
\begin{eqnarray}
A^{(0,0)} & = & \left({1\over L\,B_1^{ab}}\right)^{-1+\sum_{\iota =\pm1}\Phi_{\iota}^2}
\prod_{\iota =\pm 1}e^{-f_0^{ab} + f_2^{ab}\left(2{\tilde{\Phi}}_{\iota}\right)^2 - f_4^{ab}\left(2{\tilde{\Phi}}_{\iota}\right)^4} 
\hspace{0.40cm}{\rm and}
\nonumber \\
a (m_{+1},m_{-1}) & = & \prod_{\iota =\pm 1}{
(m_{\iota}+\Phi_{\iota}^2/4)^{-1+ \Phi_{\iota}^2}\over\Gamma (\Phi_{\iota}^2)} \, .
\label{Aamm}
\end{eqnarray}
Here,
\begin{equation}
{\tilde{\Phi}}_{\iota} = \Phi_{\iota} -\iota\,\delta N_{1,\iota}^F = - {i\over 2\pi}\ln S_{1} (\iota k_{F\downarrow}) \, ,
\label{functional2}
\end{equation}
is the scattering part of the important general functional
$\Phi_{\iota} = \iota\,\delta N_{1,\iota}^F  - {i\over 2\pi}\ln S_{1} (\iota k_{F\downarrow})$, Eq. (\ref{functional}),
whose values are real. The value of the constant $0<B_1^{ab}\leq 1$ in Eq. (\ref{Aamm})
depends on $\eta$ and $m$ and those of the three $l=0,2,4$ constants $0<f_l^{ab}<1$    
depend on $\eta$. Both $B_1^{ab}$ and $f_l^{ab}$ are independent of $L$. 
Such constants have different values for each $ab = +-,-+,zz$ dynamical structure factor component. 

In the thermodynamic limit, the matrix elements square in Eq. (\ref{SDSF}) that contribute to the line shape
near a sharp peak then read,
\begin{eqnarray}
\vert\langle \nu\vert\hat{S}^{a}_k\vert GS\rangle\vert^2 & = & 
\vert\langle k',m_{+1},m_{-1}\vert\hat{S}^{a}_k\vert GS\rangle\vert^2
= A^{(0,0)}\prod_{\iota =\pm 1}\frac{\Gamma (m_{\iota} +
\Phi_{\iota}^2)}{\Gamma (m_{\iota}+1)\,\Gamma (\Phi_{\iota}^2)} 
\nonumber \\
& = & \left({1\over L\,B_1^{ab}}\right)^{-1+\sum_{\iota =\pm1}\Phi_{\iota}^2}
\prod_{\iota =\pm 1}e^{-f_0^{ab} + f_2^{ab}\left(2{\tilde{\Phi}}_{\iota}\right)^2 - f_4^{ab}\left(2{\tilde{\Phi}}_{\iota}\right)^4} 
\frac{\Gamma (m_{\iota} + \Phi_{\iota}^2)}{\Gamma (m_{\iota}+1)\,
\Gamma (\Phi_{\iota}^2)} 
\nonumber \\
& = & \left({1\over L\,B_1^{ab}}\right)^{-1+\sum_{\iota =\pm1}\Phi_{\iota}^2}
\prod_{\iota =\pm 1}{e^{-f_0^{ab} + f_2^{ab}\left(2{\tilde{\Phi}}_{\iota}\right)^2 - f_4^{ab}\left(2{\tilde{\Phi}}_{\iota}\right)^4}  \over\Gamma (\Phi_{\iota}^2)} 
\left(m_{\iota}+\Phi_{\iota}^2/4\right)^{-1+ \Phi_{\iota}^2}
\nonumber \\
& = & \left({1\over L\,B_1^{ab}}\right)^{-1+\sum_{\iota =\pm1}\Phi_{\iota}^2}
\nonumber \\
& \times & 
\prod_{\iota =\pm 1}{e^{-f_0^{ab} + f_2^{ab}\left(2{\tilde{\Phi}}_{\iota}\right)^2 - f_4^{ab}\left(2{\tilde{\Phi}}_{\iota}\right)^4}  \over\Gamma (\Phi_{\iota}^2)} 
\left({L\over 4\pi\,v_1}(\omega' - {\bar{\omega}}^{ab}_n (k) +\iota\,v_1\,(k'-k))\right)^{-1+ \Phi_{\iota}^2} \, .
\label{ME}
\end{eqnarray}
Here $\vert\nu\rangle = \vert k',m_{+1},m_{-1}\rangle$ denotes the excited energy eigenstate generated
from the reference state $\vert \tilde{k},0,0\rangle$ by $m_{+1}=1,2,3,...$.
$m_{-1}=1,2,3,...$ elementary particle-hole processes of momentum values $\pm {2\pi\over L}$ around 
the two $\iota = \pm 1$ $1$-band reference-state Fermi points, Eq. (\ref{RSFP}).

The two corresponding equalities,
\begin{equation}
m_{\iota} + \Phi_{\iota}^2/4 = {L\over 4\pi\,v_1}(\omega' - {\bar{\omega}}^{ab}_n (k) +\iota\,v_1\,(k'-k))
\hspace{0.40cm}{\rm for}\hspace{0.40cm}\iota = \pm 1 \, ,
\label{miota}
\end{equation}
imposed by the two $\delta$-functions in Eq. (\ref{BQ-gen}) have been used to arrive to the
fourth expression in Eq. (\ref{ME}). 

Such $\delta$-functions also select the specific two $m_{+1}$ and $m_{-1}$ values given in Eq. (\ref{miota}) within the sum 
$\sum_{m_{+1};m_{-1}}$ in Eq. (\ref{BQ-gen}) that correspond to the fixed $k'$ and $\omega'$ values of 
the $1$-particle spectral function $B_1 (k',\omega')$. It follows that for the general case in which the two $\iota =\pm 1$ functionals 
$\Phi_{\iota}\in [-1,1]$ are finite, that spectral function, Eq. (\ref{BQ-gen}), can then be written as,
\begin{eqnarray}
B_1 (k',\omega') & = & {1\over L\,B_1^{ab}}\,
\prod_{\iota =\pm 1}\,\Theta (\omega' - {\bar{\omega}}^{ab}_n (k) + \iota\,v_1\,(k'-k))
{e^{-f_0^{ab} + f_2^{ab}\left(2{\tilde{\Phi}}_{\iota}\right)^2 - f_4^{ab}\left(2{\tilde{\Phi}}_{\iota}\right)^4} \over \Gamma (\Phi_{\iota}^2)}
\nonumber \\
& \times & \Bigl({\omega' - {\bar{\omega}}^{ab}_n (k) +\iota\,v_1\,(k'-k)\over 4\pi\,B_1^{ab}\,v_1}\Bigr)^{-1+\Phi_{\iota}^2} \, .
\label{B-J-i-sum-GG}
\end{eqnarray}
To reach this expression, which in the thermodynamic limit is exact, Eqs. 
(\ref{BQ-gen}), (\ref{Aamm}), (\ref{ME}), and (\ref{miota}) were used. 

Our goal is to arrive to expression, Eq. (\ref{MPSs}), for the dynamical structure factor components
at a fixed excitation momentum $k$ and excitation energy such that the deviation
$(\omega - {\bar{\omega}}^{ab}_{n} (k))$ is small, for each reference energy eigenstate
corresponding to a given $\omega'$ value in the argument of $B_1 (k',\omega')$. To reach it
one selects a suitable set of tower excited states such that in average
$\sum_{\iota =\pm1}\iota\,(m_{\iota}+\Phi_{\iota}^2/4) = 0$
whereas $\sum_{\iota =\pm1} (m_{\iota}+\Phi_{\iota}^2/4)$ is finite.

One then performs a suitable summation of $1$-particle spectral functions 
that corresponds to a summation over reference energy eigenstates with
different energy $\omega'$ values. That summation is equivalent to integrate that 
function at $k'=k$ over the energy variable $z = (\omega' - {\bar{\omega}}^{ab}_n (k))/(4\pi\,B_1^{ab}\,v_1)$ 
in suitable intervals that are determined by values of the quantity
$-1 + \sum_{\iota =\pm 1}\Phi_{\iota}^2$. It turns out to be the
exponent $\zeta_{n}^{ab} (k)$, Eq.(\ref{zetaabk}),
being positive, negative, or vanishing, respectively.

It is then useful to express $B_1 (k',\omega')\vert_{k'=k}$ as a function of the
variable $z = (\omega' - {\bar{\omega}}^{ab}_n (k))/(4\pi\,B_1^{ab}\,v_1)$ by 
considering the related function,
\begin{equation}
{\bar{B}}_1 (z) = B_1 (k,\omega')\vert_{\omega' = 4\pi\,B_1^{ab}\,v_1\,z + {\bar{\omega}}^{ab}_n (k)} =
{1\over L\,B_1^{ab}}{\Theta (z)\over z}\,z^{\zeta_{n}^{ab} (k)}\,
\prod_{\iota =\pm 1}\,
{e^{-f_0^{ab} + f_2^{ab}\left(2{\tilde{\Phi}}_{\iota}\right)^2 - f_4^{ab}\left(2{\tilde{\Phi}}_{\iota}\right)^4} \over \Gamma (\Phi_{\iota}^2)} \, .
\label{B-J-i-sum-GG2}
\end{equation}

The above expressions are valid for $\iota =\pm 1$ functional values in the
interval $\Phi_{\iota}\in [-1,1]$. In the studies of this paper we
are mostly interested in the $(k,\omega)$-plane line shape near sharp peaks for which
$\zeta_{n}^{ab} (k) <0$ in Eq. (\ref{MPSs}). However, provided
there is no spectral weight below the corresponding 
lower threshold continuum or there is near no such a weight below it,
the dynamical correlation function expression in that equation
is valid also for $0<\zeta_{n}^{ab} (k)<1$ provided that
$\Phi_{\iota}^2\leq 1$ for $\iota = \pm 1$. In that case the power-law
line shape expression, Eq. (\ref{MPSs}), does not refer to a sharp peak.

Upon performing the reference-state summation through an integration over $z$, $S^{ab} (k,\omega)$ is given 
by the following expression,
\begin{equation}
S^{ab} (k,\omega) = c_*\left({L\over 4\pi v_1}\times 4\pi\,B_1^{ab}\,v_1\right) \int_{z_*}^{{\omega - {\bar{\omega}}^{ab}_n (k)\over 4\pi\,B_1^{ab}\,v_1}} d z\,{\bar{B}}_1 (z)
= c_*L\,B_1^{ab}\int_{z_*}^{{\omega - {\bar{\omega}}^{ab}_n (k)\over 4\pi\,B_1^{ab}\,v_1}} d z\,{\bar{B}}_1 (z) \, .
\label{MPSs0}
\end{equation}
The values $c_*= 1$ or $c_*= - 1$ of $c_*$ and those of the integration limit $z_*$ specific to the three cases 
$0<\zeta_{n}^{ab} (k)<1$, $-1<\zeta_{n}^{ab} (k)<0$, and $\zeta_{n}^{ab} (k) = 0$, respectively, are given in the following.

Both for $0<\zeta_{n}^{ab} (k)<1$ and $-1<\zeta_{n}^{ab} (k)<0$ the obtained general expression
for $S^{ab} (k,\omega)$ is the same. However, the limits of integration in Eq. (\ref{MPSs0}) are different. 
For $0<\zeta_{n}^{ab} (k)<1$, such limits are such that the integral runs in the interval 
$z \in [0,(\omega - {\bar{\omega}}^{ab}_n (k))/(4\pi\,B_1^{ab}\,v_1)]$. 
This corresponds to $c^* = 1$ and $z_*=0$ in Eq. (\ref{MPSs0}). For 
small values of the energy deviation $(\omega - {\bar{\omega}}^{ab}_n (k))>0$,
$\omega'$ runs in this case from $\omega'={\bar{\omega}}^{ab}_n (k)$ to
$\omega'=\omega$. This is a small interval just below $\omega$.

Consistently with the requirement of $S^{ab} (k,\omega)$ being positive,
for $-1<\zeta_{n}^{ab} (k)<0$ the integral in Eq. (\ref{MPSs0}) must run in a $z$ interval 
that corresponds to $\omega'$ values near $\omega$ but just above it.
This generates a singular leading order term, Eq. (\ref{MPSs}), which
is that physically relevant. The second term associated with the 
integration limit $z_*$ in Eq. (\ref{MPSs0}) is independent of $\omega$ and has no
physical meaning. Indeed, the present method only captures the
leading-order term, Eq. (\ref{MPSs}). Since for $-1<\zeta_{n}^{ab} (k)<0$ the 
primitive function associated with the integrand in Eq. (\ref{MPSs0}) decreases very quickly 
upon increasing $z$, the suitable integration $z$ interval 
in Eq. (\ref{MPSs0}) is $z \in [(\omega - {\bar{\omega}}^{ab}_n (k))/(4\pi\,B_1^{ab}\,v_1),\infty]$.
This corresponds to $c^* = - 1$ and $z_*= \infty$ in Eq. (\ref{MPSs0}).
Indeed, this choice renders the second unphysical term to be zero.

Both such state summations suitable to $0<\zeta_{n}^{ab} (k)<1$ and $-1<\zeta_{n}^{ab} (k)<0$,
respectively, lead to exactly the same general expression for 
$S^{ab} (k,\omega)$,
\begin{equation}
S^{ab} (k,\omega) = \left[{1\over \vert\zeta_{n}^{ab} (k)\vert}\prod_{\iota =\pm 1}
{e^{-f_0^{ab} + f_2^{ab}\left(2{\tilde{\Phi}}_{\iota}\right)^2 - f_4^{ab}\left(2{\tilde{\Phi}}_{\iota}\right)^4} \over \Gamma (\Phi_{\iota}^2)}\right] 
\left({\omega - {\bar{\omega}}^{ab}_n (k)\over 4\pi\,B_1^{ab}\,v_1}\right)^{\zeta_{n}^{ab} (k)} \, .
\label{MPSs1}
\end{equation}
This is the expression given in Eq. (\ref{MPSs}) with $\zeta_{n}^{ab} (k)$ and $C_{ab}^n (k)$ provided in 
Eqs. (\ref{zetaabk}) and (\ref{Cabn}), respectively. It is valid for small values of the energy deviations 
$(\omega - {\bar{\omega}}^{ab}_n (k))> 0$. (The same applies to the expression, Eq. (\ref{Sotherline}).)

Importantly, this expression is not valid when $\zeta_{n}^{ab} (k) = -1 + \sum_{\iota =\pm 1}\Phi_{\iota}^2 =0$
and thus $\sum_{\iota =\pm 1}\Phi_{\iota}^2 = 1$. Again consistently with the requirement of $S^{ab} (k,\omega)$ 
being positive, for $\zeta_{n}^{ab} (k) = 0$ the integral in Eq. (\ref{MPSs0}) runs in a $z$ interval 
that corresponds to $\omega'$ values near $\omega$ but just above it. Then the physically relevant leading-order term 
contains a logarithmic rather than power-law singularity whereas the second $\omega$ independent 
term has again no physical meaning. To render that term zero, the suitable integration $z$ interval 
in Eq. (\ref{MPSs0}) is now $z \in [(\omega - {\bar{\omega}}^{ab}_n (k))/(4\pi\,B_1^{ab}\,v_1),1]$. 
This corresponds to $c^* = - 1$ and $z_*= 1$ in Eq. (\ref{MPSs0}) and leads to,
\begin{equation}
S^{ab} (k,\omega) = 
\left[{e^{\sum_{\iota =\pm 1}(-f_0^{ab} + f_2^{ab}\left(2{\tilde{\Phi}}_{\iota}\right)^2 - f_4^{ab})\left(2{\tilde{\Phi}}_{\iota}\right)^4}
\over \sum_{\iota =\pm 1}{\pi\over 2\sin (\pi\Phi_{\iota}^2 (q))}}\right]
\vert\ln \left({\omega - {\bar{\omega}}^{ab}_n (k)\over 4\pi\,B_1^{ab}\,v_1}\right)\vert \, .
\label{MPSs2}
\end{equation}
Since $\sum_{\iota =\pm 1}\Phi_{\iota}^2 = 1$, here that $\Gamma (x)\Gamma (1-x) = \pi/\sin (\pi x)$ for $x\neq 0,\pm 1,...$ was used.
This gives $\Gamma (\Phi_{+1}^2)\Gamma (\Phi_{-1}^2) = \pi/\sin (\pi \Phi_{+1}^2) = \pi/\sin (\pi \Phi_{-1}^2)$
and thus also an equivalent more symmetrical expression 
$\Gamma (\Phi_{+1}^2)\Gamma (\Phi_{-1}^2) = \sum_{\iota =\pm 1}\pi/[2\sin (\pi \Phi_{\iota}^2)]$.

The phase shifts in the general expression given in Eq. (\ref{functional}) of the $\iota =\pm 1$ functionals $\Phi_{\iota}$ 
appearing here in Eqs. (\ref{BQ-gen})-(\ref{MPSs2}) are associated with the $S$ matrices, Eqs. (\ref{Smatrix}) 
and (\ref{functional}). They determine the momentum dependence of the exponents, Eq. (\ref{zetaabk}), through 
such $\iota =\pm 1$ functionals $\Phi_{\iota}$. Such phase shifts play an important role in the dynamical properties 
studied in Sec. \ref{SECV} and are given by,
\begin{equation}
2\pi\,\Phi_{n,n'}(q,q') = 2\pi\,\bar{\Phi }_{n,n'} \left(\varphi,\varphi'\right) 
\hspace{0.40cm}{\rm where}\hspace{0.40cm}\varphi = \varphi_{n}(q)
\hspace{0.40cm}{\rm and}\hspace{0.40cm}\varphi' = \varphi_{n'}(q') \, ,
\label{Phi-barPhi}
\end{equation}
for $n\geq 1$ and $n'\geq 1$.

As justified in Sec. \ref{SECV}, in the case of the excited energy eigenstates involved in our studies,
only the phase shifts $2\pi\,\Phi_{1,n}(q,q')$ where $n\geq 1$ play an active role.
The corresponding rapidity phase shifts 
$2\pi\,\bar{\Phi }_{1,n} \left(\varphi,\varphi'\right)$ are in units
of $2\pi$ defined by the following integral equations,
\begin{equation}
\bar{\Phi }_{1,1} \left(\varphi,\varphi'\right) = 
{1\over \pi}\arctan\left(\coth (\eta)\tan\left({\varphi - \varphi'\over 2}\right)\right)
+ \int_{-B}^{B} d\varphi''\,G_1 (\varphi - \varphi'')\,\bar{\Phi }_{1,1} \left(\varphi'',\varphi'\right) \, ,
\label{Phis11}
\end{equation}
and
\begin{equation}
\bar{\Phi }_{1,n} \left(\varphi,\varphi'\right) = {1\over \pi}\sum_{\iota=\pm 1}
\arctan\left(\coth\left({(n + \iota)\,\eta\over 2}\right)\tan\left({\varphi - \varphi'\over 2}\right)\right)
+ \int_{-B}^{B} d\varphi''\,G_1 (\varphi - \varphi'')\,\bar{\Phi }_{1,n} \left(\varphi'',\varphi'\right) \, ,
\label{Phis1n}
\end{equation}
for $n>1$ where the kernel $G_1 (\varphi)$ is given in Eq. (\ref{Gne1}) of Appendix \ref{B}.

The following quantities play an important role in the dynamical properties studied in Sec. \ref{SECV}.
The phase shifts in units of $2\pi$ that appear in the expressions of the momentum-dependent exponents given
in Appendix \ref{D} are the following,
\begin{eqnarray}
\Phi_{1,1}\left(\iota k_{F\downarrow},q\right) & = & \bar{\Phi }_{1,1} \left(\iota B,\varphi_1 (q)\right) 
\hspace{0.2cm}{\rm for}\hspace{0.2cm}\iota = \pm 1
\nonumber \\
\Phi_{1,n}\left(\iota k_{F\downarrow},q\right) & = & \bar{\Phi }_{1,n} \left(\iota B,\varphi_{n} (q)\right) 
\hspace{0.2cm}{\rm for}\hspace{0.2cm}\iota = \pm 1\hspace{0.2cm}{\rm and}\hspace{0.2cm}n = 2,3  \, .
\label{Phis-all-qq}
\end{eqnarray}
The related $1$-band phase-shift parameter $\xi_{1\,1}$ in Eq. (\ref{dXxi11}) that also appears in such exponent expressions 
plays an important role as well in the spin-density curve $h (m)$, Eq. (\ref{magcurve}).
As in other integrable models \cite{Carmelo_94}, it also appears in operator descriptions 
of Virasoro algebras. From manipulations of the phase-shift integral equation, Eq. (\ref{Phis11}),
one finds that it can be expressed as,
\begin{equation}
\xi_{1\,1} = \xi_{1\,1} (B) \hspace{0.40cm}{\rm where}\hspace{0.40cm}
\xi_{1\,1} (\varphi) = 1 + \int_{-B}^{B} d\varphi'\,G_1 (\varphi - \varphi')\,\xi_{1\,1} (\varphi') \, .
\label{xi1all}
\end{equation}
The kernel $G_1 (\varphi)$ in the integral equation obeyed by the auxiliary function $\xi_{1\,1} (\varphi)$
is defined in Eq. (\ref{Gne1}) of Appendix \ref{B}.
Also the phase-shift related parameters $\xi_{1\,n}^0$ and $\xi^{\iota,\pm}_n$ where $n=2,3$ and $\iota =\pm 1$, Eq. (\ref{xpmnx0n}),
appear in the expressions of some momentum-dependent exponents given in Appendix \ref{D}. They can be expressed 
in terms of the rapidity phase shifts defined by Eqs. (\ref{Phis11}) and (\ref{Phis1n}) as $\xi^0_n =  2\bar{\Phi }_{1,n}(B,0)$ and 
$\xi^{\iota,\pm}_n = \bar{\Phi }_{1,n}(\iota B,\pm\pi)$.
For $\Delta >1$ and at $m= 0$ and at $m= 1$ the parameters $\xi_{1\,1}$, $\xi_{1\,n}^0$, and $\xi^{\iota,\pm}_n$ are given by,
\begin{eqnarray}
\xi_{1\,1} & = & {1\over 2} \, ; \hspace{0.40cm}\xi_{1\,n}^0 = 2\, ; \hspace{0.40cm}\xi^{\iota,\pm}_n = \iota \mp {1\over 2}
\hspace{0.40cm}{\rm at}\hspace{0.40cm}m=0
\nonumber \\
\xi_{1\,1} & = & 1 \, ;\hspace{0.40cm}\xi_{1\,n}^0 = 0 \, ; \hspace{0.40cm}\xi^{\iota,\pm}_n = \mp 1
\hspace{0.40cm}{\rm at}\hspace{0.40cm}m=1 \, .
\label{xi-ss-qqm1}
\end{eqnarray}

\section{Lower threshold spectra and momentum-dependent exponents}
\label{D}

The one-parametric spectra of the gapped lower thresholds of the $n=2$ continua shown
in Figs. \ref{figure4NPB},\ref{figure5NPB}\ref{figure8NPB},\ref{figure9NPB} 
and of the $n=3$ continuum shown in Figs. \ref{figure4NPB},\ref{figure5NPB}
have a different form for three spin density intervals, 
$m\in ]0,\bar{m}_n]$, $m\in ]\bar{m}_n,\tilde{m}_n]$, and $m\in [\tilde{m}_n,1[$, 
respectively. Here $\bar{m}_n$ and $\tilde{m}_n$ such that
$\bar{m}_n<1/3$ and $\tilde{m}_n >\bar{m}_n$ are for $n=2$ and $n=3$ two $\eta$-dependent 
spin densities at which the following equality holds,
\begin{equation}
W_{n}^h = - \varepsilon_{1} (2k_{F\downarrow}-k_{F\uparrow})
\hspace{0.40cm}{\rm at}\hspace{0.40cm}m = \bar{m}_n < 1/3\hspace{0.40cm}{\rm and}\hspace{0.40cm}
{\rm at}\hspace{0.40cm}m = \tilde{m}_n > \bar{m}_n \hspace{0.40cm}{\rm for}\hspace{0.40cm} n=2,3 \, .
\label{mtilde}
\end{equation}
The $1$-particle energy dispersion $\varepsilon_{1} (q)$ and the energy bandwidth $W_n = W_{n}^h$ 
for $n=2,3$ appearing here are defined in Eqs. (\ref{equA4})-(\ref{equA4B}) and (\ref{Wn}) of Appendix \ref{B},
respectively. One has that $\lim_{\eta\rightarrow 0} \bar{m}_n =0$.

Some specific excitation momentum $k$'s values separate momentum intervals of 
the lower thresholds of the $S^{+-} (k,\omega)$ $n=2,3$ continua in Figs. \ref{figure4NPB},\ref{figure5NPB}
and $S^{zz} (k,\omega)$ $n=2$ continuum in Figs. \ref{figure8NPB},\ref{figure9NPB} 
that refer to different types of momentum dependences. Such specific $k$ values
either equal a momentum denoted here by $\tilde{k}_n$ or their expression involves $\tilde{k}_n$
where $n=2,3$. That momentum is defined by the following relations, 
\begin{eqnarray}
\varepsilon_{n} (\tilde{k}_n)  & = & \varepsilon_{n} (0) - \varepsilon_{1} (k_{F\downarrow}-\tilde{k}_n)
\hspace{0.40cm}{\rm for} \hspace{0.40cm}\tilde{k}_n\leq (k_{F\uparrow} - k_{F\downarrow})\hspace{0.40cm}
{\rm and}\hspace{0.40cm}m\in ]0,\bar{m}_n] 
\nonumber \\
W_{n}^h & = & \varepsilon_{1} (k_{F\uparrow}-\tilde{k}_n) -  \varepsilon_{1} (k_{F\downarrow}-\tilde{k}_n)
\hspace{0.40cm}{\rm for} \hspace{0.40cm}\tilde{k}_n\geq (k_{F\uparrow} - k_{F\downarrow})\hspace{0.40cm}
{\rm and}\hspace{0.40cm}m\in [\bar{m}_n,\tilde{m}_n]
\nonumber \\
\varepsilon_{n} (\tilde{k}_n)  & = & \varepsilon_{n} (0) - \varepsilon_{1} (k_{F\downarrow}-\tilde{k}_n)
\hspace{0.40cm}{\rm for} \hspace{0.40cm}\tilde{k}_n\leq (k_{F\uparrow} - k_{F\downarrow})\hspace{0.40cm}
{\rm and}\hspace{0.40cm}m\in [\tilde{m}_n,1[ \, ,
\label{ktilde}
\end{eqnarray}
where the energy dispersion $\varepsilon_{n} (q)$ is for $n=2,3$ given in 
Eq. (\ref{equA4n}) of Appendix \ref{B}.

The momentum $\tilde{k}_n$ reads $\tilde{k}_n=(k_{F\uparrow} - k_{F\downarrow})$ both
at $m = \bar{m}_n$ and $m = \tilde{m}_n$. For $m\in [0,\bar{m}_n]$ it
increases upon increasing $m$ from $\tilde{k}_n=0$ for $m\rightarrow 0$ to 
$\tilde{k}_n=(k_{F\uparrow} - k_{F\downarrow})$ at $m = \bar{m}_n$. For
$m\in [\bar{m}_n,\tilde{m}_n]$ it first increases relative to 
$(k_{F\uparrow} - k_{F\downarrow})$ and after decreases relative to
it, until being again given by $\tilde{k}_n=(k_{F\uparrow} - k_{F\downarrow})$ at $m = \tilde{m}_n$. 
Finally, for $m\in [\tilde{m}_n,1[$ it
decreases upon increasing $m$ from $\tilde{k}_n=(k_{F\uparrow} - k_{F\downarrow})$
at $m = \bar{m}_n$ to $\tilde{k}_n=0$ for $m\rightarrow 1$.

The one-parametric spectra given in the following also involve the energy dispersions $\varepsilon_{1} (q)$ 
plotted in Fig. \ref{figure1NPB} and $\varepsilon_{n} (q)$ plotted in Figs. \ref{figure2NPB} and \ref{figure3NPB} 
for $n=2$ and $n=3$, respectively. Such energy dispersions are defined by Eqs. (\ref{equA4})-(\ref{equA4n10}) of Appendix \ref{B}.
In the cases of the gapped lower thresholds of the $n=2$ continua shown in 
Figs. \ref{figure4NPB},\ref{figure5NPB},\ref{figure8NPB},\ref{figure9NPB} 
and of the $n=3$ continuum shown in Figs. \ref{figure4NPB},\ref{figure5NPB}, the
one-parametric spectra under consideration obey the following relations,
\begin{eqnarray}
{\bar{\omega}}^{zz}_{n} (k) & =  & {\bar{\omega}}^{+-}_{n} (\pi - k)
\hspace{0.40cm}{\rm for}\hspace{0.40cm}n=2,3\hspace{0.40cm} 
{\rm and} \hspace{0.40cm}k\in [0,\pi] 
\nonumber \\
{\bar{\omega}}^{+-}_{3} (k) & =  & {\bar{\omega}}^{+-}_{2} (\pi - k)
\hspace{0.40cm}{\rm for}\hspace{0.40cm}k\in [0,\pi] 
\nonumber \\
{\bar{\omega}}^{zz}_{3} (k) & =  & {\bar{\omega}}^{zz}_{2} (\pi - k)
\hspace{0.40cm}{\rm for}\hspace{0.40cm}k\in [0,\pi] \, .
\label{GappLTlong}
\end{eqnarray}

In the following we provide the expressions of the one-parametric lower threshold spectra 
${\bar{\omega}}^{ab}_{n} (k)$ of the $n=1,2,3$ continua for $ab=+-$, $n=1$ continuum 
for $ab=-+$, and $n=1,2$ continua for $ab=zz$ and corresponding momentum-dependent
exponents $\zeta_{n}^{ab} (k)$ in Eq. (\ref{MPSs}). The branch-line spectrum
${\tilde{\omega}}^{+-}_{1} (k)$ and corresponding exponent
${\tilde{\zeta}}_{1}^{+-} (k)$ in Eq. (\ref{Sotherline}) are also given.

\subsection{Lower threshold spectra and exponents of $S^{+-} (k,\omega)$ for $m>0$}
\label{D1}

The $n=1$ lower threshold spectrum ${\bar{\omega}}^{+-}_{1} (k)$ 
in the expression of $S^{+-} (k,\omega)$, Eq. (\ref{MPSs}) for $ab=+-$ and $n=1$,
is for $m>0$ divided into the following two branch-line intervals,
\begin{eqnarray}
{\bar{\omega}}^{+-}_1 (k) & = & \varepsilon_1 (k - k_{F\uparrow}) \hspace{0.2cm}{\rm and}
\hspace{0.40cm} k = k_{F\uparrow} + q\hspace{0.2cm}{\rm where}
\nonumber \\
k & \in & [0, (k_{F\uparrow}-k_{F\downarrow})]
\hspace{0.2cm}{\rm for}\hspace{0.40cm}
q \in [-k_{F\uparrow},-k_{F\downarrow}] \, ,
\nonumber \\
{\bar{\omega}}^{+-}_1 (k) & = & - \varepsilon_1 (k_{F\uparrow}-k) \hspace{0.2cm}{\rm and}
\hspace{0.40cm} k = k_{F\uparrow} - q\hspace{0.2cm}{\rm where}
\nonumber \\
k & \in & [(k_{F\uparrow}-k_{F\downarrow}),\pi] 
\hspace{0.2cm}{\rm for}\hspace{0.40cm}
q \in [-k_{F\downarrow},k_{F\downarrow}] \, .
\label{OkMPRs}
\end{eqnarray}

The spectrum ${\tilde{\omega}}^{+-}_{1} (k)$ in the expression of $S^{+-} (k,\omega)$, Eq. (\ref{Sotherline}), 
of the branch line shown in Figs. \ref{figure4NPB},\ref{figure5NPB} that for small spin densities coincides with
the lower threshold of that component $n=1$ continuum for $k\in [2k_{F\downarrow},\pi]$ and for intermediate 
and large spin densities does not coincide with it, is given by,
\begin{equation}
{\tilde{\omega}}^{+-}_1 (k) = \varepsilon_1 (k - \pi - k_{F\downarrow}) \hspace{0.2cm}{\rm and}
\hspace{0.40cm} k = \pi + k_{F\downarrow} + q\hspace{0.2cm}{\rm where}
\hspace{0.40cm}k \in [2k_{F\downarrow},\pi]
\hspace{0.2cm}{\rm for}\hspace{0.40cm}
q \in [-k_{F\uparrow},-k_{F\downarrow}] \, .
\label{OkPMstar}
\end{equation}

The $2$-string gapped lower threshold spectrum ${\bar{\omega}}^{+-}_{2} (k)$ 
in the expression of $S^{+-} (k,\omega)$, Eq. (\ref{MPSs}) for $ab=+-$ and $n=2$,
is for $m>0$ divided into the following branch-line intervals,
\begin{eqnarray}
{\bar{\omega}}^{+-}_{2} (k) & = & \varepsilon_{2} (k) \hspace{0.40cm}{\rm and}\hspace{0.40cm}k = q
\hspace{0.40cm}{\rm where}
\nonumber \\
k & \in & [0,{\tilde{k}}_2[\hspace{0.40cm}{\rm and}\hspace{0.40cm} q\in [0,{\tilde{k}}_2[\hspace{0.40cm}
\hspace{0.40cm}{\rm for}\hspace{0.40cm}m\in]0,\bar{m}_2]
\nonumber \\
k & \in & [0,(k_{F\uparrow}-k_{F\downarrow})[\hspace{0.40cm}{\rm and}
\hspace{0.40cm}q\in [0,(k_{F\uparrow}-k_{F\downarrow})[
\hspace{0.40cm}{\rm for}\hspace{0.40cm}m\in [\bar{m}_2,\tilde{m}_2]
\nonumber \\
k & \in & [0,{\tilde{k}}_2[\hspace{0.40cm}{\rm and}\hspace{0.40cm} q\in [0,{\tilde{k}}_2[\hspace{0.40cm}
\hspace{0.40cm}{\rm for}\hspace{0.40cm}m\in[\tilde{m}_2,1[ \, ,
\label{Ds2}
\end{eqnarray}
\begin{eqnarray}
{\bar{\omega}}^{+-}_{2} (k) & = & \varepsilon_{2} (k_{F\uparrow} - k_{F\downarrow}) - \varepsilon_{1} (k_{F\uparrow}-k)
\hspace{0.40cm}{\rm and}
\hspace{0.40cm}k = k_{F\uparrow} - q
\hspace{0.40cm}{\rm where} 
\nonumber \\
k & \in & ](k_{F\uparrow}-k_{F\downarrow}),\tilde{k}_2[\hspace{0.40cm}{\rm and}
\hspace{0.40cm}q\in ](k_{F\uparrow}-\tilde{k}_2),k_{F\downarrow}[
\hspace{0.40cm}{\rm for}\hspace{0.40cm}m\in [\bar{m}_2,\tilde{m}_2] \, ,
\label{Dsppp1}
\end{eqnarray}
\begin{eqnarray}
{\bar{\omega}}^{+-}_{2} (k) & = & \varepsilon_{2} (0) - \varepsilon_{1} (k_{F\downarrow}-k) 
\hspace{0.40cm}{\rm and}\hspace{0.40cm}k = k_{F\downarrow}- q
\hspace{0.40cm}{\rm where} 
\nonumber \\
k & \in & \in ]{\tilde{k}}_2,2k_{F\downarrow}[\hspace{0.40cm}{\rm and}
\hspace{0.40cm}q\in ]-k_{F\downarrow},(k_{F\downarrow} - {\tilde{k}}_2)[
\hspace{0.40cm}{\rm for}\hspace{0.40cm}m\in ]0,1[ \, ,
\label{Dsppp2}
\end{eqnarray}
and
\begin{eqnarray}
{\bar{\omega}}^{+-}_{2} (k) & = & \varepsilon_{2} (k-2k_{F\downarrow}) \hspace{0.40cm}{\rm and}\hspace{0.40cm}
k = 2k_{F\downarrow} + q\hspace{0.40cm}{\rm where}
\nonumber \\
k & \in & ]2k_{F\downarrow},\pi[\hspace{0.40cm}{\rm and}
\hspace{0.40cm}q\in ]0,(k_{F\uparrow}-k_{F\downarrow})[
\hspace{0.40cm}{\rm for}\hspace{0.40cm}m\in ]0,1[ \, .
\label{Ds2p}
\end{eqnarray}

Finally, the $3$-string gapped lower threshold spectrum $\omega^{+-}_{3,l} (k)$
in the expression of $S^{+-} (k,\omega)$, Eq. (\ref{MPSs}) for $ab=+-$ and $n=3$, includes
for $m>0$ the following branch-line intervals,
\begin{eqnarray}
{\bar{\omega}}^{+-}_{3} (k) & = & \varepsilon_{3} (k - (k_{F\uparrow}-k_{F\downarrow})) 
\hspace{0.40cm}{\rm and}
\hspace{0.40cm}k = (k_{F\uparrow}-k_{F\downarrow}) + q
\hspace{0.40cm}{\rm where} 
\nonumber \\
k & \in & ]0,(k_{F\uparrow}-k_{F\downarrow})[
\hspace{0.40cm}{\rm and}\hspace{0.40cm}
q \in ]-(k_{F\uparrow}-k_{F\downarrow}),0[
\hspace{0.40cm}{\rm for}\hspace{0.40cm}m\in]0,1[ \, ,
\label{Ds3pP}
\end{eqnarray}
\begin{eqnarray}
{\bar{\omega}}^{+-}_{3} & = & \varepsilon_{3} (0) - \varepsilon_{1} \left(k_{F\uparrow} - k\right)
\hspace{0.40cm}{\rm and}\hspace{0.40cm}k = k_{F\uparrow}- q
\hspace{0.40cm}{\rm where} 
\nonumber \\
k & \in & ](k_{F\uparrow}-k_{F\downarrow}),(\pi - {\tilde{k}}_3)[
\hspace{0.40cm}{\rm and}\hspace{0.40cm}
q \in ]-(k_{F\downarrow} - {\tilde{k}}_3),k_{F\downarrow}[
\hspace{0.40cm}{\rm for}\hspace{0.40cm}m\in ]0,1[ \, ,
\label{Dsppp3P}
\end{eqnarray}
\begin{eqnarray}
{\bar{\omega}}^{+-}_{3}  & = & \varepsilon_{3} (k_{F\uparrow}-k_{F\downarrow})  - \varepsilon_{1} (k_{F\downarrow}-k)
\hspace{0.40cm}{\rm and}
\hspace{0.40cm}k = k_{F\downarrow} - q\hspace{0.40cm}{\rm where} 
\nonumber \\
k & \in & ](\pi - {\tilde{k}}_3),2k_{F\downarrow}[ 
\hspace{0.40cm}{\rm and}
\hspace{0.40cm}q \in ]-k_{F\downarrow},- (k_{F\uparrow}-\tilde{k}_3)[
\hspace{0.40cm}{\rm for}\hspace{0.40cm}m\in [\bar{m}_3,\tilde{m}_3] \, ,
\label{Dsppp1P3}
\end{eqnarray}
and
\begin{eqnarray}
{\bar{\omega}}^{+-}_{3}  (k) & = & \varepsilon_{3} (k-\pi) \hspace{0.40cm}{\rm and}\hspace{0.40cm}k = \pi + q
\hspace{0.40cm}{\rm where} 
\nonumber \\
k & \in & ](\pi - {\tilde{k}}_3),\pi[ 
\hspace{0.40cm}{\rm and}\hspace{0.40cm}q\in ]-{\tilde{k}}_3,0[
\hspace{0.40cm}{\rm for}\hspace{0.40cm}m\in]0,\bar{m}_3] 
\nonumber \\
k & \in & ]2k_{F\downarrow},\pi[
\hspace{0.40cm}{\rm and}\hspace{0.40cm}q\in ]-(k_{F\uparrow}-k_{F\downarrow}),0[
\hspace{0.40cm}{\rm for}\hspace{0.40cm}m\in [\bar{m}_3,\tilde{m}_3]
\nonumber \\
k & \in & ](\pi - {\tilde{k}}_3),\pi[ 
\hspace{0.40cm}{\rm and}\hspace{0.40cm}q\in ]-{\tilde{k}}_3,0[
\hspace{0.40cm}{\rm for}\hspace{0.40cm}m\in[\tilde{m}_3,1[ \, .
\label{Ds3P}
\end{eqnarray}

The corresponding $k$ dependent exponents that appear in the expression
of $S^{+-} (k,\omega)$, Eq. (\ref{MPSs}) for $ab=+-$ are given by,
\begin{eqnarray}
\zeta_{1}^{+-} (k) & = & -1 + \sum_{\iota =\pm 1}\left(- {\xi_{1\,1}\over 2} 
+ \Phi_{1,1}(\iota k_{F\downarrow},q)\right)^2 
\nonumber \\
&& {\rm for}\hspace{0.40cm}q=k-k_{F\uparrow}
\hspace{0.40cm}{\rm and}\hspace{0.40cm}k\in ]0,(k_{F\uparrow}-k_{F\downarrow})[
\nonumber \\
\zeta_{1}^{+-} (k) & = & -1 + \sum_{\iota =\pm 1}\left({\iota\over\xi_{1\,1}} - {\xi_{1\,1}\over 2} 
- \Phi_{1,1}(\iota k_{F\downarrow},q)\right)^2 
\nonumber \\
&& {\rm for}\hspace{0.40cm}q=k_{F\uparrow}-k
\hspace{0.40cm}{\rm and}\hspace{0.40cm}k\in ](k_{F\uparrow}-k_{F\downarrow}),\pi[ \, ,
\label{expsPM1}
\end{eqnarray}
for $n=1$,
\begin{eqnarray}
\zeta_{2}^{+-} (k) & = & -1 + \sum_{\iota=\pm 1}\left(- {\iota\over 2\xi_{1\,1}} 
+ \Phi_{1,2}(\iota k_{F\downarrow},q)\right)^2  
\hspace{0.40cm}{\rm for}\hspace{0.40cm}q=k\hspace{0.40cm}{\rm where}
\nonumber \\
k & \in & ]0,{\tilde{k}}_2[\hspace{0.40cm}{\rm for}\hspace{0.40cm}m\in]0,\bar{m}_2]
\nonumber \\
k & \in & ]0,(k_{F\uparrow}-k_{F\downarrow})[
\hspace{0.40cm}{\rm for}\hspace{0.40cm}m\in [\bar{m}_2,\tilde{m}_2]
\hspace{0.40cm}{\rm and}
\nonumber \\
k & \in & ]0,{\tilde{k}}_2[\hspace{0.40cm}{\rm for}\hspace{0.40cm}m\in[\tilde{m}_2,1[ 
\nonumber \\
\zeta_{2}^{+-} (k) & = & -1 +
\sum_{\iota=\pm 1}\Bigl({\xi_{1\,1}\over 2} 
+ \xi^{\iota,+}_2 - \Phi_{1,1}(\iota k_{F\downarrow},q)\Bigr)^2
\hspace{0.40cm}{\rm for}\hspace{0.40cm}q=k_{F\uparrow}-k
\hspace{0.40cm}{\rm where}
\nonumber \\
k & \in & ](k_{F\uparrow}-k_{F\downarrow}),\tilde{k}_2[
\hspace{0.40cm}{\rm for}\hspace{0.40cm}m\in [\bar{m}_2,\tilde{m}_2] 
\nonumber \\
\zeta_{2}^{+-} (k) & = & -1 + \sum_{\iota=\pm 1}
\left(\iota {\xi_{1\,2}^0\over 2} + {\xi_{1\,1}\over 2} - \Phi_{1,1}(\iota k_{F\downarrow},q)\right)^2
\hspace{0.40cm}{\rm for}\hspace{0.40cm}q=k_{F\downarrow}-k
\hspace{0.40cm}{\rm where}
\nonumber \\
k & \in & ]{\tilde{k}}_2,2k_{F\downarrow}[
\hspace{0.40cm}{\rm for}\hspace{0.40cm}m\in ]0,1[
\nonumber \\
\zeta_{2}^{+-} (k) & = & -1 + \sum_{\iota=\pm 1}
\left(- {\iota\over 2\xi_{1\,1}} + \xi_{1\,1}
+ \Phi_{1,2}(\iota k_{F\downarrow},q)\right)^2 
\hspace{0.40cm}{\rm for}\hspace{0.40cm}q=k-2k_{F\downarrow}
\hspace{0.40cm}{\rm where}
\nonumber \\
k & \in & ]2k_{F\downarrow},\pi[ 
\hspace{0.40cm}{\rm for}\hspace{0.40cm}m\in ]0,1[ \, ,
\label{expG+-}
\end{eqnarray}
for $n=2$ and,
\begin{eqnarray}
\zeta_{3}^{+-} (k) & = & -1 + \sum_{\iota=\pm 1}
\left(- {\iota\over\xi_{1\,1}} + \xi_{1\,1}
+ \Phi_{1,3}(\iota k_{F\downarrow},q)\right)^2 
\hspace{0.40cm}{\rm for}\hspace{0.40cm}q=k-k_{F\uparrow}+k_{F\downarrow}
\hspace{0.40cm}{\rm where}
\nonumber \\
k & \in & ]0,(k_{F\uparrow}-k_{F\downarrow})[
\hspace{0.40cm}{\rm for}\hspace{0.40cm}m\in ]0,1[ 
\nonumber \\
\zeta_{3}^{+-} (k) & = & -1 + \sum_{\iota=\pm 1}
\left(- {\iota\over 2\xi_{1\,1}} + \iota {\xi_{1\,3}^0\over 2} + {\xi_{1\,1}\over 2} 
- \Phi_{1,1}(\iota k_{F\downarrow},q)\right)^2
\hspace{0.40cm}{\rm for}\hspace{0.40cm}q=k_{F\uparrow}-k
\hspace{0.40cm}{\rm where}
\nonumber \\
k & \in &  ](k_{F\uparrow}-k_{F\downarrow}),(\pi - {\tilde{k}}_3)[
\hspace{0.40cm}{\rm for}\hspace{0.40cm}m\in ]0,1[
\nonumber \\
\zeta_{3}^{+-} (k) & = & -1 +
\sum_{\iota=\pm 1}\Bigl(- {\iota\over 2\xi_{1\,1}} + {\xi_{1\,1}\over 2} 
+ \xi^{\iota,+}_3 - \Phi_{1,1}(\iota k_{F\downarrow},q)\Bigr)^2
\hspace{0.40cm}{\rm for}\hspace{0.40cm}q=k_{F\downarrow}-k
\hspace{0.40cm}{\rm where}
\nonumber \\
k & \in & ](\pi - {\tilde{k}}_3),2k_{F\downarrow}[
\hspace{0.40cm}{\rm for}\hspace{0.40cm}m\in [\bar{m}_3,\tilde{m}_3] 
\nonumber \\
\zeta_{3}^{+-} (k) & = & -1 + \sum_{\iota=\pm 1}\left(- {\iota\over\xi_{1\,1}} 
+ \Phi_{1,3}(\iota k_{F\downarrow},q)\right)^2  
\hspace{0.40cm}{\rm for}\hspace{0.40cm}q=k-\pi
\hspace{0.40cm}{\rm where}
\nonumber \\
k & \in & ](\pi - {\tilde{k}}_3),\pi[
\hspace{0.40cm}{\rm for}\hspace{0.40cm}m\in ]0,\bar{m}_3]
\nonumber \\
k & \in & ]2k_{F\downarrow},\pi[
\hspace{0.40cm}{\rm for}\hspace{0.40cm}m\in [\bar{m}_3,\tilde{m}_3]
\hspace{0.40cm}{\rm and}
\nonumber \\
k & \in & ](\pi - {\tilde{k}}_3),\pi[
\hspace{0.40cm}{\rm for}\hspace{0.40cm}m\in[\tilde{m}_3,1[ \, ,
\label{expG3+-}
\end{eqnarray}
for $n=3$. 

Finally, the branch-line exponent in the expression of $S^{+-} (k,\omega)$, Eq. (\ref{Sotherline}), reads,
\begin{equation}
{\tilde\zeta}_{1}^{+-} (k) = -1 + \sum_{\iota =\pm 1}\left({\xi_{1\,1}\over 2} 
+ \Phi_{1,1}(\iota k_{F\downarrow},q)\right)^2 
\hspace{0.40cm}{\rm for}\hspace{0.40cm}q=k-\pi-k_{F\downarrow}
\hspace{0.40cm}{\rm and}\hspace{0.40cm}k\in ]2k_{F\downarrow},\pi[ \, .
\label{expstar}
\end{equation}

The phase shifts and the parameters $\xi_{1\,1}$, $\xi_{1\,n}^0$, $\xi^{\iota,\pm}_n$ where $n=2,3$ and $\iota =\pm 1$ appearing in the
above expressions and in the following are defined in Eq. (\ref{xpmnx0n}) and Eqs. (\ref{Phis-all-qq})-(\ref{xi1all}) of Appendix \ref{C}.

The exponents in Eqs. (\ref{expsPM1}), (\ref{expG+-}), (\ref{expG3+-}), and (\ref{expstar}) are plotted as a function of $k$
in Figs. \ref{figure4NPB},\ref{figure5NPB} for the $k$ intervals for which they are negative.

\subsection{Lower threshold spectra and exponents of $S^{-+} (k,\omega)$ for $m>0$}
\label{D2}

The $n=1$ lower threshold spectrum ${\bar{\omega}}^{-+}_{1} (k)$ 
in the expression of $S^{-+} (k,\omega)$, Eq. (\ref{MPSs}) for $ab=-+$ and $n=1$, reads,
\begin{equation}
{\bar{\omega}}^{-+}_1 (k) = - \varepsilon_1 (k_{F\uparrow}-k) \hspace{0.2cm}{\rm and}
\hspace{0.40cm} k = k_{F\uparrow} - q\hspace{0.2cm}{\rm where}\hspace{0.2cm}
k \in [(k_{F\uparrow}-k_{F\downarrow}),\pi]
\hspace{0.2cm}{\rm for}\hspace{0.40cm}
q \in ]-k_{F\downarrow},k_{F\downarrow}] \, .
\label{OkPMRs}
\end{equation}

The lower threshold spectra of the $n=2$ and $n=3$ continua of
$S^{-+} (k,\omega)$ are similar to those of $S^{+-} (k,\omega)$ given above. However, 
since the corresponding exponents are positive,
such continua have for $S^{-+} (k,\omega)$ nearly no spectral weight.
Only the exponent $\zeta_{1}^{-+} (k)$ is negative. 

In spite of their positivity for $n=2,3$, we give here the expressions of the exponents $\zeta_{n}^{-+} (k)$ for $n=1,2,3$.
They read,
\begin{equation}
\zeta_{1}^{-+} (k) = -1 + \sum_{\iota =\pm 1}\left(- {\xi_{1\,1}\over 2} 
- \Phi_{1,1}(\iota k_{F\downarrow},q)\right)^2 
\hspace{0.40cm}{\rm for}\hspace{0.40cm}q=k_{F\uparrow}-k
\hspace{0.40cm}{\rm where}\hspace{0.40cm}
k \in ](k_{F\uparrow}-k_{F\downarrow}),\pi[ \, ,
\label{expsMP1}
\end{equation}
for $n=1$,
\begin{eqnarray}
\zeta_{2}^{-+} (k) & = & -1 + \sum_{\iota=\pm 1}\left(- {\iota 3\over 2\xi_{1\,1}} 
+ \Phi_{1,2}(\iota k_{F\downarrow},q)\right)^2  
\hspace{0.40cm}{\rm for}\hspace{0.40cm}q=k\hspace{0.40cm}{\rm where}
\nonumber \\
k & \in & ]0,{\tilde{k}}_2[\hspace{0.40cm}{\rm for}\hspace{0.40cm}m\in]0,\bar{m}_2]
\nonumber \\
k & \in & ]0,(k_{F\uparrow}-k_{F\downarrow})[
\hspace{0.40cm}{\rm for}\hspace{0.40cm}m\in [\bar{m}_2,\tilde{m}_2]
\hspace{0.40cm}{\rm and}
\nonumber \\
k & \in & ]0,{\tilde{k}}_2[\hspace{0.40cm}{\rm for}\hspace{0.40cm}m\in[\tilde{m}_2,1[ 
\nonumber \\
\zeta_{2}^{-+} (k) & = & -1 +
\sum_{\iota=\pm 1}\Bigl(- {\iota\over\xi_{1\,1}} + {\xi_{1\,1}\over 2} 
+ \xi^{\iota,+}_2 - \Phi_{1,1}(\iota k_{F\downarrow},q)\Bigr)^2
\hspace{0.40cm}{\rm for}\hspace{0.40cm}q=k_{F\uparrow}-k
\hspace{0.40cm}{\rm where}
\nonumber \\
k & \in & ](k_{F\uparrow}-k_{F\downarrow}),\tilde{k}_2[
\hspace{0.40cm}{\rm for}\hspace{0.40cm}m\in [\bar{m}_2,\tilde{m}_2] 
\nonumber \\
\zeta_{2}^{-+} (k) & = & -1 + \sum_{\iota=\pm 1}
\left(-{\iota\over\xi_{1\,1}} + \iota {\xi_{1\,2}^0\over 2} + {\xi_{1\,1}\over 2} 
- \Phi_{1,1}(\iota k_{F\downarrow},q)\right)^2
\hspace{0.40cm}{\rm for}\hspace{0.40cm}q=k_{F\downarrow}-k
\hspace{0.40cm}{\rm where}
\nonumber \\
k & \in & ]{\tilde{k}}_2,2k_{F\downarrow}[
\hspace{0.40cm}{\rm for}\hspace{0.40cm}m\in ]0,1[
\nonumber \\
\zeta_{2}^{-+} (k) & = & -1 + \sum_{\iota=\pm 1}
\left(- {\iota 3\over 2\xi_{1\,1}} + \xi_{1\,1}
+ \Phi_{1,2}(\iota k_{F\downarrow},q)\right)^2 
\hspace{0.40cm}{\rm for}\hspace{0.40cm}q=k-2k_{F\downarrow}
\hspace{0.40cm}{\rm where}
\nonumber \\
k & \in & ]2k_{F\downarrow},\pi[ \hspace{0.40cm}{\rm for}\hspace{0.40cm}m\in ]0,1[ \, ,
\label{expG-+}
\end{eqnarray}
for $n=2$ and,
\begin{eqnarray}
\zeta_{3}^{-+} (k) & = & -1 + \sum_{\iota=\pm 1}
\left(- {\iota 2\over\xi_{1\,1}} + \xi_{1\,1}
+ \Phi_{1,3}(\iota k_{F\downarrow},q)\right)^2  
\hspace{0.40cm}{\rm for}\hspace{0.40cm}q=k-k_{F\uparrow}+k_{F\downarrow}
\hspace{0.40cm}{\rm where}
\nonumber \\
k & \in & ]0,(k_{F\uparrow}-k_{F\downarrow})[
\hspace{0.40cm}{\rm for}\hspace{0.40cm}m\in ]0,1[ 
\nonumber \\
\zeta_{3}^{-+} (k) & = & -1 + \sum_{\iota=\pm 1}
\left(-{\iota\,3\over 2\xi_{1\,1}} + \iota {\xi_{1\,3}^0\over 2} + {\xi_{1\,1}\over 2} 
- \Phi_{1,1}(\iota k_{F\downarrow},q)\right)^2
\hspace{0.40cm}{\rm for}\hspace{0.40cm}q=k_{F\uparrow}-k
\hspace{0.40cm}{\rm where}
\nonumber \\
k & \in &  ](k_{F\uparrow}-k_{F\downarrow}),(\pi - {\tilde{k}}_3)[
\hspace{0.40cm}{\rm for}\hspace{0.40cm}m\in ]0,1[
\nonumber \\
\zeta_{3}^{-+} (k) & = & -1 +
\sum_{\iota=\pm 1}\Bigl(- {\iota\,3\over 2\xi_{1\,1}} + {\xi_{1\,1}\over 2} 
+ \xi^{\iota,+}_3 - \Phi_{1,1}(\iota k_{F\downarrow},q)\Bigr)^2
\hspace{0.40cm}{\rm for}\hspace{0.40cm}q=k_{F\downarrow}-k
\hspace{0.40cm}{\rm where}
\nonumber \\
k & \in & ](\pi - {\tilde{k}}_3),2k_{F\downarrow}[
\hspace{0.40cm}{\rm for}\hspace{0.40cm}m\in [\bar{m}_3,\tilde{m}_3] 
\nonumber \\
\zeta_{3}^{-+} (k) & = & -1 + \sum_{\iota=\pm 1}\left(- {\iota 2\over\xi_{1\,1}} 
+ \Phi_{1,n}(\iota k_{F\downarrow},q)\right)^2 
\hspace{0.40cm}{\rm for}\hspace{0.40cm}q=k-\pi
\hspace{0.40cm}{\rm where}
\nonumber \\
k & \in & ](\pi - {\tilde{k}}_3),\pi[
\hspace{0.40cm}{\rm for}\hspace{0.40cm}m\in ]0,\bar{m}_3]
\nonumber \\
k & \in & ]2k_{F\downarrow},\pi[
\hspace{0.40cm}{\rm for}\hspace{0.40cm}m\in [\bar{m}_3,\tilde{m}_3]
\hspace{0.40cm}{\rm and}
\nonumber \\
k & \in & ](\pi - {\tilde{k}}_3),\pi[
\hspace{0.40cm}{\rm for}\hspace{0.40cm}m\in[\tilde{m}_3,1[ \, .
\label{expG3-+}
\end{eqnarray}

The exponent $\zeta_{1}^{-+} (k)$, Eq. (\ref{expsMP1}), is plotted as a function of $k$
in Figs. \ref{figure6NPB},\ref{figure7NPB}.

\subsection{Lower threshold spectra and exponents of $S^{zz} (k,\omega)$ for $m>0$}
\label{D3}

Although the $n=3$ continuum of $S^{zz} (k,\omega)$ does not contain a significant amount of spectral weight,
here we provide the $n=1$, $n=2$, and $n=3$ lower threshold's spectra ${\bar{\omega}}^{zz}_{n} (k)$ 
in the expression of that dynamical structure factor component, Eq. (\ref{MPSs}) for $ab=zz$ and $n=1,2,3$. They
include several branch-line parts and are given by,
\begin{eqnarray}
{\bar{\omega}}^{zz}_{1} (k) & = & - \varepsilon_1 (k_{F\downarrow} - k) \hspace{0.2cm}{\rm and}
\hspace{0.40cm} k = k_{F\downarrow} - q \hspace{0.2cm}{\rm where}
\nonumber \\
k & \in &[0,2k_{F\downarrow}]\hspace{0.2cm}{\rm for}
\hspace{0.40cm} q \in [-k_{F\downarrow},k_{F\downarrow}] \, ,
\nonumber \\
{\bar{\omega}}^{zz}_{1} (k) & = & \varepsilon_1 (k - k_{F\downarrow}) \hspace{0.2cm}{\rm and}
\hspace{0.40cm} k = k_{F\downarrow} + q \hspace{0.2cm}{\rm where}
\nonumber \\
k & \in &[2k_{F\downarrow},\pi]
\hspace{0.2cm}{\rm for}\hspace{0.40cm}
q \in [k_{F\downarrow},k_{F\uparrow}] \, ,
\label{OkPMRsL}
\end{eqnarray}
\begin{eqnarray}
{\bar{\omega}}^{zz}_{2} (k) & = & \varepsilon_{2} (k - (k_{F\uparrow}-k_{F\downarrow})) 
\hspace{0.40cm}{\rm and}\hspace{0.40cm}k = (k_{F\uparrow}-k_{F\downarrow}) + q
\hspace{0.40cm}{\rm where} 
\nonumber \\
k & \in & ]0,(k_{F\uparrow}-k_{F\downarrow})[
\hspace{0.40cm}{\rm and}\hspace{0.40cm}
q \in ]-(k_{F\uparrow}-k_{F\downarrow}),0[
\hspace{0.40cm}{\rm for}\hspace{0.40cm}m\in]0,1[ \, ,
\label{Ds2pL}
\end{eqnarray}
\begin{eqnarray}
{\bar{\omega}}^{zz}_{2} & = & \varepsilon_{2} (0) - \varepsilon_{1} \left(k_{F\uparrow} - k\right)
\hspace{0.40cm}{\rm and}\hspace{0.40cm}k = k_{F\uparrow}- q
\hspace{0.40cm}{\rm where} 
\nonumber \\
k & \in & ](k_{F\uparrow}-k_{F\downarrow}),(\pi - {\tilde{k}}_2)[
\hspace{0.40cm}{\rm and}\hspace{0.40cm}
q \in ]-(k_{F\downarrow} - {\tilde{k}}_2),k_{F\downarrow}[
\hspace{0.40cm}{\rm for}\hspace{0.40cm}m\in ]0,1[ \, ,
\label{Dsppp2L}
\end{eqnarray}
\begin{eqnarray}
{\bar{\omega}}^{zz}_{2}  & = & \varepsilon_{2} (-(k_{F\uparrow}-k_{F\downarrow}))  - \varepsilon_{1} (k_{F\downarrow}-k)
\hspace{0.40cm}{\rm and}\hspace{0.40cm}k = k_{F\downarrow} - q\hspace{0.40cm}{\rm where} 
\nonumber \\
k & \in & ](\pi - {\tilde{k}}_2),2k_{F\downarrow}[ 
\hspace{0.40cm}{\rm and}\hspace{0.40cm}
q \in ]-k_{F\downarrow},- (k_{F\uparrow}-\tilde{k}_2)[
\hspace{0.40cm}{\rm for}\hspace{0.40cm}m\in [\bar{m}_2,\tilde{m}_2] \, ,
\label{Dsppp1L}
\end{eqnarray}
and
\begin{eqnarray}
{\bar{\omega}}^{zz}_{2}  (k) & = & \varepsilon_{2} (k-\pi) \hspace{0.40cm}{\rm and}\hspace{0.40cm}k = \pi + q
\hspace{0.40cm}{\rm where} 
\nonumber \\
k & \in & ](\pi - {\tilde{k}}_2),\pi[ 
\hspace{0.40cm}{\rm and}\hspace{0.40cm}q\in ]-{\tilde{k}}_2,0[
\hspace{0.40cm}{\rm for}\hspace{0.40cm}m\in]0,\bar{m}_2] 
\nonumber \\
k & \in & ]2k_{F\downarrow},\pi[
\hspace{0.40cm}{\rm and}\hspace{0.40cm}q\in ]-(k_{F\uparrow}-k_{F\downarrow}),0[
\hspace{0.40cm}{\rm for}\hspace{0.40cm}m\in [\bar{m}_2,\tilde{m}_2]
\nonumber \\
k & \in & ](\pi - {\tilde{k}}_2),\pi[ 
\hspace{0.40cm}{\rm and}\hspace{0.40cm}q\in ]-{\tilde{k}}_2,0[
\hspace{0.40cm}{\rm for}\hspace{0.40cm}m\in[\tilde{m}_2,1[ \, ,
\label{Ds2L}
\end{eqnarray}

\begin{eqnarray}
{\bar{\omega}}^z_{3} (k) & = & \varepsilon_{3} (k) \hspace{0.40cm}{\rm and}\hspace{0.40cm}k = q
\hspace{0.40cm}{\rm where}
\nonumber \\
k & \in & [0,{\tilde{k}}_3[\hspace{0.40cm}{\rm and}\hspace{0.40cm} q\in [0,{\tilde{k}}_3[\hspace{0.40cm}
\hspace{0.40cm}{\rm for}\hspace{0.40cm}m\in]0,\bar{m}_3]
\nonumber \\
k & \in & [0,(k_{F\uparrow}-k_{F\downarrow})[\hspace{0.40cm}{\rm and}
\hspace{0.40cm}q\in [0,(k_{F\uparrow}-k_{F\downarrow})[
\hspace{0.40cm}{\rm for}\hspace{0.40cm}m\in [\bar{m}_3,\tilde{m}_3]
\nonumber \\
k & \in & [0,{\tilde{k}}_3[\hspace{0.40cm}{\rm and}\hspace{0.40cm} q\in [0,{\tilde{k}}_3[\hspace{0.40cm}
\hspace{0.40cm}{\rm for}\hspace{0.40cm}m\in[\tilde{m}_3,1[ \, ,
\label{DsL3}
\end{eqnarray}
\begin{eqnarray}
{\bar{\omega}}^z_{3} (k) & = & \varepsilon_{3} (k_{F\uparrow} - k_{F\downarrow}) - \varepsilon_{1} (k_{F\uparrow}-k)
\hspace{0.40cm}{\rm and}\hspace{0.40cm}k = k_{F\uparrow} - q
\hspace{0.40cm}{\rm where} 
\nonumber \\
k & \in & ](k_{F\uparrow}-k_{F\downarrow}),\tilde{k}_3[\hspace{0.40cm}{\rm and}
\hspace{0.40cm}q\in ](k_{F\uparrow}-\tilde{k}_3),k_{F\downarrow}[
\hspace{0.40cm}{\rm for}\hspace{0.40cm}m\in [\bar{m}_3,\tilde{m}_3] \, ,
\label{Dspp3L1}
\end{eqnarray}
\begin{eqnarray}
{\bar{\omega}}^z_{3} (k) & = & \varepsilon_{3} (0) - \varepsilon_{1} (k_{F\downarrow}-k) 
\hspace{0.40cm}{\rm and}\hspace{0.40cm}k = k_{F\downarrow}- q
\hspace{0.40cm}{\rm where} 
\nonumber \\
k & \in & \in ]{\tilde{k}}_3,2k_{F\downarrow}[\hspace{0.40cm}{\rm and}
\hspace{0.40cm}q\in ]-k_{F\downarrow},(k_{F\downarrow} - {\tilde{k}}_3)[
\hspace{0.40cm}{\rm for}\hspace{0.40cm}m\in ]0,1[ \, ,
\label{DsppL3}
\end{eqnarray}
\begin{eqnarray}
{\bar{\omega}}^z_{3} (k) & = & \varepsilon_{3} (k-2k_{F\downarrow}) \hspace{0.40cm}{\rm and}\hspace{0.40cm}
k = 2k_{F\downarrow} + q\hspace{0.40cm}{\rm where}
\nonumber \\
k & \in & ]2k_{F\downarrow},\pi[\hspace{0.40cm}{\rm and}
\hspace{0.40cm}q\in ]0,(k_{F\uparrow}-k_{F\downarrow})[
\hspace{0.40cm}{\rm for}\hspace{0.40cm}m\in ]0,1[ \, .
\label{Ds3L}
\end{eqnarray}

In spite of the corresponding $k$ dependent exponent $\zeta_{3}^{zz} (k)$ being positive for its
whole $k$ interval and the exponent $\zeta_{2}^{zz} (k)$ becoming
positive upon increasing $m$ and $h$, we provide in the following the expressions
of the exponents $\zeta_{n}^{zz} (k)$ in the expression of $S^{zz} (k,\omega)$, Eq. (\ref{MPSs}) for $ab=zz$ and $n=1,2,3$, 
independently of their values. They are found to read,
\begin{eqnarray}
\zeta_{1}^{zz} (k) & = & -1 +
\sum_{\iota =\pm 1}\left({\iota\over 2\xi_{1\,1}}  + {\xi_{1\,1}\over 2} 
- \Phi_{1,1}(\iota k_{F\downarrow},q)\right)^2
\hspace{0.40cm}{\rm for}\hspace{0.40cm}q=k_{F\downarrow}-k
\hspace{0.40cm}{\rm where}
\nonumber \\
k & \in & ]0,2k_{F\downarrow}[
\nonumber \\
\zeta_{1}^{zz} (k) & = & -1 + \sum_{\iota =\pm 1}\left(- {\iota\over 2\xi_{1\,1}} + {\xi_{1\,1}\over 2} 
+ \Phi_{1,1}(\iota k_{F\downarrow},q)\right)^2 
\hspace{0.40cm}{\rm for}\hspace{0.40cm}q=k-k_{F\downarrow}
\hspace{0.40cm}{\rm where}
\nonumber \\
k & \in & ]2k_{F\downarrow},\pi[ \, ,
\label{expszz1}
\end{eqnarray}
for $n=1$,
\begin{eqnarray}
\zeta_{2}^{zz} (k) & = & -1 + \sum_{\iota=\pm 1}
\left(- {\iota\over\xi_{1\,1}} - \xi_{1\,1}
+ \Phi_{1,2}(\iota k_{F\downarrow},q)\right)^2 
\hspace{0.40cm}{\rm for}\hspace{0.40cm}q=k-k_{F\uparrow}+k_{F\downarrow}
\hspace{0.40cm}{\rm where}
\nonumber \\
k & \in & ]0,(k_{F\uparrow}-k_{F\downarrow})[
\hspace{0.40cm}{\rm for}\hspace{0.40cm}m\in ]0,1[ 
\nonumber \\
\zeta_{2}^{zz} (k) & = & -1 + \sum_{\iota=\pm 1}
\left(- {\iota\over 2\xi_{1\,1}} 
+ \iota {\xi_{1\,2}^0\over 2} - {\xi_{1\,1}\over 2} 
- \Phi_{1,1}(\iota k_{F\downarrow},q)\right)^2
\hspace{0.40cm}{\rm for}\hspace{0.40cm}q=k_{F\uparrow}-k
\hspace{0.40cm}{\rm where}
\nonumber \\
k & \in &  ](k_{F\uparrow}-k_{F\downarrow}),(\pi - {\tilde{k}}_2)[
\hspace{0.40cm}{\rm for}\hspace{0.40cm}m\in ]0,1[
\nonumber \\
\zeta_{2}^{zz} (k) & = & -1 + \sum_{\iota=\pm 1}
\Bigl(-{\iota\over 2\xi_{1\,1}} - {\xi_{1\,1}\over 2}
+ \xi^{\iota,-}_2 - \Phi_{1,1}(\iota k_{F\downarrow},q)\Bigr)^2 
\hspace{0.40cm}{\rm for}\hspace{0.40cm}q=k_{F\downarrow}-k
\hspace{0.40cm}{\rm where}
\nonumber \\
k & \in & ](\pi - {\tilde{k}}_2),2k_{F\downarrow}[
\hspace{0.40cm}{\rm for}\hspace{0.40cm}m\in [\bar{m}_2,\tilde{m}_2] 
\nonumber \\
\zeta_{2}^{zz} (k) & = & -1 + \sum_{\iota=\pm 1}\left(- {\iota\over\xi_{1\,1}} 
+ \Phi_{1,2}(\iota k_{F\downarrow},q)\right)^2 
\hspace{0.40cm}{\rm for}\hspace{0.40cm}q=k-\pi
\hspace{0.40cm}{\rm where}
\nonumber \\
k & \in & ](\pi - {\tilde{k}}_2),\pi[
\hspace{0.40cm}{\rm for}\hspace{0.40cm}m\in ]0,\bar{m}_2]
\nonumber \\
k & \in & ]2k_{F\downarrow},\pi[
\hspace{0.40cm}{\rm for}\hspace{0.40cm}m\in [\bar{m}_2,\tilde{m}_2]
\hspace{0.40cm}{\rm and}
\nonumber \\
k & \in & ](\pi - {\tilde{k}}_2),\pi[
\hspace{0.40cm}{\rm for}\hspace{0.40cm}m\in[\tilde{m}_2,1[ \, ,
\label{exps2pL}
\end{eqnarray}
for $n=2$ and,
\begin{eqnarray}
\zeta_{3}^{zz} (k) & = & -1 + \sum_{\iota=\pm 1}\left(- {\iota\,3\over 2\xi_{1\,1}} 
+ \Phi_{1,3}(\iota k_{F\downarrow},q)\right)^2 
\hspace{0.40cm}{\rm for}\hspace{0.40cm}q=k\hspace{0.40cm}{\rm where}
\nonumber \\
k & \in & ]0,{\tilde{k}}_3[\hspace{0.40cm}{\rm for}\hspace{0.40cm}m\in]0,\bar{m}_3]
\nonumber \\
k & \in & ]0,(k_{F\uparrow}-k_{F\downarrow})[
\hspace{0.40cm}{\rm for}\hspace{0.40cm}m\in [\bar{m}_3,\tilde{m}_3]
\hspace{0.40cm}{\rm and}
\nonumber \\
k & \in & ]0,{\tilde{k}}_3[\hspace{0.40cm}{\rm for}\hspace{0.40cm}m\in[\tilde{m}_3,1[ 
\nonumber \\
\zeta_{3}^{zz} (k) & = & -1 - \sum_{\iota=\pm 1}
\Bigl({\iota\over\xi_{1\,1}} + {\xi_{1\,1}\over 2} 
+ \xi^{\iota,+}_3 - \Phi_{1,1}(\iota k_{F\downarrow},q)\Bigr)^2 
\hspace{0.40cm}{\rm for}\hspace{0.40cm}q=k_{F\uparrow}-k
\hspace{0.40cm}{\rm where}
\nonumber \\
k & \in & ](k_{F\uparrow}-k_{F\downarrow}),\tilde{k}_3[
\hspace{0.40cm}{\rm for}\hspace{0.40cm}m\in [\bar{m}_3,\tilde{m}_3] 
\nonumber \\
\zeta_{3}^{zz} (k) & = & -1 + \sum_{\iota=\pm 1}
\left(- {\iota\over\xi_{1\,1}} 
+ \iota {\xi_{1\,3}^0\over 2} - {\xi_{1\,1}\over 2} 
- \Phi_{1,1}(\iota k_{F\downarrow},q)\right)^2
\hspace{0.40cm}{\rm for}\hspace{0.40cm}q=k_{F\downarrow}-k
\hspace{0.40cm}{\rm where}
\nonumber \\
k & \in & ]{\tilde{k}}_3,2k_{F\downarrow}[
\hspace{0.40cm}{\rm for}\hspace{0.40cm}m\in ]0,1[
\nonumber \\
\zeta_{3}^{zz} (k) & = & -1 + \sum_{\iota=\pm 1}
\left(- {\iota\,3\over 2\xi_{1\,1}} - \xi_{1\,1}
+ \Phi_{1,3}(\iota k_{F\downarrow},q)\right)^2 
\hspace{0.40cm}{\rm for}\hspace{0.40cm}q=k-2k_{F\downarrow}
\hspace{0.40cm}{\rm where}
\nonumber \\
k & \in & ]2k_{F\downarrow},\pi[ 
\hspace{0.40cm}{\rm for}\hspace{0.40cm}m\in ]0,1[ \, ,
\label{exps3pL}
\end{eqnarray}
for $n=3$.

The exponent $\zeta_{1}^{zz} (k)$ in Eq. (\ref{expszz1}) and $\zeta_{2}^{zz} (k)$ in Eq. (\ref{exps2pL}) are plotted as a function of $k$
in Figs. \ref{figure8NPB},\ref{figure9NPB} for the $k$ intervals for which they are negative.

\end{document}